# Next-to-leading-order QCD calculations of $B \to A$ form factors with higher-twist corrections

Fang-Zhou Di [*] and Yu-Hao Liu [†]

*School of Physics, Nankai University, 300071 Tianjin, China*

Applying the light-cone sum rules approach, the next-to-leading-order corrections to $B \to A$ form factors are calculated with the twist-two and twist-three $B$-meson light-cone distribution amplitudes (LCDAs) at leading power in $\Lambda/m_b$. At one-loop level, the vacuum-to-$B$-meson correlation functions defined with interpolating currents for the $p$-wave axial-vector meson are factorized into short-distance coefficients and the distribution amplitudes of the $B$ meson, among which the hard coefficients for A0-type currents and $\mu$-dependent jet functions entering correlation functions $\Pi^{(T+\tilde{T})}_{\mu,\parallel}$, $\Pi^{(V-A)}_{\delta\mu,\perp}$, and $\Pi^{(T+\tilde{T})}_{\delta\mu,\perp}$ are identical to the corresponding ones of $B \to V$ in soft-collinear effective theory (SCET). In addition, $\Pi^{(V-A)}_{\mu,\parallel}$ and $\Pi^{(T+\tilde{T})}_{\mu,\parallel}$ coincide with the results of $B \to \pi, K$ under the limit $m_q \to 0$ up to one-loop accuracy. Then the subleading-power corrections to $B \to A$ form factors are computed with the higher-twist $B$-meson LCDAs at tree level up to the twist-six accuracy. Furthermore, we predict the $q^2$ dependence of $B \to A$ form factors via the Bourrely-Caprini-Lellouch $z$-series expanding parametrization to compute several physical observables including branching ratios of semileptonic $B \to A\ell\bar{\nu}_\ell$ and $B \to A\nu_\ell\bar{\nu}_\ell$ processes, transverse asymmetries, forward-backward asymmetries, and lepton-flavor universality observables employing the given mixing angles $\theta_K = -34°$, $\theta_{1_{P_1}} = 28°$, and $\theta_{3_{P_1}} = 23°$. We also compare our theory calculations with the results from other methods.

## I. INTRODUCTION

Precision calculations of the semileptonic $B \to A$ form factors play an indispensable role in determinating the Cabibbo-Kobayashi-Maskawa (CKM) matrix elements exclusively and the theory descriptions of the electroweak penguin $B \to A\ell\bar{\ell}$ in QCD. The probe of heavy-to-light $B$-meson decays provides a unique aspect to further understand the strong interaction dynamics of the heavy-hadron system and pursue the hints of physics beyond the Standard Model. On the one hand, the semileptonic processes of the $B$ meson decaying into either pseudoscalar or vector mesons have been the focus of extensive research. On the other hand, there is a notable paucity of investigations into the orbital excitations or other excited states generated by $B$-meson decays. Such studies will offer valuable insights beyond the extensively explored $B \to P$ and $B \to V$ transitions, where $P$ and $V$ represent pseudoscalar and vector mesons, respectively. The axial-vector meson is regarded as one of the most significant excited-state light mesons so that the exploration of the semileptonic $B \to A$ ($A$ denotes light axial-vector mesons) transitions is of considerable necessity and significance.

In the quark model, there are two types of $p$-wave axial-vector mesons, namely, $1^3P_1$ and $1^1P_1$ states due to different spin couplings of the two quarks corresponding to $J^{PC} = 1^{++}$ and $1^{+-}$, respectively. In this study, we focus on the $1^{++}$ mesons $a_1(1260)$, $f_1(1285)$, $f_1(1420)$, and $K_{1A}$, and the $1^{+-}$ mesons $b_1(1235)$, $h_1(1170)$, $h_1(1415)$, and $K_{1B}$, among which we adopt the $u\bar{u}$ and $s\bar{s}$ contents of $f_1(1285)$, $h_1(1170)$ and $f_1(1420)$, $h_1(1415)$ [1,2], respectively, since the data of decay modes for $h_1$ are not available in [3]. The physical mass eigenstates $K_1(1270)$, $K_1(1400)$ are obtained from the mixture of $K_{1A}$ and $K_{1B}$ at the mixing angle $\theta_K$, whose value has been discussed in detail in a number of previous studies [4–12]. We will choose $\theta_K = -34°$, the $f_1(1285) - f_1(1420)$ mixing angle $\theta_{3_{P_1}} = 23°$, and the $h_1(1170) - h_1(1415)$ mixing angle $\theta_{1_{P_1}} = 28°$ because this set of angles is consistent with our assumptions about the abovementioned compositions of $h_1$ and $f_1$ and close to the ranges predicted by most relative investigations [5–10,12–15].

[*] Contact author: difangzhou@mail.nankai.edu.cn
[†] Contact author: liuyh@mail.nankai.edu.cn

The primary objective of probing the semileptonic decays of $B$ mesons into axial-vector mesons is to calculate the $B \to A$ transition form factors. There is much previous research on evaluating the $B \to A$ transition form factors with light-cone sum rules (LCSR) [16–20] where three-particle $B$-meson distribution amplitudes (DAs) are utilized in [16], and axial-vector DAs up to twist-four are employed in [17–20]. The LCSR approach has been rapidly developed as a result of its distinctive advantage in analyzing heavy-to-light $B$-meson form factors [21–31] in the large recoil region. Another available methodology for exploring the heavy-to-light $B$-meson form factors is the perturbative QCD (pQCD) approach, which is based on transverse-momentum-dependent (TMD) QCD factorization for hard processes and developed from [32,33]. Several studies have accomplished next-to-leading-order (NLO) twist-two [34,35] and twist-three [36] corrections to $B \to \pi$ form factors applying the TMD factorization approach and [37] have proposed a new definition of a TMD wave function with simpler soft subtraction for $k_T$ factorization of hard exclusive processes, albeit a complete understanding of TMD factorization for exclusive processes with large momentum transfer has not been achieved conceptually until now.

In this work, we will calculate the semileptonic $B \to A$ transition form factors in order to examine the semileptonic decays $B \to A\ell\nu$ and $B \to A\nu\bar{\nu}$ by applying the LCSR method [38–41] with the light-cone distribution amplitude (LCDA) of $B$ mesons. Higher-order perturbative corrections to correlation functions as well as contributions from subleading-power effects have been systematically calculated within the framework of heavy-to-light $B$-meson decay form factors [42–49], heavy-to-heavy $B$-meson decay form factors [45,50–52], and semileptonic heavy-baryon decay form factors [53–55] with the aid of this method. The fundamental strategy to achieve the calculation of the $B \to A$ transition form factors with higher-twist corrections resembles that of Refs. [42–44,46–49]. Taking this strategy into account, we will adopt the method of regions to compute the QCD corrections to the correlation function by extracting the hard functions and jet functions, and calculate the tree-level two-particle and three-particle corrections with higher-twist $B$-meson DAs up to next-to-leading-power (NLP) accuracy.

The layout of this paper is as follows: In Sec. II, we review the definitions of the $B \to A$ form factors and construct the LCSR for these form factors at tree level with the vacuum-to-$B$-meson correlation functions. In Sec. III, we establish the factorization of the correlation functions and calculate short-distance coefficients using the method of regions at NLO at leading power. We show the next-to-leading-logarithmic (NLL) resummation improved LCSR for the $B \to A$ form factors in Sec. IV. The subleading-power contributions to the $B \to A$ form factors are computed in detail with the higher-twist $B$-meson LCDAs at tree level in Sec. V. We manage to match our LCSR calculations with the $z$-series parametrizations in Sec. VI in order to display the predictions for the $B \to A$ form factors in the entire kinematic region. In this section, we also present some predicted results including branching ratios and relevant observables. Finally, we will discuss our main observations and perspectives on future developments in Sec. VII.

## II. THE $B$-MESON LCSR F $B \to A$ FORM FACTORS AT TREE LEVEL

### A. Definition of $B \to A$ form factors

According to previous studies [2,56], we can readily write down the QCD matrix elements of the heavy-to-light currents in terms of the semileptonic $B \to A$ form factors as (see [57] for alternative definitions of the $B \to A$ form factors)

$$
\begin{aligned}
\langle A(p,\epsilon^*)|\bar{q}\gamma_\mu\gamma_5 b|\bar{B}(p_B)\rangle &= -\frac{2iA(q^2)}{m_B - m_A}\epsilon_{\mu\nu\rho\sigma}\epsilon^{*\nu}p_B^\rho p^\sigma, \\
\langle A(p,\epsilon^*)|\bar{q}\gamma_\mu b|\bar{B}(p_B)\rangle &= -2m_A V_0(q^2)\frac{\epsilon^*\cdot q}{q^2}q_\mu - (m_B - m_A)V_1(q^2)\left[\epsilon^*_\mu - \frac{\epsilon^*\cdot q}{q^2}q_\mu\right] \\
&\quad + V_2(q^2)\frac{\epsilon^*\cdot q}{m_B - m_A}\left[(p_B + p)_\mu - \frac{m_B^2 - m_A^2}{q^2}q_\mu\right], \\
\langle A(p,\epsilon^*)|\bar{q}\sigma_{\mu\nu}\gamma_5 q^\nu b|\bar{B}(p_B)\rangle &= -2T_1(q^2)\epsilon_{\mu\nu\rho\sigma}\epsilon^{*\nu}p_B^\rho p^\sigma, \\
\langle A(p,\epsilon^*)|\bar{q}\sigma_{\mu\nu}q^\nu b|\bar{B}(p_B)\rangle &= -iT_2(q^2)[(m_B^2 - m_A^2)\epsilon^*_\mu - (\epsilon^*\cdot q)(p_B + p)_\mu] \\
&\quad - iT_3(q^2)(\epsilon^*\cdot q)\left[q_\mu - \frac{q^2}{m_B^2 - m_A^2}(p_B + p)_\mu\right],
\end{aligned}
\tag{2.1}
$$

where $m_A$, $p$, and $\epsilon^*$ represent the mass, momentum, and polarization vector of light axial-vector mesons, respectively, and $q$ is the transfer momentum of weak transition current. We adopt the convention $\epsilon_{0123} = -1$. The pseudoscalar matrix element $\langle A(p,\epsilon^*)|\bar{q}\gamma_5 b|\bar{B}(p_B)\rangle$ vanishes due to the fact that

$$q^\mu \langle A(p, \epsilon^*)|\bar{q}\gamma_\mu \gamma_5 b|\bar{B}(p_B)\rangle = 0. \tag{2.2}$$

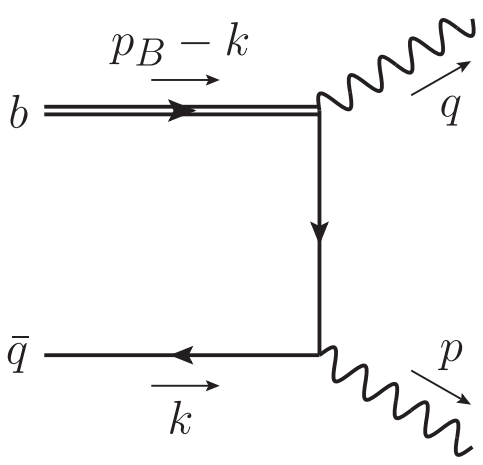

FIG. 1. Diagrammatic representations of correlation functions at tree level.

Similarly, one can calculate

$$q^\mu \langle A(p, \epsilon^*)|\bar{q}\gamma_\mu b|\bar{B}(p_B)\rangle = -2m_A(\epsilon^* \cdot q)V_0(q^2), \tag{2.3}$$

which indicates that the definition for the scalar matrix element $\langle A(p, \epsilon^*)|\bar{q}b|\bar{B}(p_B)\rangle$ has been included in the definition of the vector matrix element $\langle A(p, \epsilon^*)|\bar{q}\gamma_\mu b|\bar{B}(p_B)\rangle$. At the maximal recoil point $q^2 = 0$, two relations are obtained from (2.1) as follows:

$$\frac{m_B - m_A}{2m_A} V_1(0) - \frac{m_B + m_A}{2m_A} V_2(0) = V_0(0), \qquad T_1(0) = T_2(0). \tag{2.4}$$

### B. The $B$-meson LCSR of form factors at tree level

The tree diagram is shown in Fig. 1. After reviewing the standard process in [42,44], we first establish the vacuum-to-$B$-meson correlation functions in the rest frame of the $B$ meson as

$$\begin{aligned} \Pi^{(a)}_{\mu,\parallel}(p, q) &= \int d^4x e^{ip\cdot x} \langle 0|\mathrm{T}\{j^A_\parallel(x), \bar{q}(0)\Gamma^{(a)}_\mu b(0)\}|\bar{B}(p+q)\rangle, \\ \Pi^{(a)}_{\delta\mu,\perp}(p, q) &= \int d^4x e^{ip\cdot x} \langle 0|\mathrm{T}\{j^A_{\delta,\perp}(x), \bar{q}(0)\Gamma^{(a)}_\mu b(0)\}|\bar{B}(p+q)\rangle, \end{aligned} \tag{2.5}$$

where superscript $a$ denotes $V - A$ and $T + \tilde{T}$, and the interpolating QCD currents for the axial-vector mesons with the longitudinal and transverse polarization are defined as

$$j^A_\parallel(x) = \bar{q}'(x)\frac{\not{n}}{2}\gamma_5 q(x), \qquad j^A_{\delta,\perp}(x) = \bar{q}'(x)\frac{\not{n}}{2}\gamma_{\delta\perp}\gamma_5 q(x), \tag{2.6}$$

and the Dirac structures of the heavy-to-light transition currents are

$$\Gamma^{(a)}_\mu \in \{\gamma_\mu(1 - \gamma_5), \qquad i\sigma_{\mu\nu}(1 + \gamma_5)q^\nu\}. \tag{2.7}$$

Substituting the QCD interpolating and heavy-to-light transition currents (2.6) and (2.7) in (2.5), we can construct the LCSR of the $B \to A$ form factors in terms of the correlation functions and the specific Lorentz structures as

$$\begin{aligned} \Pi^{(V-A)}_{\mu,\parallel}(p, q) &= \Pi_1(n \cdot p, \bar{n} \cdot p)n_\mu + \tilde{\Pi}_1(n \cdot p, \bar{n} \cdot p)\bar{n}_\mu, \\ \Pi^{(T+\tilde{T})}_{\mu,\parallel}(p, q) &= \tilde{\Pi}_2(n \cdot p, \bar{n} \cdot p)[n \cdot q\bar{n}_\mu - \bar{n} \cdot qn_\mu], \\ \Pi^{(V-A)}_{\delta\mu,\perp}(p, q) &= \tilde{\Pi}_3(n \cdot p, \bar{n} \cdot p)[2g_{\delta\mu\perp} + i\epsilon_{\delta\mu\perp}], \\ \Pi^{(T+\tilde{T})}_{\delta\mu,\perp}(p, q) &= \tilde{\Pi}_4(n \cdot p, \bar{n} \cdot p)[2g_{\delta\mu\perp} + i\epsilon_{\delta\mu\perp}]\bar{n} \cdot q, \end{aligned} \tag{2.8}$$

with the convention $\epsilon_{\delta\mu\perp} = \epsilon_{\rho\sigma\delta\mu}n^\rho\bar{n}^\sigma$. The two lightlike vectors $n$ and $\bar{n}$ satisfy $n \cdot \bar{n} = 2, n \cdot n = 0, \bar{n} \cdot \bar{n} = 0$. We introduce the velocity vector $v^\mu = (n^\mu + \bar{n}^\mu)/2$ and $\gamma^\mu_\perp = \gamma^\mu - \not{n}\bar{n}^\mu/2 - \not{\bar{n}}n^\mu/2$. $n \cdot p$ is referred to as the large component, and we adopt the power counting scheme $n\cdot p \sim m_b$. Applying the definitions of the semileptonic $B\to A$ form factors (2.1) and the longitudinal and transverse decay constants of the axial vectors

$$\langle 0|\bar{q}_2 \frac{\not{n}}{2}\gamma_5 q_1|A(p,\epsilon^*)\rangle = -\frac{i}{2} f_A m_A \epsilon\cdot n, \qquad \langle 0|\bar{q}_2 \frac{\not{n}}{2}\gamma_{\delta\perp}\gamma_5 q_1|A(p,\epsilon^*)\rangle = -\frac{i}{2} f_A^T \epsilon_{\delta\perp} n\cdot p, \tag{2.9}$$

one can derive the desired hadronic representations of the correlation functions (2.5)

$$\begin{aligned}
\Pi^{(V-A)}_{\mu,\parallel}(p,q) &= -\frac{i}{2}\frac{f_A m_A n\cdot p}{m_A^2-p^2}\left(\frac{n\cdot p}{2m_A}\right)^2\left\{\frac{m_B}{m_B-n\cdot p}n_\mu\right.\\
&\quad\times\left[\left(-\frac{2m_A}{n\cdot p}V_0(q^2)\right)+\left(\frac{m_B-m_A}{n\cdot p}V_1(q^2)-\frac{m_B+m_A}{m_B}V_2(q^2)\right)\right]\\
&\quad\left.-\bar{n}^\mu\left[\left(-\frac{2m_A}{n\cdot p}V_0(q^2)\right)+\left(\frac{m_B-m_A}{n\cdot p}V_1(q^2)-\frac{m_B+m_A}{m_B}V_2(q^2)\right)\right]\right\}\\
&\quad+\int_{\omega_s}^{\infty}d\omega'\frac{1}{\omega'-\bar{n}\cdot p-i0}\left[n_\mu\varrho^{(V-A)}_{n,\parallel}(\omega',n\cdot p)+\bar{n}_\mu\varrho^{(V-A)}_{\bar{n},\parallel}(\omega',n\cdot p)\right],\\
\Pi^{(T+\tilde{T})}_{\mu,\parallel}(p,q) &= -\frac{i}{2}\frac{f_A m_A n\cdot p}{m_A^2-p^2}\left(\frac{n\cdot p}{2m_A}\right)^2[n\cdot q\bar{n}_\mu-\bar{n}\cdot q n_\mu]\left[\frac{m_B}{n\cdot p}T_2(q^2)-T_3(q^2)\right]\\
&\quad+\int_{\omega_s}^{\infty}d\omega'\frac{1}{\omega'-\bar{n}\cdot p-i0}[n\cdot q\bar{n}_\mu-\bar{n}\cdot q n_\mu]\varrho^{(T+\tilde{T})}_{\parallel}(\omega',n\cdot p),\\
\Pi^{(V-A)}_{\delta\mu,\perp}(p,q) &= -\frac{i}{4}\frac{f_A^T(n\cdot p)^2}{m_A^2-p^2}\left[2g_{\delta\mu\perp}\left(\frac{m_B-m_A}{n\cdot p}V_1(q^2)\right)+i\epsilon_{\delta\mu\perp}\left(\frac{m_B}{m_B-m_A}A(q^2)\right)\right]\\
&\quad+\int_{\omega_s}^{\infty}d\omega'\frac{1}{\omega'-\bar{n}\cdot p-i0}\left[2g_{\delta\mu\perp}\varrho^{(V-A)}_{\perp,V_1}(\omega',n\cdot p)+i\epsilon_{\delta\mu\perp}\varrho^{(V-A)}_{\perp,A}(\omega',n\cdot p)\right],\\
\Pi^{(T+\tilde{T})}_{\delta\mu,\perp}(p,q) &= \frac{i}{4}\frac{f_A^T m_B(n\cdot p)^2}{m_A^2-p^2}\left[2g_{\delta\mu\perp}\left(\frac{m_B}{n\cdot p}T_2(q^2)\right)+i\epsilon_{\delta\mu\perp}T_1(q^2)\right]\\
&\quad+\int_{\omega_s}^{\infty}d\omega'\frac{1}{\omega'-\bar{n}\cdot p-i0}\left[2g_{\delta\mu\perp}\varrho^{(T+\tilde{T})}_{\perp,T_2}(\omega',n\cdot p)+i\epsilon_{\delta\mu\perp}\varrho^{(T+\tilde{T})}_{\perp,T_1}(\omega',n\cdot p)\right],
\end{aligned}\tag{2.10}$$

where $\omega_s$ is the hadronic threshold for the axial-vector-meson channel, and $\varrho$ denotes hadronic spectral density. In order to obtain the factorization formulas of $\Pi^{(a)}_{\mu,\parallel}(p,q)$ and $\Pi^{(a)}_{\delta\mu,\perp}(p,q)$ at one-loop level, we replace the external state $|\bar{B}(p+q)\rangle$ by $|b(p_B-k)\bar{q}(k)\rangle$ with $p_B\equiv p+q$ and start with the tree-level partonic correlation function

$$\Pi^{(a,0)}_{i,b\bar{q}}(n\cdot p,\bar{n}\cdot p) = \int d\omega' T^{(a,0)}_{i,\alpha\beta}(n\cdot p,\bar{n}\cdot p,\omega')\Phi^{(0)\alpha\beta}_{b\bar{q}}(\omega'), \tag{2.11}$$

where the leading-order (LO) hard-scattering kernel $T^{(a,0)}_{i,\alpha\beta}$ ($i=\mu,\parallel$ or $\delta\mu,\perp$) can be written as

$$\begin{aligned}
T^{(V-A,0)}_{\mu,\parallel,\alpha\beta} &= \frac{i}{4}\frac{1}{\bar{n}\cdot p-\omega'+i0}[\not{n}\gamma_5\not{\bar{n}}\gamma_\mu]_{\alpha\beta},\\
T^{(T+\tilde{T},0)}_{\mu,\parallel,\alpha\beta} &= \frac{i}{4}\frac{1}{\bar{n}\cdot p-\omega'+i0}[\not{n}\gamma_5\not{\bar{n}}i\sigma_{\mu\nu}q^\nu]_{\alpha\beta},\\
T^{(V-A,0)}_{\delta\mu,\perp,\alpha\beta} &= \frac{i}{4}\frac{1}{\bar{n}\cdot p-\omega'+i0}[\not{n}\gamma_{\delta\perp}\gamma_5\not{\bar{n}}\gamma_\mu(1-\gamma_5)]_{\alpha\beta},\\
T^{(T+\tilde{T},0)}_{\delta\mu,\perp\alpha\beta} &= \frac{i}{4}\frac{1}{\bar{n}\cdot p-\omega'+i0}[\not{n}\gamma_{\delta\perp}\gamma_5\not{\bar{n}}i\sigma_{\mu\nu}(1+\gamma_5)q^\nu]_{\alpha\beta}.
\end{aligned}\tag{2.12}$$

The partonic DA of the $B$ meson is defined as

$$\Phi^{\alpha\beta}_{b\bar{q}}(\omega') = \int \frac{d\tau}{2\pi} \mathrm{e}^{i\omega'\tau} \langle 0|\bar{q}_\beta(\tau\bar{n})[\tau\bar{n}, 0] b_\alpha(0)|b(p_B - k)\bar{q}(k)\rangle, \tag{2.13}$$

with $[\tau\bar{n}, 0]$ being a finite-distance Wilson line and the tree-level contribution

$$\Phi^{(0)\alpha\beta}_{b\bar{q}}(\omega') = \delta(\bar{n}\cdot k - \omega')\bar{q}_\beta(k) b_\alpha(p_B - k). \tag{2.14}$$

One can compute the final expression of the partonic correlation function with the aid of the light-cone projector of the $B$ meson in momentum space [21,41]

$$M_{\beta\alpha} = -\frac{i\tilde{f}_B(\mu) m_B}{4}\left\{\frac{1+\not{v}}{2}\left[\phi^+_B(\omega)\not{n} + \phi^-_B(\omega)\not{\bar{n}} - \frac{2\omega}{D-2}\phi^-_B(\omega)\gamma^\rho_\perp \frac{\partial}{\partial k^\rho_\perp}\right]\gamma_5\right\}_{\alpha\beta} \tag{2.15}$$

in $D$ dimensions with the replacement $\phi^\pm_B(\omega') \to \phi^\pm_{b\bar{q}}(\omega')$, where $\tilde{f}_B(\mu)$ is the $B$-meson decay constant in the static limit, and the relation between the former and the QCD decay constant is

$$f_B = \tilde{f}_B(\mu)\left[1 + \frac{\alpha_s C_F}{4\pi}\left(-3\ln\frac{\mu}{m_b} - 2\right)\right]. \tag{2.16}$$

After calculating the tree diagram and considering the physical $B$ state, we derive the manifest tree-level expressions of correlation functions as

$$\begin{aligned}
\Pi^{(V-A,0)}_{\mu,\parallel} &= \frac{i}{2}\tilde{f}_B(\mu) m_B \int_0^\infty d\omega \frac{\phi^-_B(\omega)}{\omega - \bar{n}\cdot p + i0}\bar{n}_\mu, \\
\Pi^{(T+\tilde{T},0)}_{\mu,\parallel} &= -\frac{i}{4}\tilde{f}_B(\mu) m_B \int_0^\infty d\omega \frac{\phi^-_B(\omega)}{\omega - \bar{n}\cdot p + i0}[n\cdot q\bar{n}_\mu - \bar{n}\cdot q n_\mu], \\
\Pi^{(V-A,0)}_{\delta\mu,\perp} &= \frac{i}{4}\tilde{f}_B(\mu) m_B \int_0^\infty d\omega \frac{\phi^-_B(\omega)}{\omega - \bar{n}\cdot p + i0}[2g_{\delta\mu\perp} + i\epsilon_{\delta\mu\perp}], \\
\Pi^{(T+\tilde{T},0)}_{\delta\mu,\perp} &= -\frac{i}{4}\tilde{f}_B(\mu) m_B \int_0^\infty d\omega \frac{\phi^-_B(\omega)}{\omega - \bar{n}\cdot p + i0}\bar{n}\cdot q[2g_{\delta\mu\perp} + i\epsilon_{\delta\mu\perp}],
\end{aligned} \tag{2.17}$$

where the tree-level result $\Pi^{(V-A,0)}_{\mu,\parallel}$ is identical to the corresponding one of $B \to \pi$ in [42] due to the coincident interpolating particle current $\bar{q}'\not{n}\gamma_5 q$ and the heavy-to-light transition currents $\bar{q}\gamma_\mu b$ in the vacuum-to-$B$ correlating functions (2.5) in both cases. We can therefore predict that the NLO results of $\Pi^{(V-A,1)}_{\mu,\parallel}$ will also be the same as the corresponding ones of $B \to \pi$ in [42]. It is also evident that the $B$-meson DA $\phi^-_B(\omega)$ instead of $\phi^+_B(\omega)$ appears in all correlation functions at tree level. Matching the tree-level correlation functions with the hadronic ones and taking advantage of Boral transformation $p^2 \to M^2$, we finally obtain the tree-level LCSR of the semileptonic $B \to A$ form factors

$$\begin{aligned}
\frac{2m_A}{n\cdot p}V_0(q^2) &= \frac{2\tilde{f}_B(\mu) m_B m_A}{f_A (n\cdot p)^2} e^{m_A^2/(n\cdot p\omega_M)} \int_0^{\omega_s} d\omega\, e^{-\omega/\omega_M}\phi^-_B(\omega) + \mathcal{O}(\alpha_s), \\
\frac{m_B - m_A}{n\cdot p}V_1(q^2) - \frac{m_B + m_A}{m_B}V_2(q^2) &= \frac{2\tilde{f}_B(\mu) m_B m_A}{f_A (n\cdot p)^2} e^{m_A^2/(n\cdot p\omega_M)} \int_0^{\omega_s} d\omega\, e^{-\omega/\omega_M}\phi^-_B(\omega) + \mathcal{O}(\alpha_s), \\
\frac{m_B - m_A}{n\cdot p}V_1(q^2) &= -\frac{\tilde{f}_B(\mu) m_B}{f^T_A n\cdot p} e^{m_A^2/(n\cdot p\omega_M)} \int_0^{\omega_s} d\omega\, e^{-\omega/\omega_M}\phi^-_B(\omega) + \mathcal{O}(\alpha_s),
\end{aligned}$$

$$\frac{m_B}{m_B-m_A}A(q^2) = -\frac{\tilde{f}_B(\mu)m_B}{f_A^T n\cdot p}e^{m_A^2/(n\cdot p\omega_M)}\int_0^{\omega_s}d\omega\, e^{-\omega/\omega_M}\phi_B^-(\omega)+\mathcal{O}(\alpha_s),$$
$$T_1(q^2) = -\frac{\tilde{f}_B(\mu)m_B}{f_A^T n\cdot p}e^{m_A^2/(n\cdot p\omega_M)}\int_0^{\omega_s}d\omega\, e^{-\omega/\omega_M}\phi_B^-(\omega)+\mathcal{O}(\alpha_s),$$
$$\frac{m_B}{n\cdot p}T_2(q^2) = -\frac{\tilde{f}_B(\mu)m_B}{f_A^T n\cdot p}e^{m_A^2/(n\cdot p\omega_M)}\int_0^{\omega_s}d\omega\, e^{-\omega/\omega_M}\phi_B^-(\omega)+\mathcal{O}(\alpha_s),$$
$$\frac{m_B}{n\cdot p}T_2(q^2)-T_3(q^2) = \frac{2\tilde{f}_B(\mu)m_B m_A}{f_A(n\cdot p)^2}e^{m_A^2/(n\cdot p\omega_M)}\int_0^{\omega_s}d\omega\, e^{-\omega/\omega_M}\phi_B^-(\omega)+\mathcal{O}(\alpha_s), \tag{2.18}$$

where the Borel mass $\omega_M$ and the threshold parameter $\omega_s$ are defined as $\omega_M = M^2/n\cdot p$ and $\omega_s = s_0/n\cdot p$, respectively. $M^2$ and $s_0$ stand for the Borel parameter and the effective threshold, respectively.

## III. FACTORIZATION OF THE CORRELATION FUNCTION AT $\mathcal{O}(\alpha_s)$

In this section, our aim is to construct factorization formulas for correlation functions $\Pi^{(a)}_{\mu,\|}$ and $\Pi^{(a)}_{\delta\mu,\perp}$ in QCD at one-loop level. We employ the diagrammatic factorization method to expand the correlators $\Pi^{(a)}_{i,b\bar{q}}$ ($i=\mu,\|$ or $\delta\mu,\perp$), the short-distance functions $T_i$, and the partonic DA of the $B$ meson $\Phi_{b\bar{q}}$ in perturbation theory. We can schematically perform this process as

$$\begin{aligned}\Pi^{(a)}_{i,b\bar{q}} &= \Pi^{(a,0)}_{i,b\bar{q}}+\Pi^{(a,1)}_{i,b\bar{q}}+\cdots = \Phi_{b\bar{q}}\otimes T_i^{(a)}\\ &= \Phi^{(0)}_{b\bar{q}}\otimes T_i^{(a,0)}+\left[\Phi^{(0)}_{b\bar{q}}\otimes T_i^{(a,1)}+\Phi^{(1)}_{b\bar{q}}\otimes T_i^{(a,0)}\right]\\ &\quad+\cdots,\end{aligned} \tag{3.1}$$

where $\otimes$ indicates the convolution in the variable $\omega'$ defined in Eq. (2.13), and 0, 1 in the superscripts denote the order of $\alpha_s$. The matching condition determining the hard-scattering kernels at $\alpha_s$ is

$$\Phi^{(0)}_{b\bar{q}}\otimes T_i^{(a,1)} = \Pi^{(a,1)}_{i,b\bar{q}}-\Phi^{(1)}_{b\bar{q}}\otimes T_i^{(a,0)}, \tag{3.2}$$

where the second term is the infrared (soft) subtraction. We must verify that only the hard and/or hard-collinear regions generate contributions at leading power in $\Lambda/m_b$ since the soft contributions to $\Pi^{(a,1)}_{i,b\bar{q}}$ can cancel $\Phi^{(1)}_{b\bar{q}}\otimes T_i^{(a,0)}$, which is a significant aspect of demonstrating the factorization of $\Pi^{(a)}_{i,b\bar{q}}$. It is also worth noting that the collinear region [with the momentum scaling $l_\mu \sim (1,\lambda^2,\lambda)$] does not contribute to the correlation functions at leading power because the $B$-meson LCDAs can only cover the soft QCD dynamics of $\Pi^{(a)}_{i,b\bar{q}}$.

According to the previous strategy in [42], we will take advantage of the method of regions to calculate the master formula of $T_i^{(a,1)}$ in Eq. (3.2) diagram by diagram in order to obtain the hard coefficient functions ($C$) and the jet functions ($J$), from which leads to the factorization formula

$$\Pi^{(a)}_{i,b\bar{q}} = \Phi_{b\bar{q}}\otimes T_i^{(a)} = C_k\cdot J_k\otimes\Phi_{b\bar{q}}, \qquad (k=1,2,3,4). \tag{3.3}$$

The outline of our calculation strategy is summarized as follows: (i) We start with ascertaining the leading regions of the scalar integral for each diagram. (ii) We then simplify the Dirac algebra in the numerator for an identified leading region and compute the relevant integrals applying the method of regions. (iii) Next, we calculate the contributions of hard and hard-collinear regions with the light-cone projector of the $B$ meson in momentum space. (iv) We display the equivalence of the soft subtraction term and the soft-region correlation function. (v) We eventually add up the hard-region and hard-collinear-region contributions separately.

### A. Weak vertex correction

Taking the combination of the QCD interpolating axial current and heavy-to-light transition $V-A$ current into account, the weak vertex correction in Fig. 2(a) reads

$$\begin{aligned}\Pi^{(V-A,1)}_{\mu,\|,\mathrm{weak}} &= \frac{ig_s^2C_F\mu^{2\epsilon}}{4(\bar{n}\cdot p-\omega)}\int\frac{d^Dl}{(2\pi)^D}\frac{1}{[(p-k+l)^2+i0][(m_bv+l)^2-m_b^2+i0][l^2+i0]}\\ &\quad\times\bar{q}(k)\not{n}\gamma_5\not{\bar{n}}\gamma_\rho(\not{p}-\not{k}+\not{l})\gamma_\mu(m_b\not{v}+\not{l}+m_b)\gamma^\rho b(v),\end{aligned} \tag{3.4}$$

where we will thereafter omit the subscript $b\bar{q}$ in $\Pi_{b\bar{q}}$ and $D=4-2\epsilon$, and adopt the scaling behaviors as

$$n \cdot p \sim m_b, \qquad \bar{n} \cdot p \sim \Lambda, \qquad k_\mu \sim \Lambda, \tag{3.5}$$

with hadronic scale $\Lambda$ of order $\Lambda_{\mathrm{QCD}}$. The leading-power contributions of the loop integral for the weak vertex correction

$$I_1 = \int \frac{d^D l}{(2\pi)^D} \frac{1}{[(p-k+l)^2 + i0][(m_b v + l)^2 - m_b^2 + i0][l^2 + i0]} \tag{3.6}$$

can be extracted from the hard, hard-collinear, and soft regions with the power $I_1 \sim \lambda^0$, which indicates that only the leading-power terms in the numerator will lead to the contribution. Though the semihard region also has the power $\lambda^0$, the corresponding contribution leads to a scaleless integral (see [48,58] for detailed discussions).

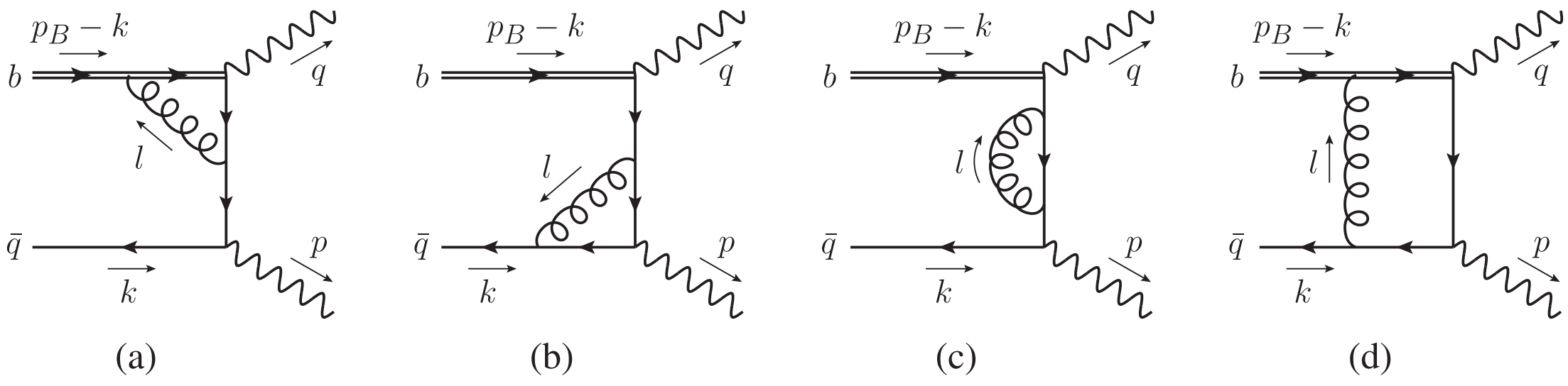

FIG. 2. Diagrammatic representations of correlation functions at one-loop level.

Employing the partonic light-cone projector, we obtain the hard-region contribution

$$\begin{aligned}\Pi^{(V-A,1),\mathrm{h}}_{\mu,\|,\mathrm{weak}} = {}& -\frac{1}{2} g_s^2 C_F \tilde{f}_B(\mu) m_B \frac{\phi^-_{b\bar{q}}(\omega)}{\bar{n} \cdot p - \omega} \int \frac{d^D l}{(2\pi)^D} \frac{1}{[l^2 + n \cdot p \bar{n} \cdot l + i0][l^2 + 2 m_b v \cdot l + i0][l^2 + i0]} \\ & \times \left\{ \bar{n}_\mu \left[ 2 m_b (n \cdot p + n \cdot l) + (D-2) l_\perp^2 \right] - n_\mu (D-2) (\bar{n} \cdot l)^2 \right\}.\end{aligned} \tag{3.7}$$

We also use the program [59] from Shi to verify that our simplification results of the Dirac structure in one-loop diagrams are identical to the ones from the program. We can compute the integral result

$$\begin{aligned}\Pi^{(V-A,1),\mathrm{h}}_{\mu,\|,\mathrm{weak}} = {}& \frac{i \alpha_s C_F}{8\pi} \tilde{f}_B(\mu) m_B \frac{\phi^-_{b\bar{q}}(\omega)}{\bar{n} \cdot p - \omega} \left\{ \left[ \frac{1}{\epsilon^2} + \frac{1}{\epsilon} + \frac{2}{\epsilon} \ln \frac{\mu}{n \cdot p} + 2 \ln^2 \frac{\mu}{n \cdot p} + 2 \ln \frac{\mu}{m_b} - \ln^2 r \right. \right. \\ & \left. \left. - 2 \mathrm{Li}_2 \left( -\frac{\bar{r}}{r} \right) - \frac{2-r}{\bar{r}} \ln r + \frac{\pi^2}{12} + 3 \right] \bar{n}_\mu + \left[ \frac{1}{r-1} \left( 1 + \frac{r}{\bar{r}} \ln r \right) \right] n_\mu \right\},\end{aligned} \tag{3.8}$$

with $r = \frac{n \cdot p}{m_b}$ and $\bar{r} = 1 - r$. Following the same approach, the contribution of the hard-collinear region at leading power is

$$\Pi^{(V-A,1),\mathrm{hc}}_{\mu,\|,\mathrm{weak}} = g_s^2 C_F \tilde{f}_B(\mu) m_B \mu^{2\epsilon} \frac{\phi^-_{b\bar{q}}(\omega)}{\omega - \bar{n} \cdot p + i0} \int \frac{d^D l}{(2\pi)^D} \frac{m_b (n \cdot p + n \cdot l) \bar{n}_\mu}{[(p-k+l)^2 + i0][l^2 + 2 m_b v \cdot l + i0][l^2 + i0]}, \tag{3.9}$$

where we compute the integral result as

$$\begin{aligned}\Pi^{(V-A,1),\mathrm{hc}}_{\mu,\|,\mathrm{weak}} = {}& -\frac{i \alpha_s C_F}{8\pi} \tilde{f}_B(\mu) m_B \frac{\phi^-_{b\bar{q}}(\omega)}{\bar{n} \cdot p - \omega} \bar{n}_\mu \left[ \frac{2}{\epsilon^2} + \frac{2}{\epsilon} \left( \ln \frac{\mu^2}{n \cdot p (\omega - \bar{n} \cdot p)} + 1 \right) + \ln^2 \frac{\mu^2}{n \cdot p (\omega - \bar{n} \cdot p)} \right. \\ & \left. + 2 \ln \frac{\mu^2}{n \cdot p (\omega - \bar{n} \cdot p)} - \frac{\pi^2}{6} + 4 \right].\end{aligned} \tag{3.10}$$

The contribution extracted from the soft region is

$$\begin{aligned}\Pi^{(a,1),\mathrm{s}}_{i,\mathrm{weak}} = {}& \frac{g_s^2 C_F}{4(\bar{n} \cdot p - \omega)} \int \frac{d^D l}{(2\pi)^D} \frac{1}{[\bar{n} \cdot (p-k+l) + i0][v \cdot l + i0][l^2 + i0]} \\ & \times \bar{q}(k) \{ \not{n} \gamma_5, \not{n} \gamma_{\delta\perp} \gamma_5 \} \not{\bar{n}} \{ \gamma_\mu (1-\gamma_5), i \sigma_{\mu\nu} q^\nu (1+\gamma_5) \} b(v).\end{aligned} \tag{3.11}$$

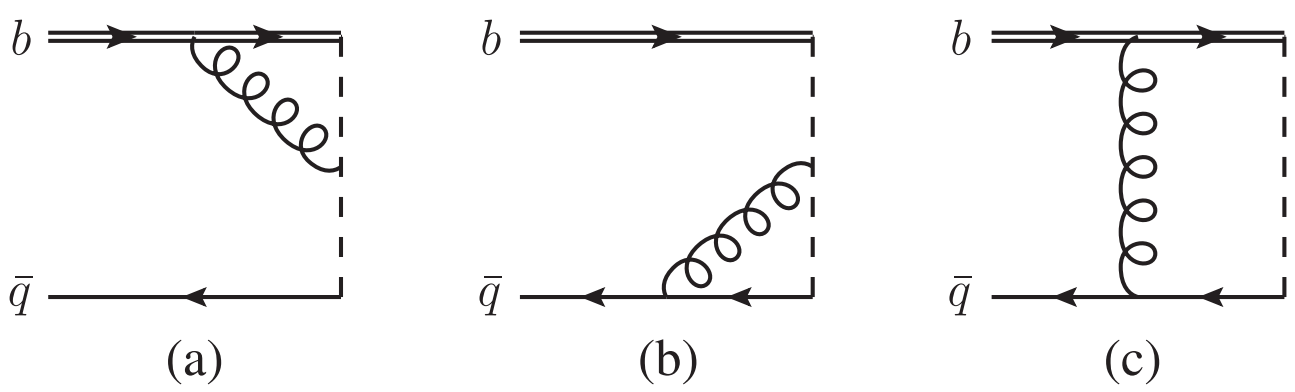

FIG. 3. Diagrammatic representations of the $B$-meson DA $\Phi^{\alpha\beta}_{b\bar{q}}(\omega')$ at one-loop level.

Considering the Wilson-line Feynman rules, we calculate the contribution from partonic DA in Fig. 3(a) as

$$\Phi^{\alpha\beta,(1)}_{b\bar{q},a}(\omega,\omega') = ig_s^2 C_F \int \frac{d^D l}{(2\pi)^D} \frac{1}{[\bar{n}\cdot l + i0][v\cdot l + i0][l^2 + i0]} [\delta(\omega' - \omega - \bar{n}\cdot l) - \delta(\omega' - \omega)][\bar{q}(k)]_\alpha [b(v)]_\beta. \tag{3.12}$$

According to Eq. (3.12), we proceed to write down the infrared subtraction term

$$\begin{aligned}\Phi^{(1)}_{b\bar{q},a} \otimes T_i^{(a,0)} = \frac{g_s^2 C_F}{4(\bar{n}\cdot p - \omega)} \int \frac{d^D l}{(2\pi)^D} \frac{1}{[\bar{n}\cdot (p - k + l) + i0][v\cdot l + i0][l^2 + i0]} \\ \times \bar{q}(k)\{\not{n}\gamma_5, \not{n}\gamma_{\delta\perp}\gamma_5\}\not{\bar{n}}\{\gamma_\mu(1-\gamma_5), i\sigma_{\mu\nu}q^\nu(1+\gamma_5)\}b(v),\end{aligned} \tag{3.13}$$

which leads to the complete cancellation to the soft-region contribution of the weak vertex correction.

Applying the identical partonic light-cone projector and the results of the integral, the rest hard-region contributions are derived as

$$\begin{aligned}\Pi^{(T+\tilde{T},1),\mathrm{h}}_{\mu,\|,\mathrm{weak}} = &-\frac{i\alpha_s C_F}{16\pi} \tilde{f}_B(\mu) m_B \frac{\phi^-_{b\bar{q}}(\omega)}{\bar{n}\cdot p - \omega} [n\cdot q \bar{n}_\mu - \bar{n}\cdot q n_\mu] \left[\frac{1}{\epsilon^2} + \frac{2}{\epsilon}\left(\ln\frac{\mu}{n\cdot p} + 1\right) + 2\ln^2\frac{\mu}{n\cdot p} + 4\ln\frac{\mu}{m_b} - \ln^2 r \right. \\ &\left. - 2\mathrm{Li}_2\left(-\frac{\bar{r}}{r}\right) - 2\frac{\bar{r} - r}{\bar{r}}\ln r + \frac{\pi^2}{12} + 4\right],\end{aligned} \tag{3.14}$$

$$\begin{aligned}\Pi^{(V-A,1),\mathrm{h}}_{\delta\mu,\perp,\mathrm{weak}} = &\frac{i\alpha_s C_F}{16\pi} \tilde{f}_B(\mu) m_B \frac{\phi^-_{b\bar{q}}(\omega)}{\bar{n}\cdot p - \omega} [2g_{\delta\mu\perp} + i\epsilon_{\delta\mu\perp}] \left[\frac{1}{\epsilon^2} + \frac{1}{\epsilon} + \frac{2}{\epsilon}\ln\frac{\mu}{n\cdot p} + 2\ln^2\frac{\mu}{n\cdot p} + 2\ln\frac{\mu}{m_b} - \ln^2 r \right. \\ &\left. - 2\mathrm{Li}_2\left(-\frac{\bar{r}}{r}\right) - \frac{3\bar{r} - 1}{\bar{r}}\ln r + \frac{\pi^2}{12} + 4\right],\end{aligned} \tag{3.15}$$

$$\begin{aligned}\Pi^{(T+\tilde{T},1),\mathrm{h}}_{\delta\mu,\perp,\mathrm{weak}} = &-\frac{i\alpha_s C_F}{16\pi} \tilde{f}_B(\mu) m_B \frac{\phi^-_{b\bar{q}}(\omega)}{\bar{n}\cdot p - \omega} \bar{n}\cdot q[2g_{\delta\mu\perp} + i\epsilon_{\delta\mu\perp}] \left[\frac{1}{\epsilon^2} + \frac{2}{\epsilon}\left(\ln\frac{\mu}{n\cdot p} + 1\right) + 2\ln^2\frac{\mu}{n\cdot p} + 4\ln\frac{\mu}{m_b} - \ln^2 r \right. \\ &\left. - 2\mathrm{Li}_2\left(-\frac{\bar{r}}{r}\right) - 2\ln r + \frac{\pi^2}{12} + 4\right],\end{aligned} \tag{3.16}$$

as well as the rest contributions from the hard-collinear region

$$\begin{aligned}\Pi^{(T+\tilde{T},1),\mathrm{hc}}_{\mu,\|,\mathrm{weak}} = &\frac{i\alpha_s C_F}{16\pi} \tilde{f}_B(\mu) m_B \frac{\phi^-_{b\bar{q}}(\omega)}{\bar{n}\cdot p - \omega} [n\cdot q \bar{n}_\mu - \bar{n}\cdot q n_\mu] \left[\frac{2}{\epsilon^2} + \frac{2}{\epsilon}\left(\ln\frac{\mu^2}{n\cdot p(\omega - \bar{n}\cdot p)} + 1\right) + \ln^2\frac{\mu^2}{n\cdot p(\omega - \bar{n}\cdot p)} \right. \\ &\left. + 2\ln\frac{\mu^2}{n\cdot p(\omega - \bar{n}\cdot p)} - \frac{\pi^2}{6} + 4\right],\end{aligned} \tag{3.17}$$

$$\Pi^{(V-A,1),\mathrm{hc}}_{\delta\mu,\perp,\mathrm{weak}} = -\frac{i\alpha_s C_F}{16\pi} \tilde{f}_B(\mu) m_B \frac{\phi^-_{b\bar{q}}(\omega)}{\bar{n}\cdot p - \omega} [2g_{\delta\mu\perp} + i\epsilon_{\delta\mu\perp}] \left[\frac{2}{\epsilon^2} + \frac{2}{\epsilon}\left(\ln\frac{\mu^2}{n\cdot p(\omega - \bar{n}\cdot p)} + 1\right) + \ln^2\frac{\mu^2}{n\cdot p(\omega - \bar{n}\cdot p)} + 2\ln\frac{\mu^2}{n\cdot p(\omega - \bar{n}\cdot p)} - \frac{\pi^2}{6} + 4\right], \tag{3.18}$$

$$\Pi^{(T+\tilde{T},1),\mathrm{hc}}_{\delta\mu,\perp,\mathrm{weak}} = \frac{i\alpha_s C_F}{16\pi} \tilde{f}_B(\mu) m_B \frac{\phi^-_{b\bar{q}}(\omega)}{\bar{n}\cdot p - \omega} \bar{n}\cdot q [2g_{\delta\mu\perp} + i\epsilon_{\delta\mu\perp}] \left[\frac{2}{\epsilon^2} + \frac{2}{\epsilon}\left(\ln\frac{\mu^2}{n\cdot p(\omega - \bar{n}\cdot p)} + 1\right) + \ln^2\frac{\mu^2}{n\cdot p(\omega - \bar{n}\cdot p)} + 2\ln\frac{\mu^2}{n\cdot p(\omega - \bar{n}\cdot p)} - \frac{\pi^2}{6} + 4\right]. \tag{3.19}$$

### B. Axial-vector vertex correction

The QCD correction to the axial-vector vertex [the diagram in Fig. 2(b)] is derived as

$$\begin{aligned}\Pi^{(V-A,1)}_{\mu,\parallel,\mathrm{axial}} &= -\frac{ig_s^2 C_F}{2n\cdot p(\bar{n}\cdot p - \omega)} \int \frac{d^D l}{(2\pi)^D} \frac{1}{[(p-l)^2 + i0][(l-k)^2 + i0][l^2 + i0]} \bar{q}(k)\gamma_5\gamma_\rho \not{l}\not{n}\gamma_5(\not{p} - \not{l})\gamma^\rho(\not{p} - \not{k})\gamma_\mu b(v), \\ &= -\frac{ig_s^2 C_F}{2n\cdot p(\bar{n}\cdot p - \omega)} \bar{q}(k)\gamma_\rho\gamma_\alpha\not{n}\gamma_5\gamma_\beta\gamma^\rho(\not{p} - \not{k})\gamma_\mu\gamma_5 b(v) \frac{i}{(4\pi)^2} I_2^{\alpha\beta},\end{aligned} \tag{3.20}$$

where $I_2^{\alpha\beta}$ is given by [42]

$$I_2^{\alpha\beta} = \int \frac{d^D l}{(2\pi)^D} \frac{l^\alpha (p-l)^\beta}{[(p-l)^2 + i0][(l-k)^2 + i0][l^2 + i0]}. \tag{3.21}$$

We straightforward write down the corresponding scalar integral

$$I_2 = \int \frac{d^D l}{(2\pi)^D} \frac{1}{[(l-k)^2 + i0][(p-l)^2 + i0][l^2 + i0]}, \tag{3.22}$$

whose power is $\lambda^{-1}$ and leading-power contributions are extracted from the hard-collinear and soft regions. According to the previous investigations of $B \to S$ [48] and $B \to \pi$ [42], the contributions from the soft and hard regions vanish because of the scaleless integrals in dimensional regularization. Therefore, we can safely compute the contribution from the hard-collinear region by implementing the partonic light-cone projector and obtain

$$\begin{aligned}\Pi^{(V-A,1)}_{\mu,\parallel,\mathrm{axial}} = \frac{i\alpha_s C_F}{8\pi} \tilde{f}_B(\mu) m_B \frac{1}{\bar{n}\cdot p - \omega} \Bigg\{ & n_\mu \phi^+_{b\bar{q}}(\omega) \left[\frac{\bar{n}\cdot p - \omega}{\omega} \ln\frac{\bar{n}\cdot p - \omega}{\bar{n}\cdot p}\right] \\ & + \bar{n}_\mu \phi^-_{b\bar{q}}(\omega) \Bigg[\left(\frac{1}{\epsilon} + \ln\left(-\frac{\mu^2}{p^2}\right)\right)\left(\frac{2\bar{n}\cdot p}{\omega}\ln\frac{\bar{n}\cdot p - \omega}{\bar{n}\cdot p} + 1\right) \\ & - \frac{\bar{n}\cdot p}{\omega}\ln\frac{\bar{n}\cdot p - \omega}{\bar{n}\cdot p}\left(\ln\frac{\bar{n}\cdot p - \omega}{\bar{n}\cdot p} + \frac{2\omega}{\bar{n}\cdot p} - 4\right) + 4\Bigg]\Bigg\}.\end{aligned} \tag{3.23}$$

We then turn to calculate the soft contribution

$$\Pi^{(a,1),\mathrm{s}}_{i,\mathrm{axial}} = -\frac{g_s^2 C_F}{4(n\cdot p - \omega)} \int \frac{d^D l}{(2\pi)^D} \frac{1}{[n\cdot(p-l) + i0][(l-k)^2 + i0][l^2 + i0]} \bar{q}(k)\not{\bar{n}}\not{l}\{\not{n}\gamma_5, \not{n}\gamma_{\delta\perp}\gamma_5\}\not{\bar{n}}\{\gamma_\mu\gamma_5, i\sigma_{\mu\nu}q^\nu\gamma_5\} b(v). \tag{3.24}$$

The corresponding contribution to the partonic DA [see the diagram in Fig. 3(b)] reads

$$\Phi^{\alpha\beta,(1)}_{b\bar{q},b}(\omega,\omega') = ig_s^2 C_F \int \frac{d^D l}{(2\pi)^D} \frac{1}{[\bar{n}\cdot l + i0][(k+l) + i0][l^2 + i0]} [\delta(\omega' - \omega - \bar{n}\cdot l) - \delta(\omega' - \omega)][\bar{q}(k)\not{\bar{n}}(\not{k} + \not{l})]_\alpha [b(v)]_\beta. \tag{3.25}$$

The soft subtraction term is deduced as

$$\Phi^{(1)}_{b\bar{q},b} \otimes T_i^{(a,0)} = -\frac{g_s^2 C_F}{4(\bar{n}\cdot p - \omega)} \int \frac{d^D l}{(2\pi)^D} \frac{1}{[\bar{n}\cdot(p-k-l) + i0][(k+l)^2 + i0][l^2 + i0]} \times \bar{q}(k) \not{\bar{n}} \not{l} \{\not{n}\gamma_5, \not{n}\gamma_{\delta\perp}\gamma_5\} \not{\bar{n}} \{\gamma_\mu(1-\gamma_5), i\sigma_{\mu\nu} q^\nu (1+\gamma_5)\} b(v), \tag{3.26}$$

which can counter the one-loop correction to the axial vertex.

Other contributions from the hard-collinear region are given as follows:

$$\Pi^{(T+\tilde{T},1)}_{\mu,\parallel,\mathrm{axial}} = -\frac{i\alpha_s C_F}{16\pi} \frac{\tilde{f}_B(\mu) m_B}{\bar{n}\cdot p - \omega} [\bar{n}_\mu n\cdot q - \bar{n}\cdot q n_\mu] \Bigg\{ \phi^+_{b\bar{q}}(\omega) \left[ (1+\eta) \frac{\ln(1+\eta)}{\eta} \right] + \phi^-_{b\bar{q}}(\omega) \Bigg[ \frac{1}{\epsilon} + \ln\left(-\frac{\mu^2}{p^2}\right) - \frac{1+\eta}{\eta} \ln(1+\eta) + 1 + \ln(1+\eta) - \frac{\eta - \ln(1+\eta)}{\eta} \left( \frac{2}{\epsilon} + 2\ln\left(-\frac{\mu^2}{p^2}\right) - 2\ln(1+\eta) + 5 \right) + \frac{\ln^2(1+\eta)}{\eta} \Bigg] \Bigg\}, \tag{3.27}$$

$$\Pi^{(V-A,1)}_{\delta\mu,\perp,\mathrm{axial}} = \frac{i\alpha_s C_F}{16\pi} \tilde{f}_B(\mu) m_B \frac{\phi^-_{b\bar{q}}(\omega)}{\bar{n}\cdot p - \omega} [2g_{\delta\mu\perp} + i\epsilon_{\delta\mu\perp}] \times \left\{ \frac{\eta - \ln(1+\eta)}{\eta} \left[ \frac{2}{\epsilon} + 2\ln\left(-\frac{\mu^2}{p^2}\right) - 2\ln(1+\eta) + 4 \right] - \frac{\ln^2(1+\eta)}{\eta} \right\}, \tag{3.28}$$

$$\Pi^{(T+\tilde{T},1)}_{\delta\mu,\perp,\mathrm{axial}} = -\frac{i\alpha_s C_F}{16\pi} \tilde{f}_B(\mu) m_B \frac{\phi^-_{b\bar{q}}(\omega)}{\bar{n}\cdot p - \omega} \bar{n}\cdot q [2g_{\delta\mu\perp} + i\epsilon_{\delta\mu\perp}] \times \left\{ \frac{\eta - \ln(1+\eta)}{\eta} \left[ \frac{2}{\epsilon} + 2\ln\left(-\frac{\mu^2}{p^2}\right) - 2\ln(1+\eta) + 4 \right] - \frac{\ln^2(1+\eta)}{\eta} \right\}, \tag{3.29}$$

with $\eta = -\omega/\bar{n}\cdot p$.

### C. Wave function renormalization

One can write the self-energy correction to the intermediate quark propagator [see the diagram in Fig. 2(c)] as

$$\Pi^{(V-A,1)}_{\mu,\parallel,\mathrm{wfc}} = \frac{i}{2} \frac{g_s^2 C_F \mu^{2\epsilon}}{(n\cdot p)^2 (\bar{n}\cdot p - \omega)^2} \int \frac{d^D l}{(2\pi)^D} \frac{1}{[(p-k+l)^2 + i0][l^2 + i0]} \bar{q}(k) \not{n}\gamma_5 (\not{p} - \not{k}) \gamma_\rho (\not{p} - \not{k} + \not{l}) \gamma^\rho (\not{p} - \not{k}) \gamma_\mu b(v). \tag{3.30}$$

Because of the nonexistence of the soft and collinear divergences, it is straightforward to obtain the result

$$\Pi^{(V-A,1)}_{\mu,\parallel,\mathrm{wfc}} = \frac{i\alpha_s C_F}{8\pi} \frac{\tilde{f}_B(\mu) m_B}{\bar{n}\cdot p - \omega} \phi^-_{b\bar{q}}(\omega) \bar{n}_\mu \left[ \frac{1}{\epsilon} + \ln\frac{\mu^2}{n\cdot p(\omega - \bar{n}\cdot p)} + 1 \right]. \tag{3.31}$$

According to the previous work [42], there is no contribution to the hard-scattering kernel from the massless quark. The wave function renormalization of the $b$-quark in QCD is derived as

$$\Pi^{(V-A,1)}_{\mu,\parallel,\mathrm{bwf}} = -\frac{\alpha_s C_F}{8\pi} \left[ \frac{3}{\epsilon} + 3\ln\frac{\mu^2}{m_b^2} + 4 \right] \Pi^{(V-A,0)}_{\mu,\parallel}, \tag{3.32}$$

with $\Pi^{(V-A,0)}_{\mu,\parallel}$ given in (2.17). In heavy quark effective theory (HQET), the wave function renormalization of the $b$-quark reads

$$\Phi^{(1)}_{b\bar{q},\mathrm{bwf}} \otimes T_i^{(a,0)} = 0 \tag{3.33}$$

because of the scaleless integral. Evaluating Eqs. (3.32) and (3.33), we will obtain

$$\Pi^{(V-A,1)}_{\mu,\parallel,\mathrm{bwf}} - \Phi^{(1)}_{b\bar{q},\mathrm{bwf}} \otimes T^{(V-A,0)}_{\mu,\parallel} = -\frac{\alpha_s C_F}{8\pi}\left[\frac{3}{\epsilon} + 3\ln\frac{\mu^2}{m_b^2} + 4\right]\Pi^{(V-A,0)}_{\mu,\parallel}. \tag{3.34}$$

Applying the same method, the rest results are displayed as follows:

$$\Pi^{(T+\tilde{T},1)}_{\mu,\parallel,\mathrm{wfc}} = -\frac{i\alpha_s C_F}{16\pi}\frac{\tilde{f}_B(\mu)m_B}{\bar{n}\cdot p - \omega}\phi^-_{b\bar{q}}(\omega)[n\cdot q\bar{n}_\mu - \bar{n}\cdot q n_\mu]\left[\frac{1}{\epsilon} + \ln\frac{\mu^2}{n\cdot p(\omega - \bar{n}\cdot p)} + 1\right], \tag{3.35}$$

$$\Pi^{(V-A,1)}_{\delta\mu,\perp,\mathrm{wfc}} = \frac{i\alpha_s C_F}{16\pi}\frac{\tilde{f}_B(\mu)m_B}{\bar{n}\cdot p - \omega}\phi^-_{b\bar{q}}(\omega)[2g_{\delta\mu\perp} + i\epsilon_{\delta\mu\perp}]\left[\frac{1}{\epsilon} + \ln\frac{\mu^2}{n\cdot p(\omega - \bar{n}\cdot p)} + 1\right], \tag{3.36}$$

$$\Pi^{(T+\tilde{T},1)}_{\delta\mu,\perp,\mathrm{wfc}} = -\frac{i\alpha_s C_F}{16\pi}\frac{\tilde{f}_B(\mu)m_B}{\bar{n}\cdot p - \omega}\phi^-_{b\bar{q}}(\omega)\bar{n}\cdot q[2g_{\delta\mu\perp} + i\epsilon_{\delta\mu\perp}]\left[\frac{1}{\epsilon} + \ln\frac{\mu^2}{n\cdot p(\omega - \bar{n}\cdot p)} + 1\right], \tag{3.37}$$

$$\Pi^{(T+\tilde{T},1)}_{\mu,\parallel,\mathrm{bwf}} - \Phi^{(1)}_{b\bar{q},\mathrm{bwf}} \otimes T^{(T+\tilde{T},0)}_{\mu,\parallel} = -\frac{\alpha_s C_F}{8\pi}\left[\frac{3}{\epsilon} + 3\ln\frac{\mu^2}{m_b^2} + 4\right]\Pi^{(T+\tilde{T},0)}_{\mu,\parallel}, \tag{3.38}$$

$$\Pi^{(V-A,1)}_{\delta\mu,\perp,\mathrm{bwf}} - \Phi^{(1)}_{b\bar{q},\mathrm{bwf}} \otimes T^{(V-A,0)}_{\delta\mu,\perp} = -\frac{\alpha_s C_F}{8\pi}\left[\frac{3}{\epsilon} + 3\ln\frac{\mu^2}{m_b^2} + 4\right]\Pi^{(V-A,0)}_{\delta\mu,\perp}, \tag{3.39}$$

$$\Pi^{(T+\tilde{T},1)}_{\delta\mu,\perp,\mathrm{bwf}} - \Phi^{(1)}_{b\bar{q},\mathrm{bwf}} \otimes T^{(T+\tilde{T},0)}_{\delta\mu,\perp} = -\frac{\alpha_s C_F}{8\pi}\left[\frac{3}{\epsilon} + 3\ln\frac{\mu^2}{m_b^2} + 4\right]\Pi^{(T+\tilde{T},0)}_{\delta\mu,\perp}. \tag{3.40}$$

### D. Box diagram

We then perform the computation of the box diagram displayed in Fig. 2(d) with the expression

$$\begin{aligned}\Pi^{(V-A,1)}_{\mu,\parallel,\mathrm{box}} &= \frac{i}{2}g_s^2 C_F\int\frac{d^D l}{(2\pi)^D}\frac{-\mu^{2\epsilon}}{[(m_b v + l)^2 - m_b^2 + i0][(p-k+l)^2 + i0][(k-l)^2 + i0][l^2 + i0]}\\ &\quad\times \bar{q}(k)\gamma_\rho(\not{l} - \not{k})\not{n}\gamma_5(\not{p} - \not{k} + \not{l})\gamma_\mu(m_b\not{v} + \not{l} + m_b)\gamma^\rho b(v).\end{aligned} \tag{3.41}$$

The master integral of the box diagram is

$$I_4 = \int\frac{d^D l}{(2\pi)^D}\frac{1}{[(m_b v + l)^2 - m_b^2 + i0][(p-k+l)^2 + i0][(k-l)^2 + i0][l^2 + i0]}, \tag{3.42}$$

with the scaling behavior $I_4 \sim \lambda^{-1}(\lambda^{-2})$ in the hard-collinear and semihard (soft) regions. The contribution of the semihard region, however, leads to a scaleless integral.

Computing the contribution of the hard-collinear region with the partonic momentum-space projector, the result is

$$\begin{aligned}\Pi^{(V-A,1)}_{\mu,\parallel,\mathrm{box}} &= -\frac{1}{2}g_s^2 C_F\tilde{f}_B(\mu)\frac{m_B}{m_b}\bar{n}_\mu\mu^{2\epsilon}\int\frac{d^D l}{(2\pi)^D}[(2-d)n\cdot l\phi^+_{b\bar{q}}(\omega) + 2m_b\phi^-_{b\bar{q}}(\omega)]\\ &\quad\times\frac{n\cdot(p+l)}{[n\cdot(p+l)\bar{n}\cdot(p-k+l) + l_\perp^2 + i0][n\cdot l\bar{n}(l-k) + l_\perp^2 + i0][l^2 + i0]},\end{aligned} \tag{3.43}$$

where we calculate the loop integrals and obtain

$$\Pi^{(V-A,1)}_{\mu,\parallel,\mathrm{box}} = \frac{i\alpha_s C_F}{4\pi}\tilde{f}_B(\mu)m_B\phi^-_{b\bar{q}}(\omega)\left\{\left[\frac{r}{\omega}\ln(1+\eta)\right]n_\mu - \frac{\ln(1+\eta)}{2\omega}\left[\frac{1}{\epsilon} + \ln\frac{\mu^2}{n\cdot p(\omega - \bar{n}\cdot p)} + \frac{1}{2}\ln(1+\eta) + 1\right]\bar{n}_\mu\right\}. \tag{3.44}$$

The soft-region contribution of box diagram can be written as

$$\Pi_{i,\mathrm{box}}^{(a,1),\mathrm{s}} = -\frac{g_s^2 C_F \mu^{2\epsilon}}{4}\int \frac{d^D l}{(2\pi)^D}\frac{-1}{[v\cdot l + i0][\bar{n}\cdot(p-k+l)+i0][(k-l)^2+i0][l^2+i0]} \\ \times \bar{q}(k)\slashed{v}(\slashed{k}-\slashed{l})\{\slashed{n}\gamma_5, \slashed{n}\gamma_{\delta\perp}\gamma_5\}\slashed{\bar{n}}\{\gamma_\mu(1-\gamma_5), i\sigma_{\mu\nu}q^\nu(1+\gamma_5)\}b(v). \tag{3.45}$$

Evaluating the diagram in Fig. 3(c), the NLO contribution to the partonic DA is

$$\Phi_{b\bar{q},c}^{\alpha\beta,(1)}(\omega,\omega') = -ig_s^2 C_F\int \frac{d^D l}{(2\pi)^D}\frac{1}{[(l-k)^2+i0][v\cdot l+i0][l^2+i0]}\delta(\omega'-\omega+\bar{n}\cdot l)[\bar{q}(k)\slashed{v}(\slashed{l}-\slashed{k})]_\alpha[b(v)]_\beta. \tag{3.46}$$

Then we can derive the soft subtraction term as

$$\Phi_{b\bar{q},c}^{(1)}\otimes T_i^{(a,0)} = \frac{g_s^2 C_F \mu^{2\epsilon}}{4}\int \frac{d^D l}{(2\pi)^D}\frac{1}{[v\cdot l + i0][\bar{n}\cdot(p-k+l)+i0][(l-k)^2+i0][l^2+i0]} \\ \times \bar{q}(k)\slashed{v}(\slashed{k}-\slashed{l})\{\slashed{n}\gamma_5, \slashed{n}\gamma_{\delta\perp}\gamma_5\}\slashed{\bar{n}}\{\gamma_\mu(1-\gamma_5), i\sigma_{\mu\nu}q^\nu(1+\gamma_5)\}b(v), \tag{3.47}$$

which yields the complete cancellation to the soft contribution of the box diagram.

Employing the same calculation strategy, other hard-collinear-region results of the box diagram are given as follows:

$$\Pi_{\mu,\|,\mathrm{box}}^{(T+\tilde{T},1)} = \frac{i\alpha_s C_F}{4\pi}\tilde{f}_B(\mu)m_B\phi_{b\bar{q}}^-(\omega)[n\cdot q\bar{n}_\mu - \bar{n}\cdot q n_\mu]\frac{\ln(1+\eta)}{2\omega}\left[\frac{1}{\epsilon}+\ln\frac{\mu^2}{n\cdot p(\omega-\bar{n}\cdot p)}+\frac{1}{2}\ln(1+\eta)+1\right], \tag{3.48}$$

$$\Pi_{\delta\mu,\perp,\mathrm{box}}^{(V-A,1)} = -\frac{i\alpha_s C_F}{4\pi}\tilde{f}_B(\mu)m_B\phi_{b\bar{q}}^-(\omega)[2g_{\delta\mu\perp}+i\epsilon_{\delta\mu\perp}]\frac{\ln(1+\eta)}{2\omega}\left[\frac{1}{\epsilon}+\ln\frac{\mu^2}{n\cdot p(\omega-\bar{n}\cdot p)}+\frac{1}{2}\ln(1+\eta)+1\right], \tag{3.49}$$

$$\Pi_{\delta\mu,\perp,\mathrm{box}}^{(T+\tilde{T},1)} = \frac{i\alpha_s C_F}{4\pi}\tilde{f}_B(\mu)m_B\bar{n}\cdot q[2g_{\delta\mu\perp}+i\epsilon_{\delta\mu\perp}]\left\{-\frac{r\ln(1+\eta)}{4\omega}\phi_{b\bar{q}}^+(\omega) \\ +\frac{\ln(1+\eta)}{2\omega}\left[\frac{1}{\epsilon}+\ln\frac{\mu^2}{n\cdot p(\omega-\bar{n}\cdot p)}+\frac{1}{2}\ln(1+\eta)+1\right]\phi_{b\bar{q}}^-(\omega)\right\}. \tag{3.50}$$

### E. The hard-scattering kernel at $\mathcal{O}(\alpha_s)$

Collecting different results from Secs. III A–III D, we readily derive one-loop hard-scattering kernels from the matching condition Eq. (3.2)

$$\Phi_{b\bar{q}}^{(0)}\otimes T_i^{(a,1)} = \left[\Pi_{i,\mathrm{weak}}^{(a,1)}+\Pi_{i,\mathrm{axial}}^{(a,1)}+\Pi_{i,\mathrm{wfc}}^{(a,1)}+\Pi_{i,\mathrm{box}}^{(a,1)}+\Pi_{i,\mathrm{bwf}}^{(a,1)}+\Pi_{i,\mathrm{dwf}}^{(a,1)}\right] \\ -\left[\Phi_{b\bar{q},a}^{(1)}+\Phi_{b\bar{q},b}^{(1)}+\Phi_{b\bar{q},c}^{(1)}+\Phi_{b\bar{q},\mathrm{bwf}}^{(1)}+\Phi_{b\bar{q},\mathrm{dwf}}^{(1)}\right]\otimes T_i^{(a,0)} \\ = \left[\Pi_{i,\mathrm{weak}}^{(a,1),\mathrm{h}}+\left(\Pi_{i,\mathrm{bwf}}^{(a,1)}-\Phi_{b\bar{q},\mathrm{bwf}}^{(a,1)}\otimes T_i^{(a,0)}\right)\right]+\left[\Pi_{i,\mathrm{weak}}^{(a,1),\mathrm{hc}}+\Pi_{i,\mathrm{axial}}^{(a,1),\mathrm{hc}}+\Pi_{i,\mathrm{wfc}}^{(a,1),\mathrm{hc}}\right], \tag{3.51}$$

where the terms in the first and second square brackets of the second equality correspond to the hard coefficient functions and the jet functions at $\mathcal{O}(\alpha_s)$. The one-loop factorization formulas of the correlation functions defined in (2.8) at leading power in $\Lambda/m_b$ can be written as

$$
\begin{aligned}
\Pi_1 &= \frac{1}{2}\tilde{f}_B(\mu) m_B \sum_{k=\pm} C_1^{(k)}(n\cdot p,\mu) \int_0^\infty \frac{d\omega}{\omega - \bar{n}\cdot p} J_1^{(k)}\left(\frac{\mu^2}{n\cdot p\omega}, \frac{\omega}{\bar{n}\cdot p}\right) \phi_B^{(k)}(\omega,\mu),\\
\tilde{\Pi}_1 &= \frac{1}{2}\tilde{f}_B(\mu) m_B \sum_{k=\pm} \tilde{C}_1^{(k)}(n\cdot p,\mu) \int_0^\infty \frac{d\omega}{\omega - \bar{n}\cdot p} \tilde{J}_1^{(k)}\left(\frac{\mu^2}{n\cdot p\omega}, \frac{\omega}{\bar{n}\cdot p}\right) \phi_B^{(k)}(\omega,\mu),\\
\tilde{\Pi}_2 &= -\frac{1}{4}\tilde{f}_B(\mu) m_B \sum_{k=\pm} \tilde{C}_2^{(k)}(n\cdot p,\mu) \int_0^\infty \frac{d\omega}{\omega - \bar{n}\cdot p} \tilde{J}_2^{(k)}\left(\frac{\mu^2}{n\cdot p\omega}, \frac{\omega}{\bar{n}\cdot p}\right) \phi_B^{(k)}(\omega,\mu),\\
\tilde{\Pi}_3 &= \frac{1}{4}\tilde{f}_B(\mu) m_B \sum_{k=\pm} \tilde{C}_3^{(k)}(n\cdot p,\mu) \int_0^\infty \frac{d\omega}{\omega - \bar{n}\cdot p} \tilde{J}_3^{(k)}\left(\frac{\mu^2}{n\cdot p\omega}, \frac{\omega}{\bar{n}\cdot p}\right) \phi_B^{(k)}(\omega,\mu),\\
\tilde{\Pi}_4 &= -\frac{1}{4}\tilde{f}_B(\mu) m_B \sum_{k=\pm} \tilde{C}_4^{(k)}(n\cdot p,\mu) \int_0^\infty \frac{d\omega}{\omega - \bar{n}\cdot p} \tilde{J}_4^{(k)}\left(\frac{\mu^2}{n\cdot p\omega}, \frac{\omega}{\bar{n}\cdot p}\right) \phi_B^{(k)}(\omega,\mu).
\end{aligned} \tag{3.52}
$$

The corresponding hard matching coefficients for the four sets of correlation functions are

$$
\begin{aligned}
C_1^{(+)} &= \tilde{C}_1^{(+)} = 1,\\
C_1^{(-)} &= \frac{\alpha_s C_F}{4\pi}\frac{1}{\bar{r}}\left[\frac{r}{\bar{r}}\ln r + 1\right],\\
\tilde{C}_1^{(-)} &= 1 - \frac{\alpha_s C_F}{4\pi}\left[2\ln^2\frac{\mu}{n\cdot p} + 5\ln\frac{\mu}{m_b} - \ln^2 r - 2\mathrm{Li}_2\left(-\frac{\bar{r}}{r}\right) + \frac{2-r}{r-1}\ln r + \frac{\pi^2}{12} + 5\right],
\end{aligned} \tag{3.53}
$$

$$
\begin{aligned}
\tilde{C}_2^{(+)} &= 1,\\
\tilde{C}_2^{(-)} &= 1 - \frac{\alpha_s C_F}{4\pi}\left[2\ln^2\frac{\mu}{n\cdot p} + 7\ln\frac{\mu}{m_b} - \ln^2 r - 2\mathrm{Li}_2\left(-\frac{\bar{r}}{r}\right) - 2\frac{\bar{r}-r}{\bar{r}}\ln r + \frac{\pi^2}{12} + 6\right],
\end{aligned} \tag{3.54}
$$

$$
\begin{aligned}
\tilde{C}_3^{(+)} &= 0,\\
\tilde{C}_3^{(-)} &= 1 - \frac{\alpha_s C_F}{4\pi}\left[2\ln^2\frac{\mu}{n\cdot p} + 5\ln\frac{\mu}{m_b} - \ln^2 r - 2\mathrm{Li}_2\left(-\frac{\bar{r}}{r}\right) - \frac{3\bar{r}-1}{\bar{r}}\ln r + \frac{\pi^2}{12} + 6\right],
\end{aligned} \tag{3.55}
$$

$$
\begin{aligned}
\tilde{C}_4^{(+)} &= 1,\\
\tilde{C}_4^{(-)} &= 1 - \frac{\alpha_s C_F}{4\pi}\left[2\ln^2\frac{\mu}{n\cdot p} + 7\ln\frac{\mu}{m_b} - \ln^2 r - 2\mathrm{Li}_2\left(-\frac{\bar{r}}{r}\right) - 2\ln r + \frac{\pi^2}{12} + 6\right].
\end{aligned} \tag{3.56}
$$

The jet functions are given as follows:

$$
\begin{aligned}
J_1^{(+)} &= \frac{1}{r}\tilde{J}^{(+)} = \frac{\alpha_s C_F}{4\pi}\left(1 - \frac{\bar{n}\cdot p}{\omega}\right)\ln\left(1 - \frac{\omega}{\bar{n}\cdot p}\right),\\
J_1^{(-)} &= 1,\\
\tilde{J}_1^{(-)} &= 1 + \frac{\alpha_s C_F}{4\pi}\left[\ln^2\frac{\mu^2}{n\cdot p(\omega - \bar{n}\cdot p)} - 2\ln\frac{\bar{n}\cdot p - \omega}{\bar{n}\cdot p}\ln\frac{\mu^2}{n\cdot p(\omega - \bar{n}\cdot p)}\right.\\
&\qquad \left. - \ln^2\frac{\bar{n}\cdot p - \omega}{\bar{n}\cdot p} - \left(1 + \frac{2\bar{n}\cdot p}{\omega}\right)\ln\frac{\bar{n}\cdot p - \omega}{\bar{n}\cdot p} - \frac{\pi^2}{6} - 1\right],
\end{aligned} \tag{3.57}
$$

$$
\begin{aligned}
\tilde{J}_2^{(+)} &= -\frac{\alpha_s C_F}{4\pi}\left[\left(1-\frac{\bar{n}\cdot p}{\omega}\right)\ln\left(1-\frac{\omega}{\bar{n}\cdot p}\right)\right],\\
\tilde{J}_2^{(-)} &= 1+\frac{\alpha_s C_F}{4\pi}\left[\ln^2\frac{\mu^2}{n\cdot p(\omega-\bar{n}\cdot p)}-2\ln\frac{\bar{n}\cdot p-\omega}{\bar{n}\cdot p}\ln\frac{\mu^2}{n\cdot p(\omega-\bar{n}\cdot p)}\right.\\
&\quad\left.-\ln^2\frac{\bar{n}\cdot p-\omega}{\bar{n}\cdot p}-\left(1+\frac{2\bar{n}\cdot p}{\omega}\right)\ln\frac{\bar{n}\cdot p-\omega}{\bar{n}\cdot p}-\frac{\pi^2}{6}-1\right],
\end{aligned}
\tag{3.58}
$$

$$
\begin{aligned}
\tilde{J}_3^{(+)} &= 0,\\
\tilde{J}_3^{(-)} &= 1+\frac{\alpha_s C_F}{4\pi}\left[\ln^2\frac{\mu^2}{n\cdot p(\omega-\bar{n}\cdot p)}-2\ln\frac{\bar{n}\cdot p-\omega}{\bar{n}\cdot p}\ln\frac{\mu^2}{n\cdot p(\omega-\bar{n}\cdot p)}\right.\\
&\quad\left.-\ln\frac{\mu^2}{n\cdot p(\omega-\bar{n}\cdot p)}-\ln^2\frac{\bar{n}\cdot p-\omega}{\bar{n}\cdot p}-\left(2+\frac{2\bar{n}\cdot p}{\omega}\right)\ln\frac{\bar{n}\cdot p-\omega}{\bar{n}\cdot p}-\frac{\pi^2}{6}-1\right],
\end{aligned}
\tag{3.59}
$$

$$
\begin{aligned}
\tilde{J}_4^{(+)} &= \frac{\alpha_s C_F}{4\pi}\left[\left(1-\frac{\bar{n}\cdot p}{\omega}\right)r\ln\left(1-\frac{\omega}{\bar{n}\cdot p}\right)\right],\\
\tilde{J}_4^{(-)} &= 1+\frac{\alpha_s C_F}{4\pi}\left[\ln^2\frac{\mu^2}{n\cdot p(\omega-\bar{n}\cdot p)}-2\ln\frac{\bar{n}\cdot p-\omega}{\bar{n}\cdot p}\ln\frac{\mu^2}{n\cdot p(\omega-\bar{n}\cdot p)}\right.\\
&\quad\left.-\ln\frac{\mu^2}{n\cdot p(\omega-\bar{n}\cdot p)}-\ln^2\frac{\bar{n}\cdot p-\omega}{\bar{n}\cdot p}-\left(2+\frac{2\bar{n}\cdot p}{\omega}\right)\ln\frac{\bar{n}\cdot p-\omega}{\bar{n}\cdot p}-\frac{\pi^2}{6}-1\right].
\end{aligned}
\tag{3.60}
$$

The hard functions contain two different scales related to the UV (because the currents require renormalization) and IR which matches the jet and soft function. Since the tensor currents are not conserved, the hard coefficients $\tilde{C}_2^{(-)}$ and $\tilde{C}_4^{(-)}$ are scale dependent. Similarly, the nonconservation of the interpolating QCD current $\not{n}\gamma_{\delta\perp}\gamma_5$ implies that the jet functions $\tilde{J}_3^{(-)}$ and $\tilde{J}_4^{(-)}$ lead to an additional scale-dependent term. There also exist two relations $\tilde{J}_1^{(-)}=\tilde{J}_2^{(-)}$ and $\tilde{J}_3^{(-)}=\tilde{J}_4^{(-)}$ because $\tilde{J}_1^{(-)}$ and $\tilde{J}_2^{(-)}$ are computed from the same interpolating current of axial-vector mesons as well as $\tilde{J}_3^{(-)}$ and $\tilde{J}_4^{(-)}$. It is obvious that the hard functions (3.53) and (3.54) and jet functions (3.57) and (3.58) of semileptonic $B\to A$ are consistent with the corresponding ones of $B\to\pi,K$ [44] under the limit $m_q\to 0$ because of the same heavy-to-light transition currents and the interpolating particle currents entering the correlation functions in these decay modes. The $\mu$-dependent hard coefficients in (3.54)–(3.56) and jet functions in (3.57)–(3.60) coincide with the $B\to V$ corresponding results obtained via the A0-type $\mathrm{SCET_I}$ currents in [46]. In addition, our hard function $\tilde{C}_3^{(-)}$ also coincides with the $B\to A\gamma$ one [60] in SCET.

In order to verify the factorization-scale dependence of the correlation functions, the evolution equations are straightforward given as

$$
\begin{aligned}
&\frac{d}{d\ln\mu}\tilde{C}^{(-)}(n\cdot p,\mu) = -\frac{\alpha_s C_F}{4\pi}\left[\Gamma_{\rm cusp}^{(0)}\ln\frac{\mu}{n\cdot p}+5\right]\tilde{C}^{(-)}(n\cdot p,\mu),\\
&\frac{d}{d\ln\mu}\frac{\tilde{J}^{(-)}\left(\frac{\mu^2}{n\cdot p\omega},\frac{\omega}{\bar{n}\cdot p}\right)}{\bar{n}\cdot p-\omega+i0} = \frac{\alpha_s C_F}{4\pi}\left[\Gamma_{\rm cusp}^{(0)}\ln\frac{\mu^2}{n\cdot p\omega}\right]\frac{\tilde{J}^{(-)}\left(\frac{\mu^2}{n\cdot p\omega},\frac{\omega}{\bar{n}\cdot p}\right)}{\bar{n}\cdot p-\omega+i0}+\frac{\alpha_s C_F}{4\pi}\int_0^\infty d\omega'\Gamma(\omega,\omega',\mu)\frac{\tilde{J}^{(-)}\left(\frac{\mu^2}{n\cdot p\omega'},\frac{\omega'}{\bar{n}\cdot p}\right)}{\bar{n}\cdot p-\omega'+i0},\\
&\frac{d}{d\ln\mu}[\tilde{f}_B(\mu)\phi_B^-(\omega,\mu)] = -\frac{\alpha_s C_F}{4\pi}\left[\Gamma_{\rm cusp}^{(0)}\ln\frac{\mu}{\omega}-5\right][\tilde{f}_B(\mu)\phi_B^-(\omega,\mu)]-\frac{\alpha_s C_F}{4\pi}\int_0^\infty d\omega'\omega\Gamma(\omega,\omega',\mu)[\tilde{f}_B(\mu)\phi_B^-(\omega',\mu)],
\end{aligned}
\tag{3.61}
$$

with the function $\Gamma$ given by [61]

$$
\Gamma(\omega,\omega',\mu) = -\Gamma_{\rm cusp}^{(0)}\frac{\theta(\omega'-\omega)}{\omega'\omega}-\Gamma_{\rm cusp}^{(0)}\left[\frac{\theta(\omega'-\omega)}{\omega'(\omega'-\omega)}+\frac{\theta(\omega-\omega')}{\omega(\omega-\omega')}\right]_\oplus,
\tag{3.62}
$$

and $\Gamma_{\rm cusp}^{(0)} = 4$ because of the geometry of the Wilson line [42] up to one-loop accuracy. Observing the results mentioned above, we can straightforwardly write the scale dependence of the correlation functions as

$$\begin{aligned} &\frac{d}{d\ln\mu}\left[\Pi_1^{(1)}, \tilde{\Pi}_1^{(1)}\right] = \mathcal{O}(\alpha_s^2), \qquad \frac{d}{d\ln\mu}\tilde{\Pi}_2^{(1)} + \gamma_T \tilde{\Pi}_2^{(0)} = \mathcal{O}(\alpha_s^2), \\ &\frac{d}{d\ln\mu}\tilde{\Pi}_3^{(1)} + \gamma_A \tilde{\Pi}_3^{(0)} = \mathcal{O}(\alpha_s^2), \qquad \frac{d}{d\ln\mu}\tilde{\Pi}_4^{(1)} + (\gamma_A + \gamma_T)\tilde{\Pi}_4^{(0)} = \mathcal{O}(\alpha_s^2), \end{aligned} \tag{3.63}$$

where the anomalous dimensions of the interpolating axial-vector-meson pseudotensor current and tensor heavy-to-light transition current read

$$\gamma_A = 2\frac{\alpha_s C_F}{4\pi}, \qquad \gamma_T = 2\frac{\alpha_s C_F}{4\pi}. \tag{3.64}$$

After investigating Eqs. (3.53)–(3.56), (3.57)–(3.60), and (2.16), the appearance of the large logarithms of order $\ln m_b/\Lambda_{\rm QCD}$ is inevitable in the hard functions, jet functions, $\tilde{f}_B(\mu)$, and $B$-meson distribution amplitudes regardless of how to choose the common value $\mu$. Solving the three renormalization group (RG) equations mentioned above, we can accomplish the resummation of these logarithms to all orders of $\alpha_s$. By selecting $\mu$ at the hard-collinear scale $\mu_{hc} \sim \sqrt{m_b \Lambda}$, one obtains the evolution functions from the running of the renormalization scale from the hard scale $\mu_{h_1} \sim n \cdot p$ to $\mu_{hc}$ in $\tilde{C}_i^{(-)}$ ($i = 1, 2, 3, 4$), and $\mu_{h_2} \sim m_b$ to $\mu_{hc}$ in $\tilde{f}_B(\mu)$ as

$$\begin{aligned} \tilde{C}_i^{(-)}(n \cdot p, \mu) &= U_1(n \cdot p, \mu_{h_1}, \mu) \tilde{C}_i^{(-)}(n \cdot p, \mu_{h_1}), \\ \tilde{f}_B(\mu) &= U_2(\mu_{h_2}, \mu) \tilde{f}_B(\mu_{h_2}). \end{aligned} \tag{3.65}$$

For the sake of the indispensable NLL resummation of large logarithms in the hard coefficients $\tilde{C}_i^{(-)}$ and $\tilde{f}_B(\mu)$, the general expressions of the RG equations (3.65) are

$$\begin{aligned} \frac{d}{d\ln\mu}\tilde{C}^{(-)}(n \cdot p, \mu) &= \left[-\Gamma_{\rm cusp}(\alpha_s)\ln\frac{\mu}{n \cdot p} + \gamma(\alpha_s)\right]\tilde{C}^{(-)}(n \cdot p, \mu), \\ \frac{d}{d\ln\mu}\tilde{f}_B(\mu) &= \tilde{\gamma}(\alpha_s)\tilde{f}_B(\mu), \end{aligned} \tag{3.66}$$

with the expansion of cusp anomalous dimension $\Gamma_{\rm cusp}$, $\gamma(\alpha_s)$, and $\tilde{\gamma}(\alpha_s)$,

$$\begin{aligned} \Gamma_{\rm cusp}(\alpha_s) &= \frac{\alpha_s C_F}{4\pi}\left[\Gamma_{\rm cusp}^{(0)} + \left(\frac{\alpha_s}{4\pi}\right)\Gamma_{\rm cusp}^{(1)} + \left(\frac{\alpha_s}{4\pi}\right)^2\Gamma_{\rm cusp}^{(2)} + \cdots\right], \\ \gamma(\alpha_s) &= \frac{\alpha_s C_F}{4\pi}\left[\gamma^{(0)} + \left(\frac{\alpha_s}{4\pi}\right)\gamma^{(1)} + \cdots\right], \\ \tilde{\gamma}(\alpha_s) &= \frac{\alpha_s C_F}{4\pi}\left[\tilde{\gamma}^{(0)} + \left(\frac{\alpha_s}{4\pi}\right)\tilde{\gamma}^{(1)} + \cdots\right]. \end{aligned} \tag{3.67}$$

One can find the specific expressions of $\Gamma_{\rm cusp}$, $\gamma(\alpha_s)$, and $\tilde{\gamma}(\alpha_s)$ in [62], and solve the RG equations up to NLL accuracy to obtain

$$\begin{aligned} U_1(n \cdot p, \mu_{h_1}, \mu) &= \exp\left(\int_{\alpha_s(\mu_{h_1})}^{\alpha_s(\mu)} d\alpha_s\left[\frac{\gamma(\alpha_s)}{\beta(\alpha_s)} + \frac{\Gamma_{\rm cusp}(\alpha_s)}{\beta(\alpha_s)}\left(\ln\frac{n \cdot p}{\mu_{h_1}} - \int_{\alpha_s(\mu_{h_1})}^{\alpha_s(\mu)}\frac{d\alpha_s'}{\beta(\alpha_s')}\right)\right]\right) \\ &= \exp\left(-\frac{\Gamma_0}{4\beta_0^2}\left(\frac{4\pi}{\alpha_s(\mu_{h_1})}\left[\ln r' - 1 + \frac{1}{r'}\right] - \frac{\beta_1}{2\beta_0}\ln^2 r' + \left(\frac{\Gamma_1}{\Gamma_0} - \frac{\beta_1}{\beta_0}\right)[r' - 1 - \ln r']\right)\right) \times \left(\frac{n \cdot p}{\mu_{h_1}}\right)^{-\frac{\Gamma_0}{2\beta_0}\ln r'} r'^{-\frac{\gamma_0}{2\beta_0}} \\ &\quad \times \left[1 - \frac{\alpha_s(\mu_{h_1})}{4\pi}\frac{\Gamma_0}{4\beta_0^2}\left(\frac{\Gamma_2}{2\Gamma_0}[1 - r']^2 + \frac{\beta_2}{2\beta_0}[1 - r'^2 + 2\ln r'] - \frac{\Gamma_1\beta_1}{2\Gamma_0\beta_0}[3 - 4r' + r'^2 + 2r'\ln r']\right.\right. \\ &\quad \left.\left. + \frac{\beta_1^2}{2\beta_0^2}[1 - r'][1 - r' - 2\ln r']\right) + \frac{\alpha_s(\mu_{h_1})}{4\pi}\left(\ln\frac{n \cdot p}{\mu_{h_1}}\left(\frac{\Gamma_1}{2\beta_0} - \frac{\Gamma_0\beta_1}{2\beta_0^2}\right) + \frac{\gamma_1}{2\beta_0} - \frac{\gamma_0\beta_1}{2\beta_0^2}\right)[1 - r'] + \mathcal{O}(\alpha_s^2)\right], \end{aligned} \tag{3.68}$$

$$U_2(\mu_{h_2},\mu) = \exp\left[\int_{\alpha_s(\mu_{h_2})}^{\alpha_s(\mu)} d\alpha_s \frac{\tilde{\gamma}(\alpha_s)}{\beta(\alpha_s)}\right]$$
$$= z^{-\frac{\tilde{\gamma}_0}{2\beta_0}C_F}\left[1+\frac{\alpha_s(\mu_{h_2})C_F}{4\pi}\left(\frac{\tilde{\gamma}^{(1)}}{2\beta_0}-\frac{\tilde{\gamma}^{(0)}\beta_1}{2\beta_0^2}\right)(1-z)+\mathcal{O}(\alpha_s^2)\right], \tag{3.69}$$

with $z = \alpha_s(\mu)/\alpha_s(\mu_{h_2})$, $r' = \alpha_s(\mu)/\alpha_s(\mu_{h_1})$.

Eventually, the manifest expressions of the RG improved correlation functions at NLL accuracy are derived as follows:

$$\Pi_1 = \frac{1}{2}m_B[U_2(\mu_{h_2},\mu)\tilde{f}_B(\mu_{h_2})]\int_0^\infty \frac{d\omega}{\omega-\bar{n}\cdot p}J_1^{(+)}\left(\frac{\mu^2}{n\cdot p\omega},\frac{\omega}{\bar{n}\cdot p}\right)\phi_B^{(+)}(\omega,\mu)+\frac{1}{2}m_B[U_2(\mu_{h_2},\mu)\tilde{f}_B(\mu_{h_2})]C_1^{(-)}(n\cdot p,\mu)$$
$$\times\int_0^\infty \frac{d\omega}{\omega-\bar{n}\cdot p}\phi_B^{(-)}(\omega,\mu),$$
$$\tilde{\Pi}_1 = \frac{1}{2}m_B[U_2(\mu_{h_2},\mu)\tilde{f}_B(\mu_{h_2})]\int_0^\infty \frac{d\omega}{\omega-\bar{n}\cdot p}\tilde{J}_1^{(+)}\left(\frac{\mu^2}{n\cdot p\omega},\frac{\omega}{\bar{n}\cdot p}\right)\phi_B^{(+)}(\omega,\mu)+\frac{1}{2}m_B[U_1(n\cdot p,\mu_{h_1},\mu)U_2(\mu_{h_2},\mu)]$$
$$\times[\tilde{f}_B(\mu_{h_2})\tilde{C}_1^{(-)}(n\cdot p,\mu_{h_1})]\int_0^\infty \frac{d\omega}{\omega-\bar{n}\cdot p}\tilde{J}_1^{(-)}\left(\frac{\mu^2}{n\cdot p\omega},\frac{\omega}{\bar{n}\cdot p}\right)\phi_B^{(-)}(\omega,\mu), \tag{3.70}$$

$$\tilde{\Pi}_2 = -\frac{1}{4}m_B[U_2(\mu_{h_2},\mu)\tilde{f}_B(\mu_{h_2})]\left\{\int_0^\infty \frac{d\omega}{\omega-\bar{n}\cdot p}\tilde{J}_2^{(+)}\left(\frac{\mu^2}{n\cdot p\omega},\frac{\omega}{\bar{n}\cdot p}\right)\phi_B^{+}(\omega,\mu)\right.$$
$$\left.+[U_1(n\cdot p,\mu_{h_1},\mu)\tilde{C}_2^{(-)}(n\cdot p,\mu_{h_1})]\int_0^\infty \frac{d\omega}{\omega-\bar{n}\cdot p}\tilde{J}_2^{(-)}\left(\frac{\mu^2}{n\cdot p\omega},\frac{\omega}{\bar{n}\cdot p}\right)\phi_B^{-}(\omega,\mu)\right\}, \tag{3.71}$$

$$\tilde{\Pi}_3 = \frac{1}{4}m_B[U_1(n\cdot p,\mu_{h_1},\mu)U_2(\mu_{h_2},\mu)][\tilde{f}_B(\mu_{h_2})\tilde{C}_3^{(-)}(n\cdot p,\mu_{h_1})]\int_0^\infty \frac{d\omega}{\omega-\bar{n}\cdot p}\tilde{J}_3^{(-)}\left(\frac{\mu^2}{n\cdot p\omega},\frac{\omega}{\bar{n}\cdot p}\right)\phi_B^{(-)}(\omega,\mu), \tag{3.72}$$

$$\tilde{\Pi}_4 = -\frac{1}{4}m_B[U_2(\mu_{h_2},\mu)\tilde{f}_B(\mu_{h_2})]\left\{\int_0^\infty \frac{d\omega}{\omega-\bar{n}\cdot p}\tilde{J}_4^{(+)}\left(\frac{\mu^2}{n\cdot p\omega},\frac{\omega}{\bar{n}\cdot p}\right)\phi_B^{+}(\omega,\mu)+[U_1(n\cdot p,\mu_{h_1},\mu)\tilde{C}_4^{(-)}(n\cdot p,\mu_{h_1})]\right.$$
$$\left.\times\int_0^\infty \frac{d\omega}{\omega-\bar{n}\cdot p}\tilde{J}_4^{(-)}\left(\frac{\mu^2}{n\cdot p\omega},\frac{\omega}{\bar{n}\cdot p}\right)\phi_B^{-}(\omega,\mu)\right\}. \tag{3.73}$$

## IV. THE LCSR FOR $B \to A$ FORM FACTORS AT $\mathcal{O}(\alpha_s)$

Implementing the Borel transformation and matching the hadronic representations of the correlation functions (2.10) and renormalization group improved correlation functions (3.70)–(3.73) via the parton-hadron duality approximation yields the NLL LCSR for the semileptonic $B \to A$ form factors at leading power (LP)

$$-\frac{f_A^T n\cdot p}{m_B}e^{m_A^2/(n\cdot p\omega_M)}\left[\frac{m_B}{m_B-m_A}A^{\rm NLL}(q^2)\right]$$
$$=[U_1(n\cdot p,\mu_{h_1},\mu)U_2(\mu_{h_2},\mu)][\tilde{f}_B(\mu_{h_2})\tilde{C}_3^{(-)}(n\cdot p,\mu_{h_1})]\int_0^{\omega_s}d\omega' e^{-\omega'/\omega_M}\phi_{B,\mathrm{eff},\perp}^{-}(\omega',\mu),$$
$$\frac{f_A(n\cdot p)^2}{2m_A m_B}e^{m_A^2/(n\cdot p\omega_M)}\left[\frac{2m_A}{n\cdot p}V_0^{\rm NLL}(q^2)\right]$$
$$=[U_2(\mu_{h_2},\mu)\tilde{f}_B(\mu_{h_2})]\int_0^{\omega_s}d\omega' e^{-\omega'/\omega_M}\left\{-r\phi_{B,\mathrm{eff}}^{+}(\omega',\mu)+[U_1(n\cdot p,\mu_{h_1},\mu)\tilde{C}_1^{(-)}(n\cdot p,\mu_{h_1})]\phi_{B,\mathrm{eff},\parallel}^{-}(\omega',\mu)\right.$$
$$\left.+\frac{m_B-n\cdot p}{m_B}(-\phi_{B,\mathrm{eff}}^{+}(\omega',\mu)+C_1^{(-)}(n\cdot p,\mu)\phi_B^{-}(\omega',\mu))\right\},$$

$$
\begin{aligned}
&-\frac{f_A^T n\cdot p}{m_B} e^{m_A^2/(n\cdot p\omega_M)}\left[\frac{m_B-m_A}{n\cdot p}V_1^{\rm NLL}(q^2)\right]\\
&\quad=[U_1(n\cdot p,\mu_{h_1},\mu)U_2(\mu_{h_2},\mu)][\tilde{f}_B(\mu_{h_2})\tilde{C}_3^{(-)}(n\cdot p,\mu_{h_1})]\int_0^{\omega_s}d\omega' e^{-\omega'/\omega_M}\phi^-_{B,\rm eff,\perp}(\omega',\mu),\\
&\frac{f_A(n\cdot p)^2}{2m_A m_B} e^{m_A^2/(n\cdot p\omega_M)}\left[\frac{m_B-m_A}{n\cdot p}V_1^{\rm NLL}(q^2)-\frac{(m_B+m_A)}{m_B}V_2^{\rm NLL}(q^2)\right]\\
&\quad=[U_2(\mu_{h_2},\mu)\tilde{f}_B(\mu_{h_2})]\int_0^{\omega_s}d\omega' e^{-\omega'/\omega_M}\bigg\{-r\phi^+_{B,\rm eff}(\omega',\mu)+[U_1(n\cdot p,\mu_{h_1},\mu)\tilde{C}_1^{(-)}(n\cdot p,\mu_{h_1})]\phi^-_{B,\rm eff,\parallel}(\omega',\mu)\\
&\qquad+\frac{n\cdot p-m_B}{m_B}(-\phi^+_{B,\rm eff}(\omega',\mu)+C_1^{(-)}(n\cdot p,\mu)\phi^-_B(\omega',\mu))\bigg\},\\
&-\frac{f_A^T n\cdot p}{m_B} e^{m_A^2/(n\cdot p\omega_M)}T_1^{\rm NLL}(q^2)\\
&\quad=[U_2(\mu_{h_2},\mu)\tilde{f}_B(\mu_{h_2})]\int_0^{\omega_s}d\omega' e^{-\omega'/\omega_M}\{-r\phi^+_{B,\rm eff}(\omega',\mu)+[U_1(n\cdot p,\mu_{h_1},\mu)\tilde{C}_4^{(-)}(n\cdot p,\mu_{h_1})]\phi^-_{B,\rm eff,\perp}(\omega',\mu)\},\\
&-\frac{f_A^T n\cdot p}{m_B} e^{m_A^2/(n\cdot p\omega_M)}\left[\frac{m_B}{n\cdot p}T_2^{\rm NLL}(q^2)\right]\\
&\quad=[U_2(\mu_{h_2},\mu)\tilde{f}_B(\mu_{h_2})]\int_0^{\omega_s}d\omega' e^{-\omega'/\omega_M}\{-r\phi^+_{B,\rm eff}(\omega',\mu)+[U_1(n\cdot p,\mu_{h_1},\mu)\tilde{C}_4^{(-)}(n\cdot p,\mu_{h_1})]\phi^-_{B,\rm eff,\perp}(\omega',\mu)\},\\
&\frac{f_A(n\cdot p)^2}{2m_A m_B} e^{m_A^2/(n\cdot p\omega_M)}\left[\frac{m_B}{n\cdot p}T_2^{\rm NLL}(q^2)-T_3^{\rm NLL}(q^2)\right]\\
&\quad=[U_2(\mu_{h_2},\mu)\tilde{f}_B(\mu_{h_2})]\int_0^{\omega_s}d\omega' e^{-\omega'/\omega_M}\{\phi^+_{B,\rm eff}(\omega',\mu)+[U_1(n\cdot p,\mu_{h_1},\mu)\tilde{C}_2^{(-)}(n\cdot p,\mu_{h_1})]\phi^-_{B,\rm eff,\parallel}(\omega',\mu)\},
\end{aligned}
\tag{4.1}
$$

where the effective $B$-meson DAs are

$$
\phi^+_{B,\rm eff}(\omega',\mu)=-\frac{\alpha_s C_F}{4\pi}\int_{\omega'}^{\infty}\frac{d\omega}{\omega}\phi^+_B(\omega,\mu),
\tag{4.2}
$$

$$
\begin{aligned}
\phi^-_{B,\rm eff,\parallel}(\omega',\mu)=\phi^-_B(\omega',\mu)+\frac{\alpha_s C_F}{4\pi}\bigg\{&\int_0^{\omega'}d\omega\left[\frac{2}{\omega-\omega'}\left(\ln\frac{\mu^2}{n\cdot p\omega'}-2\ln\frac{\omega'-\omega}{\omega'}\right)\right]_{\oplus}\\
&\times\phi^-_B(\omega,\mu)-\int_{\omega'}^{\infty}d\omega\left[\ln^2\frac{\mu^2}{n\cdot p\omega'}-\left(2\ln\frac{\mu^2}{n\cdot p\omega'}+3\right)\ln\frac{\omega-\omega'}{\omega'}\right.\\
&\left.+2\ln\frac{\omega}{\omega'}+\frac{\pi^2}{6}-1\right]\frac{d\phi^-_B(\omega,\mu)}{d\omega}\bigg\},
\end{aligned}
\tag{4.3}
$$

$$
\begin{aligned}
\phi^-_{B,\rm eff,\perp}(\omega',\mu)=\phi^-_B(\omega',\mu)+\frac{\alpha_s C_F}{4\pi}\bigg\{&\int_0^{\omega'}d\omega\left[\frac{2}{\omega-\omega'}\left(\ln\frac{\mu^2}{n\cdot p\omega'}-2\ln\frac{\omega'-\omega}{\omega'}-\frac{1}{2}\right)\right]_{\oplus}\\
&\times\phi^-_B(\omega,\mu)-\int_{\omega'}^{\infty}d\omega\left[\ln^2\frac{\mu^2}{n\cdot p\omega'}-\left(3\ln\frac{\mu^2}{n\cdot p\omega'}+4\right)\ln\frac{\omega-\omega'}{\omega'}\right.\\
&\left.+2\ln\frac{\omega}{\omega'}+\frac{\pi^2}{6}-1\right]\frac{d\phi^-_B(\omega,\mu)}{d\omega}\bigg\}.
\end{aligned}
\tag{4.4}
$$

## V. THE HIGHER-TWIST CONTRIBUTIONS TO THE LCSR

We will proceed to derive the formulas of higher-twist corrected $B \to A$ form factors at tree level including both two-particle and three-particle corrections utilizing the LCSR approach. According to the standard strategy in [44,46], we start with the light-cone expansion of the quark propagator in the background gluon filled up to the gluon field strength terms without the covariant derivatives [63]

$$\langle 0|\mathrm{T}\{\bar{q}(x),q(0)\}|0\rangle \supset ig_s \int_0^\infty \frac{d^4 l}{(2\pi)^4} e^{-il\cdot x} \int_0^1 du \left[\frac{u x_\mu \gamma_\nu}{l^2-m^2} - \frac{(\not{l}+m)\sigma_{\mu\nu}}{2(l^2-m^2)^2}\right] G^{\mu\nu}(ux), \tag{5.1}$$

with $G_{\mu\nu}=G^a_{\mu\nu}T^a=[D_\mu,A_\nu]$. Adopting the parametrized vacuum-to-$B$-meson matrix element by the three-body light-ray HQET operator [64]

$$\begin{aligned}\langle 0|\bar{q}_\alpha(z_1\bar{n})g_s G_{\mu\nu}(z_2\bar{n})h_{v\beta}(0)|\bar{B}_v\rangle = \frac{\tilde{f}_B(\mu)m_B}{4}\big[(1+\not{v})\{(v_\mu\gamma_\nu - v_\nu\gamma_\mu)[\Psi_A(z_1,z_2,\mu)-\Psi_V(z_1,z_2,\mu)] - i\sigma_{\mu\nu}\Psi_V(z_1,z_2,\mu)\\ -(\bar{n}_\mu v_\nu - \bar{n}_\nu v_\mu)X_A(z_1,z_2,\mu) + (\bar{n}_\mu\gamma_\nu - \bar{n}_\nu\gamma_\mu)\big[W(z_1,z_2,\mu)+Y_A(z_1,z_2,\mu)\big]\\ + i\epsilon_{\mu\nu\alpha\beta}\bar{n}^\alpha v^\beta\gamma_5\tilde{X}_A(z_1,z_2,\mu) - i\epsilon_{\mu\nu\alpha\beta}\bar{n}^\alpha\gamma^\beta\gamma_5\tilde{Y}_A(z_1,z_2,\mu)\\ -(\bar{n}_\mu v_\nu - \bar{n}_\nu v_\mu)\not{\bar{n}}W(z_1,z_2,\mu) + (\bar{n}_\mu\gamma_\nu - \bar{n}_\nu\gamma_\mu)\not{\bar{n}}Z(z_1,z_2,\mu)\}\gamma_5\big]_{\beta\alpha},\end{aligned} \tag{5.2}$$

the three-particle higher-twist corrections to the defined correlation functions in (2.5) up to NLP accuracy obtained from the tree diagram are derived as follows:

$$\begin{aligned}\Pi^{(V-A),3\mathrm{P}}_{\mu,\parallel}(p,q) &= -\frac{i\tilde{f}_B(\mu)m_B}{2n\cdot p}\int_0^\infty d\omega_1\int_0^\infty d\omega_2\int_0^1 du\frac{1}{[\bar{n}\cdot p-\omega_1-u\omega_2+i0]^2}\Big\{\bar{n}_\mu\Big[\rho^{(V-A),3\mathrm{P}}_{\bar{n},\parallel,\mathrm{LP}}(u,\omega_1,\omega_2,\mu)\\ &\quad+\frac{m}{n\cdot p}\rho^{(V-A),3\mathrm{P}}_{\bar{n},\parallel,\mathrm{NLP}}(u,\omega_1,\omega_2,\mu)\Big] + n_\mu\Big[\rho^{(V-A),3\mathrm{P}}_{n,\parallel,\mathrm{LP}}(u,\omega_1,\omega_2,\mu)+\frac{m}{n\cdot p}\rho^{(V-A),3\mathrm{P}}_{n,\parallel,\mathrm{NLP}}(u,\omega_1,\omega_2,\mu)\Big]\Big\},\\ \Pi^{(T+\tilde{T}),3\mathrm{P}}_{\mu,\parallel}(p,q) &= \frac{i\tilde{f}_B(\mu)m_B}{4n\cdot p}\int_0^\infty d\omega_1\int_0^\infty d\omega_2\int_0^1 du\frac{1}{[\bar{n}\cdot p-\omega_1-u\omega_2+i0]^2}\\ &\quad\times[n\cdot q\bar{n}_\mu - \bar{n}\cdot q n_\mu]\Big[\rho^{(T+\tilde{T}),3\mathrm{P}}_{\parallel,\mathrm{LP}}(u,\omega_1,\omega_2)+\frac{m}{n\cdot p}\rho^{(T+\tilde{T}),3\mathrm{P}}_{\parallel,\mathrm{NLP}}(u,\omega_1,\omega_2)\Big],\\ \Pi^{(V-A),3\mathrm{P}}_{\delta\mu,\perp}(p,q) &= -\frac{i\tilde{f}_B(\mu)m_B}{4n\cdot p}\int_0^\infty d\omega_1\int_0^\infty d\omega_2\int_0^1 du\frac{1}{[\bar{n}\cdot p-\omega_1-u\omega_2+i0]^2}\\ &\quad\times\Big\{[2g_{\delta\mu\perp}+i\epsilon_{\delta\mu\perp}]\rho^{(V-A),3\mathrm{P}}_{\perp,\mathrm{LP}}(u,\omega_1,\omega_2)+\frac{m}{n\cdot p}[2g_{\delta\mu\perp}-i\epsilon_{\delta\mu\perp}]\rho^{(V-A),3\mathrm{P}}_{\perp,\mathrm{NLP}}(u,\omega_1,\omega_2)\Big\},\\ \Pi^{(T+\tilde{T}),3\mathrm{P}}_{\delta\mu,\perp}(p,q) &= \frac{i\tilde{f}_B(\mu)m_B}{4n\cdot p}\bar{n}\cdot q\int_0^\infty d\omega_1\int_0^\infty d\omega_2\int_0^1 du\frac{1}{[\bar{n}\cdot p-\omega_1-u\omega_2+i0]^2}\\ &\quad\times\Big\{[2g_{\delta\mu\perp}+i\epsilon_{\delta\mu\perp}]\rho^{(T+\tilde{T}),3\mathrm{P}}_{\perp,\mathrm{LP}}(u,\omega_1,\omega_2)+\frac{mn\cdot q}{2p\cdot q}[2g_{\delta\mu\perp}-i\epsilon_{\delta\mu\perp}]\rho^{(T+\tilde{T}),3\mathrm{P}}_{\perp,\mathrm{NLP}}(u,\omega_1,\omega_2)\Big\},\end{aligned} \tag{5.3}$$

where we can express the invariant functions entering the tree-level factorization formulas (5.3) as

$$\begin{aligned}\rho^{(V-A),3\mathrm{P}}_{\bar{n},\parallel,\mathrm{LP}} &= (1-2u)[X_A-\Psi_A-2Y_A]-\tilde{X}_A-\Psi_V+2\tilde{Y}_A,\\ \rho^{(V-A),3\mathrm{P}}_{\bar{n},\parallel,\mathrm{NLP}} &= 2[\Psi_A-\Psi_V]+4[W+Y_A+\tilde{Y}_A-2Z],\\ \rho^{(V-A),3\mathrm{P}}_{n,\parallel,\mathrm{LP}} &= 2(u-1)(\Psi_A+\Psi_V),\\ \rho^{(V-A),3\mathrm{P}}_{n,\parallel,\mathrm{NLP}} &= (\Psi_A-\Psi_V)-[X_A+\tilde{X}_A-2Y_A-2\tilde{Y}_A],\\ \rho^{(T+\tilde{T}),3\mathrm{P}}_{\parallel,\mathrm{LP}} &= (1-2u)(X_A+\Psi_V-2Y_A)-\tilde{X}_A+\Psi_A+2\tilde{Y}_A,\\ \rho^{(T+\tilde{T}),3\mathrm{P}}_{\parallel,\mathrm{NLP}} &= (\Psi_A-\Psi_V)+X_A+\tilde{X}_A+2(Y_A+\tilde{Y}_A)+4(W-2Z),\end{aligned}$$

$$
\begin{aligned}
\rho_{\perp,\mathrm{LP}}^{(V-A),3\mathrm{P}} &= (1-2u)(X_A - \Psi_A - 2Y_A) + \tilde{X}_A + \Psi_V - 2\tilde{Y}_A,\\
\rho_{\perp,\mathrm{NLP}}^{(V-A),3\mathrm{P}} &= -(\Psi_A + \Psi_V) + X_A - \tilde{X}_A - 2(Y_A - \tilde{Y}_A),\\
\rho_{\perp,\mathrm{LP}}^{(T+\tilde{T}),3\mathrm{P}} &= (1-2u)(X_A - \Psi_A - 2Y_A) + \tilde{X}_A + \Psi_V - 2\tilde{Y}_A,\\
\rho_{\perp,\mathrm{NLP}}^{(T+\tilde{T}),3\mathrm{P}} &= -(\Psi_A + \Psi_V) + X_A - \tilde{X}_A - 2(Y_A - \tilde{Y}_A).
\end{aligned}
\tag{5.4}
$$

It is apparent that the explicit expressions of the invariant functions entering the factorized correlation function $\Pi_{\mu,\|}^{(a)}(p,q)$ are identical to the corresponding ones of $B \to \pi, K$ in [44] because the interpolating particle currents and the $b \to q$ weak transition currents entering the vacuum-to-$B$ correlating functions (2.5) are coincident in these decays. Apparently, there also exist the interesting relations $\rho_{\mathrm{LP}}^{B\to A} = -\rho_{\mathrm{LP}}^{B\to V}$ and $\rho_{\mathrm{NLP}}^{B\to A} = \rho_{\mathrm{NLP}}^{B\to V}$ between our results and the relative ones of $B \to V$ in [46] as well as

$$
\rho_{\perp,\mathrm{LP}}^{(V-A),3\mathrm{P}} = \rho_{\perp,\mathrm{LP}}^{(T+\tilde{T}),3\mathrm{P}}, \qquad \rho_{\perp,\mathrm{NLP}}^{(V-A),3\mathrm{P}} = \rho_{\perp,\mathrm{NLP}}^{(T+\tilde{T}),3\mathrm{P}}.
\tag{5.5}
$$

Our next aim is to perform the computation of higher-twist two-particle contributions to vacuum-to-$B$-meson correlation functions (2.5) with higher-twist $B$-meson DAs. Keeping the truncation of light-cone correction terms at $\mathcal{O}(x^2)$, the two-particle renormalized HQET matrix element of the light-cone operator can be written as

$$
\begin{aligned}
\langle 0|(\bar{q}_s Y_s)_\beta(x)(Y_s^\dagger h_v)_\alpha(0)|\bar{B}_v\rangle = &-\frac{i\tilde{f}_B(\mu)m_B}{4}\int_0^\infty d\omega\, e^{-i\omega v\cdot x}\left[\frac{1+\not{v}}{2}\left\{2[\phi_B^+(\omega,\mu) + x^2 g_B^+(\omega,\mu)]\right.\right.\\
&\left.\left.-\frac{\not{x}}{v\cdot x}[(\phi_B^+(\omega,\mu) - \phi_B^-(\omega,\mu)) + x^2(g_B^+(\omega,\mu) - g_B^-(\omega,\mu))]\right\}\gamma_5\right]_{\alpha\beta}.
\end{aligned}
\tag{5.6}
$$

Utilizing the specific operator identities of the light-cone HQET operators [65]

$$
\begin{aligned}
&\frac{\partial}{\partial x^\mu}(\bar{q}_s Y_s)(x)\gamma^\mu\Gamma(Y_s^\dagger h_v)(0) = -i\int_0^1 du\, u(\bar{q}_s Y_s)(x)x^\alpha g_s(Y_s^\dagger G_{\alpha\mu}Y_s)(ux)\gamma^\mu\Gamma(Y_s^\dagger h_v)(0),\\
&v_\mu\frac{\partial}{\partial x_\mu}(\bar{q}_s Y_s)(x)\Gamma(Y_s^\dagger h_v)(0) = i\int_0^1 du\,\bar{u}(\bar{q}_s Y_s)(x)x^\alpha g_s(Y_s^\dagger G_{\alpha\mu}Y_s)(ux)v^\mu\Gamma(Y_s^\dagger h_v)(0) + (v\cdot\partial)(\bar{q}_s Y_s)(x)\Gamma(Y_s^\dagger h_v)(0),
\end{aligned}
\tag{5.7}
$$

it is straightforward to generalize the two DAs $g_B^+$ and $g_B^-$ with twist-four and -five, respectively, in the momentum space as [44,64]

$$
\begin{aligned}
-2\frac{d^2}{d\omega^2}g_B^+(\omega,\mu) = &\left[\frac{3}{2} + (\omega - \bar{\Lambda})\frac{d}{d\omega}\right]\phi_B^+(\omega,\mu) - \frac{1}{2}\phi_B^-(\omega,\mu) + \int_0^\infty \frac{d\omega_2}{\omega_2}\frac{d}{d\omega}\Psi_4(\omega,\omega_2,\mu)\\
&-\int_0^\infty \frac{d\omega_2}{\omega_2^2}\Psi_4(\omega,\omega_2,\mu) + \int_0^\omega \frac{d\omega_2}{\omega_2^2}\Psi_4(\omega-\omega_2,\omega_2,\mu),\\
-2\frac{d^2}{d\omega^2}g_B^-(\omega,\mu) = &\left[\frac{3}{2} + (\omega - \bar{\Lambda})\frac{d}{d\omega}\right]\phi_B^-(\omega,\mu) - \frac{1}{2}\phi_B^+(\omega,\mu) + \int_0^\infty \frac{d\omega_2}{\omega_2}\frac{d}{d\omega}\Psi_5(\omega,\omega_2,\mu)\\
&-\int_0^\infty \frac{d\omega_2}{\omega_2^2}\Psi_5(\omega,\omega_2,\mu) + \int_0^\omega \frac{d\omega_2}{\omega_2^2}\Psi_5(\omega-\omega_2,\omega_2,\mu).
\end{aligned}
\tag{5.8}
$$

The results of higher-twist two-particle corrections to correlation functions (2.5) at tree level are given as

$$
\begin{aligned}
\Pi_{\mu,\parallel}^{(V-A),\mathrm{2PHT}}(p,q) &= -\frac{2i\tilde{f}_B(\mu)m_B}{n\cdot p}\bar{n}_\mu\bigg\{\int_0^\infty \frac{d\omega}{(\bar{n}\cdot p-\omega)^2}\hat{g}_B^-(\omega,\mu) \\
&\quad -\frac{1}{2}\int_0^\infty d\omega_1\int_0^\infty d\omega_2\int_0^1 du\,\frac{\bar{u}\Psi_5(\omega_1,\omega_2,\mu)}{(\bar{n}\cdot p-\omega_1-u\omega_2)^2}\bigg\}, \\
\Pi_{\mu,\parallel}^{(T+\tilde{T}),\mathrm{2PHT}}(p,q) &= \frac{i\tilde{f}_B(\mu)m_B}{n\cdot p}[n\cdot q\bar{n}_\mu - \bar{n}\cdot q n_\mu]\bigg\{\int_0^\infty \frac{d\omega}{(\bar{n}\cdot p-\omega)^2}\hat{g}_B^-(\omega,\mu) \\
&\quad -\frac{1}{2}\int_0^\infty d\omega_1\int_0^\infty d\omega_2\int_0^1 du\,\frac{\bar{u}\Psi_5(\omega_1,\omega_2,\mu)}{(\bar{n}\cdot p-\omega_1-u\omega_2)^2}\bigg\}, \\
\Pi_{\delta\mu,\perp}^{(V-A),\mathrm{2PHT}}(p,q) &= -\frac{i\tilde{f}_B(\mu)m_B}{n\cdot p}[2g_{\delta\mu\perp}+i\epsilon_{\delta\mu\perp}]\bigg\{\int_0^\infty \frac{d\omega}{(\bar{n}\cdot p-\omega)^2}\hat{g}_B^-(\omega,\mu) \\
&\quad -\frac{1}{2}\int_0^\infty d\omega_1\int_0^\infty d\omega_2\int_0^1 du\,\frac{\bar{u}\Psi_5(\omega_1,\omega_2,\mu)}{(\bar{n}\cdot p-\omega_1-u\omega_2)^2}\bigg\}, \\
\Pi_{\delta\mu,\perp}^{(T+\tilde{T}),\mathrm{2PHT}}(p,q) &= \frac{i\tilde{f}_B(\mu)m_B}{n\cdot p}\bar{n}\cdot q[2g_{\delta\mu\perp}+i\epsilon_{\delta\mu\perp}]\bigg\{\int_0^\infty \frac{d\omega}{(\bar{n}\cdot p-\omega)^2}\hat{g}_B^-(\omega,\mu) \\
&\quad -\frac{1}{2}\int_0^\infty d\omega_1\int_0^\infty d\omega_2\int_0^1 du\,\frac{\bar{u}\Psi_5(\omega_1,\omega_2,\mu)}{(\bar{n}\cdot p-\omega_1-u\omega_2)^2}\bigg\},
\end{aligned} \tag{5.9}
$$

where the "genuine" two-particle twist-five DA $\hat{g}_B^-$ is

$$
\hat{g}_B^-(\omega,\mu) = \frac{1}{4}\int_\omega^\infty d\rho\{(\rho-\omega)[\phi_B^+(\rho)-\phi_B^-(\rho)] - 2(\bar{\Lambda}-\rho)\phi_B^-(\rho)\}. \tag{5.10}
$$

Adding up the three- [Eq. (5.3)] and two-particle [Eq. (5.9)] contributions for vacuum-to-$B$-meson correlation functions (2.5), one can obtain the explicit formulas of higher-twist corrections to these correlation functions up to NLP accuracy at tree level

$$
\begin{aligned}
\Pi_{\mu,\parallel}^{(V-A),\mathrm{HT}}(p,q) &= -\frac{i\tilde{f}_B(\mu)m_B}{2n\cdot p}\bigg\{\bar{n}_\mu\int_0^\infty \frac{d\omega}{(\bar{n}\cdot p-\omega)^2}4\hat{g}_B^-(\omega,\mu) + \int_0^\infty d\omega_1\int_0^\infty d\omega_2\int_0^1 du\,\frac{1}{(\bar{n}\cdot p-\omega_1-u\omega_2)^2} \\
&\quad\times\bigg[\bar{n}_\mu\bigg(\rho_{\bar{n},\parallel,\mathrm{LP}}^{(V-A)}(\omega_1,\omega_2,u,\mu)+\frac{m}{n\cdot p}\rho_{\bar{n},\parallel,\mathrm{NLP}}^{(V-A)}(\omega_1,\omega_2,u,\mu)\bigg) \\
&\quad + n_\mu\bigg(\rho_{n,\parallel,\mathrm{LP}}^{(V-A)}(\omega_1,\omega_2,u,\mu)+\frac{m}{n\cdot p}\rho_{n,\parallel,\mathrm{NLP}}^{(V-A)}(\omega_1,\omega_2,u,\mu)\bigg)\bigg]\bigg\}, \\
\Pi_{\mu,\parallel}^{(T+\tilde{T}),\mathrm{HT}}(p,q) &= \frac{i\tilde{f}_B(\mu)m_B}{4n\cdot p}[n\cdot q\bar{n}_\mu-\bar{n}\cdot q n_\mu]\bigg\{\int_0^\infty \frac{d\omega}{(\bar{n}\cdot p-\omega)^2}4\hat{g}_B^-(\omega,\mu) \\
&\quad + \int_0^\infty d\omega_1\int_0^\infty d\omega_2\int_0^1 du\,\frac{1}{(\bar{n}\cdot p-\omega_1-u\omega_2)^2}\bigg[\rho_{\parallel,\mathrm{LP}}^{(T+\tilde{T})}(\omega_1,\omega_2,u,\mu)+\frac{m}{n\cdot p}\rho_{\parallel,\mathrm{NLP}}^{(T+\tilde{T})}(\omega_1,\omega_2,u,\mu)\bigg]\bigg\}, \\
\Pi_{\delta\mu,\perp}^{(V-A),\mathrm{HT}}(p,q) &= -\frac{i\tilde{f}_B(\mu)m_B}{4n\cdot p}\bigg\{(2g_{\delta\mu\perp}+i\epsilon_{\delta\mu\perp})\int_0^\infty \frac{d\omega}{(\bar{n}\cdot p-\omega)^2}4\hat{g}_B^-(\omega,\mu) \\
&\quad + \int_0^\infty d\omega_1\int_0^\infty d\omega_2\int_0^1 du\,\frac{1}{(\bar{n}\cdot p-\omega_1-u\omega_2)^2}\bigg[(2g_{\delta\mu\perp}+i\epsilon_{\delta\mu\perp})\rho_{\perp,\mathrm{LP}}^{(V-A)}(\omega_1,\omega_2,u,\mu) \\
&\quad + \frac{m}{n\cdot p}(2g_{\delta\mu\perp}-i\epsilon_{\delta\mu\perp})\rho_{\perp,\mathrm{NLP}}^{(V-A)}(\omega_1,\omega_2,u,\mu)\bigg]\bigg\},
\end{aligned}
$$

$$
\begin{aligned}
\Pi^{(T+\tilde{T}),\mathrm{HT}}_{\delta\mu,\perp}(p,q) = \frac{i\tilde{f}_B(\mu) m_B}{4 n\cdot p}\, \bar{n}\cdot q \Bigg\{ & (2 g_{\delta\mu\perp} + i\epsilon_{\delta\mu\perp}) \int_0^\infty \frac{d\omega}{(\bar{n}\cdot p - \omega)^2}\, 4\hat{g}_B^-(\omega,\mu) \\
& + \int_0^\infty d\omega_1 \int_0^\infty d\omega_2 \int_0^1 du\, \frac{1}{(\bar{n}\cdot p - \omega_1 - u\omega_2)^2} \Bigg[ (2 g_{\delta\mu\perp} + i\epsilon_{\delta\mu\perp}) \rho^{(T+\tilde{T})}_{\perp,\mathrm{LP}}(\omega_1,\omega_2,u,\mu) \\
& + \frac{m n\cdot q}{2 p\cdot q} (2 g_{\delta\mu\perp} - i\epsilon_{\delta\mu\perp}) \rho^{(T+\tilde{T})}_{\perp,\mathrm{NLP}}(\omega_1,\omega_2,u,\mu) \Bigg] \Bigg\},
\end{aligned} \tag{5.11}
$$

with the nonvanishing spectral functions as follows:

$$
\begin{aligned}
&\rho^{(V-A)}_{\bar{n},\parallel,\mathrm{LP}} = \tilde{\Psi}_5 - \Psi_5, && \rho^{(V-A)}_{\bar{n},\parallel,\mathrm{NLP}} = 2\Phi_6, \\
&\rho^{(V-A)}_{n,\parallel,\mathrm{LP}} = 2(u-1)\Phi_4, && \rho^{(V-A)}_{n,\parallel,\mathrm{NLP}} = \tilde{\Psi}_5 - \Psi_5, \\
&\rho^{(T+\tilde{T})}_{\parallel,\mathrm{LP}} = 2(1-u)\Phi_4 - \Psi_5 + \tilde{\Psi}_5, && \rho^{(T|+\tilde{T})}_{\parallel,\mathrm{NLP}} = 2\Phi_6 + \Psi_5 - \tilde{\Psi}_5, \\
&\rho^{(V-A)}_{\perp,\mathrm{LP}} = -\Psi_5 - \tilde{\Psi}_5, && \rho^{(V-A)}_{\perp,\mathrm{NLP}} = \Psi_5 + \tilde{\Psi}_5, \\
&\rho^{(T+\tilde{T})}_{\perp,\mathrm{LP}} = -\Psi_5 - \tilde{\Psi}_5, && \rho^{(T+\tilde{T})}_{\perp,\mathrm{NLP}} = \Psi_5 + \tilde{\Psi}_5.
\end{aligned} \tag{5.12}
$$

Matching the hadronic representations of the vacuum-to-$B$-meson correlation functions (2.10) and the dispersion representations of the tree-level factorization formulas (5.11) via the parton-hadron duality approximation yields the sum rules for the higher-twist corrections to the semileptonic $B \to A$ form factors

$$
\begin{aligned}
f_A^T n\cdot p\, e^{m_A^2/(n\cdot p\,\omega_M)} \left[ \frac{m_B}{m_B - m_A} A^{\mathrm{HT}}(q^2) \right] = \frac{f_B(\mu) m_B}{n\cdot p} \Bigg\{ & e^{-\omega_s/\omega_M} 4\hat{g}_B^-(\omega,\mu) + \int_0^{\omega_s} d\omega' \frac{1}{\omega_M} e^{-\omega'/\omega_M} 4\hat{g}_B^-(\omega',\mu) \\
& + \int_0^{\omega_s} d\omega_1 \int_{\omega_s-\omega_1}^\infty \frac{d\omega_2}{\omega_2} e^{-\omega_s/\omega_M} \Bigg[ \rho^{(V-A)}_{\perp,\mathrm{LP}}\left( \frac{\omega_s - \omega_1}{\omega_2}, \omega_1, \omega_2, \mu \right) \\
& - \frac{m}{n\cdot p} \rho^{(V-A)}_{\perp,\mathrm{NLP}}\left( \frac{\omega_s - \omega_1}{\omega_2}, \omega_1, \omega_2, \mu \right) \Bigg] \\
& + \int_0^{\omega_s} d\omega' \int_0^{\omega'} d\omega_1 \int_{\omega'-\omega_1}^\infty \frac{d\omega_2}{\omega_2} \frac{1}{\omega_M} e^{-\omega'/\omega_M} \Bigg[ \rho^{(V-A)}_{\perp,\mathrm{LP}}\left( \frac{\omega' - \omega_1}{\omega_2}, \omega_1, \omega_2, \mu \right) \\
& - \frac{m}{n\cdot p} \rho^{(V-A)}_{\perp,\mathrm{NLP}}\left( \frac{\omega' - \omega_1}{\omega_2}, \omega_1, \omega_2, \mu \right) \Bigg] \Bigg\},
\end{aligned} \tag{5.13}
$$

$$
\begin{aligned}
\frac{f_A (n\cdot p)^2}{2 m_A} e^{m_A^2/(n\cdot p\,\omega_M)} \left[ \frac{2 m_A}{n\cdot p} V_0^{\mathrm{HT}}(q^2) \right] = -\frac{f_B(\mu) m_B}{n\cdot p} \Bigg\{ & e^{-\omega_s/\omega_M} 4\hat{g}_B^-(\omega_s,\mu) + \int_0^{\omega_s} d\omega' \frac{1}{\omega_M} e^{-\omega'/\omega_M} 4\hat{g}_B^-(\omega',\mu) \\
& + \int_0^{\omega_s} d\omega_1 \int_{\omega_s-\omega_1}^\infty \frac{d\omega_2}{\omega_2} e^{-\omega_s/\omega_M} \Bigg[ \rho^{(V-A)}_{\bar{n},\parallel,\mathrm{LP}}\left( \frac{\omega_s - \omega_1}{\omega_2}, \omega_1, \omega_2, \mu \right) \\
& + \frac{m}{n\cdot p} \rho^{(V-A)}_{\bar{n},\parallel,\mathrm{NLP}}\left( \frac{\omega_s - \omega_1}{\omega_2}, \omega_1, \omega_2, \mu \right) + \frac{m_B - n\cdot p}{m_B} \Bigg( \rho^{(V-A)}_{n,\parallel,\mathrm{LP}}\left( \frac{\omega_s - \omega_1}{\omega_2}, \omega_1, \omega_2, \mu \right) \\
& + \frac{m}{n\cdot p} \rho^{(V-A)}_{n,\parallel,\mathrm{NLP}}\left( \frac{\omega_s - \omega_1}{\omega_2}, \omega_1, \omega_2, \mu \right) \Bigg) \Bigg] \\
& + \int_0^{\omega_s} d\omega' \int_0^{\omega'} d\omega_1 \int_{\omega'-\omega_1}^\infty \frac{d\omega_2}{\omega_2} \frac{1}{\omega_M} e^{-\omega'/\omega_M} \Bigg[ \rho^{(V-A)}_{\bar{n},\parallel,\mathrm{LP}}\left( \frac{\omega' - \omega_1}{\omega_2}, \omega_1, \omega_2, \mu \right) \\
& + \frac{m}{n\cdot p} \rho^{(V-A)}_{\bar{n},\parallel,\mathrm{NLP}}\left( \frac{\omega' - \omega_1}{\omega_2}, \omega_1, \omega_2, \mu \right) + \frac{m_B - n\cdot p}{m_B} \Bigg( \rho^{(V-A)}_{n,\parallel,\mathrm{LP}}\left( \frac{\omega' - \omega_1}{\omega_2}, \omega_1, \omega_2, \mu \right) \\
& + \frac{m}{n\cdot p} \rho^{(V-A)}_{n,\parallel,\mathrm{NLP}}\left( \frac{\omega' - \omega_1}{\omega_2}, \omega_1, \omega_2, \mu \right) \Bigg) \Bigg] \Bigg\},
\end{aligned} \tag{5.14}
$$

$$
\begin{aligned}
f_A^T n\cdot p\, e^{m_A^2/(n\cdot p\,\omega_M)}\left[\frac{m_B-m_A}{n\cdot p}V_1^{\mathrm{HT}}(q^2)\right] &= \frac{f_B(\mu)m_B}{n\cdot p}\Bigg\{e^{-\omega_s/\omega_M}4\hat{g}_B^-(\omega,\mu)+\int_0^{\omega_s}d\omega'\,\frac{1}{\omega_M}e^{-\omega'/\omega_M}4\hat{g}_B^-(\omega',\mu)\\
&\quad+\int_0^{\omega_s}d\omega_1\int_{\omega_s-\omega_1}^{\infty}\frac{d\omega_2}{\omega_2}e^{-\omega_s/\omega_M}\Bigg[\rho_{\perp,\mathrm{LP}}^{(V-A)}\left(\frac{\omega_s-\omega_1}{\omega_2},\omega_1,\omega_2,\mu\right)\\
&\quad+\frac{m}{n\cdot p}\rho_{\perp,\mathrm{NLP}}^{(V-A)}\left(\frac{\omega_s-\omega_1}{\omega_2},\omega_1,\omega_2,\mu\right)\Bigg]\\
&\quad+\int_0^{\omega_s}d\omega'\int_0^{\omega'}d\omega_1\int_{\omega'-\omega_1}^{\infty}\frac{d\omega_2}{\omega_2}\frac{1}{\omega_M}e^{-\omega'/\omega_M}\Bigg[\rho_{\perp,\mathrm{LP}}^{(V-A)}\left(\frac{\omega'-\omega_1}{\omega_2},\omega_1,\omega_2,\mu\right)\\
&\quad+\frac{m}{n\cdot p}\rho_{\perp,\mathrm{NLP}}^{(V-A)}\left(\frac{\omega'-\omega_1}{\omega_2},\omega_1,\omega_2,\mu\right)\Bigg]\Bigg\},
\end{aligned}
\tag{5.15}
$$

$$
\begin{aligned}
&\frac{f_A(n\cdot p)^2}{2m_A}e^{m_A^2/(n\cdot p\,\omega_M)}\left[\frac{m_B-m_A}{n\cdot p}V_1^{\mathrm{HT}}(q^2)-\frac{m_B+m_A}{m_B}V_2^{\mathrm{HT}}(q^2)\right]\\
&\quad=-\frac{f_B(\mu)m_B}{n\cdot p}\Bigg\{e^{-\omega_s/\omega_M}4\hat{g}_B^-(\omega_s,\mu)+\int_0^{\omega_s}d\omega'\,\frac{1}{\omega_M}e^{-\omega'/\omega_M}4\hat{g}_B^-(\omega',\mu)\\
&\qquad+\int_0^{\omega_s}d\omega_1\int_{\omega_s-\omega_1}^{\infty}\frac{d\omega_2}{\omega_2}e^{-\omega_s/\omega_M}\Bigg[\rho_{\bar{n},\parallel,\mathrm{LP}}^{(V-A)}\left(\frac{\omega_s-\omega_1}{\omega_2},\omega_1,\omega_2,\mu\right)+\frac{m}{n\cdot p}\rho_{\bar{n},\parallel,\mathrm{NLP}}^{(V-A)}\left(\frac{\omega_s-\omega_1}{\omega_2},\omega_1,\omega_2,\mu\right)\\
&\qquad-\frac{m_B-n\cdot p}{m_B}\left(\rho_{n,\parallel,\mathrm{LP}}^{(V-A)}\left(\frac{\omega_s-\omega_1}{\omega_2},\omega_1,\omega_2,\mu\right)+\frac{m}{n\cdot p}\rho_{n,\parallel,\mathrm{NLP}}^{(V-A)}\left(\frac{\omega_s-\omega_1}{\omega_2},\omega_1,\omega_2,\mu\right)\right)\Bigg]\\
&\qquad+\int_0^{\omega_s}d\omega'\int_0^{\omega'}d\omega_1\int_{\omega'-\omega_1}^{\infty}\frac{d\omega_2}{\omega_2}\frac{1}{\omega_M}e^{-\omega'/\omega_M}\Bigg[\rho_{\bar{n},\parallel,\mathrm{LP}}^{(V-A)}\left(\frac{\omega'-\omega_1}{\omega_2},\omega_1,\omega_2,\mu\right)+\frac{m}{n\cdot p}\rho_{\bar{n},\parallel,\mathrm{NLP}}^{(V-A)}\left(\frac{\omega'-\omega_1}{\omega_2},\omega_1,\omega_2,\mu\right)\\
&\qquad-\frac{m_B-n\cdot p}{m_B}\left(\rho_{n,\parallel,\mathrm{LP}}^{(V-A)}\left(\frac{\omega'-\omega_1}{\omega_2},\omega_1,\omega_2,\mu\right)+\frac{m}{n\cdot p}\rho_{n,\parallel,\mathrm{NLP}}^{(V-A)}\left(\frac{\omega'-\omega_1}{\omega_2},\omega_1,\omega_2,\mu\right)\right)\Bigg]\Bigg\},
\end{aligned}
\tag{5.16}
$$

$$
\begin{aligned}
f_A^T n\cdot p\, e^{m_A^2/(n\cdot p\,\omega_M)}T_1^{\mathrm{HT}}(q^2) &= \frac{f_B(\mu)m_B}{n\cdot p}\Bigg\{e^{-\omega_s/\omega_M}4\hat{g}_B^-(\omega,\mu)+\int_0^{\omega_s}d\omega'\,\frac{1}{\omega_M}e^{-\omega'/\omega_M}4\hat{g}_B^-(\omega',\mu)\\
&\quad+\int_0^{\omega_s}d\omega_1\int_{\omega_s-\omega_1}^{\infty}\frac{d\omega_2}{\omega_2}e^{-\omega_s/\omega_M}\Bigg[\rho_{\perp,\mathrm{LP}}^{(T+\tilde{T})}\left(\frac{\omega_s-\omega_1}{\omega_2},\omega_1,\omega_2,\mu\right)\\
&\quad-\frac{mn\cdot q}{2p\cdot q}\rho_{\perp,\mathrm{NLP}}^{(T+\tilde{T})}\left(\frac{\omega_s-\omega_1}{\omega_2},\omega_1,\omega_2,\mu\right)\Bigg]\\
&\quad+\int_0^{\omega_s}d\omega'\int_0^{\omega'}d\omega_1\int_{\omega'-\omega_1}^{\infty}\frac{d\omega_2}{\omega_2}\frac{1}{\omega_M}e^{-\omega'/\omega_M}\Bigg[\rho_{\perp,\mathrm{LP}}^{(T+\tilde{T})}\left(\frac{\omega'-\omega_1}{\omega_2},\omega_1,\omega_2,\mu\right)\\
&\quad-\frac{mn\cdot q}{2p\cdot q}\rho_{\perp,\mathrm{NLP}}^{(T+\tilde{T})}\left(\frac{\omega'-\omega_1}{\omega_2},\omega_1,\omega_2,\mu\right)\Bigg]\Bigg\},
\end{aligned}
\tag{5.17}
$$

$$
\begin{aligned}
f_A^T n\cdot p\, e^{m_A^2/(n\cdot p\,\omega_M)}\left[\frac{m_B}{n\cdot p}T_2^{\mathrm{HT}}(q^2)\right] &= \frac{f_B(\mu)m_B}{n\cdot p}\Bigg\{e^{-\omega_s/\omega_M}4\hat{g}_B^-(\omega,\mu)+\int_0^{\omega_s}d\omega'\,\frac{1}{\omega_M}e^{-\omega'/\omega_M}4\hat{g}_B^-(\omega',\mu)\\
&\quad+\int_0^{\omega_s}d\omega_1\int_{\omega_s-\omega_1}^{\infty}\frac{d\omega_2}{\omega_2}e^{-\omega_s/\omega_M}\Bigg[\rho_{\perp,\mathrm{LP}}^{(T+\tilde{T})}\left(\frac{\omega_s-\omega_1}{\omega_2},\omega_1,\omega_2,\mu\right)\\
&\quad+\frac{mn\cdot q}{2p\cdot q}\rho_{\perp,\mathrm{NLP}}^{(T+\tilde{T})}\left(\frac{\omega_s-\omega_1}{\omega_2},\omega_1,\omega_2,\mu\right)\Bigg]\\
&\quad+\int_0^{\omega_s}d\omega'\int_0^{\omega'}d\omega_1\int_{\omega'-\omega_1}^{\infty}\frac{d\omega_2}{\omega_2}\frac{1}{\omega_M}e^{-\omega'/\omega_M}\Bigg[\rho_{\perp,\mathrm{LP}}^{(T+\tilde{T})}\left(\frac{\omega'-\omega_1}{\omega_2},\omega_1,\omega_2,\mu\right)\\
&\quad+\frac{mn\cdot q}{2p\cdot q}\rho_{\perp,\mathrm{NLP}}^{(T+\tilde{T})}\left(\frac{\omega'-\omega_1}{\omega_2},\omega_1,\omega_2,\mu\right)\Bigg]\Bigg\},
\end{aligned}
\tag{5.18}
$$

$$
\begin{aligned}
\frac{f_A (n\cdot p)^2}{2m_A} e^{m_A^2/(n\cdot p\,\omega_M)}\left[\frac{m_B}{n\cdot p}T_2^{\rm HT}(q^2) - T_3^{\rm HT}(q^2)\right] = -\frac{f_B(\mu) m_B}{n\cdot p}\Bigg\{ & e^{-\omega_s/\omega_M} 4\hat{g}_B^-(\omega,\mu) + \int_0^{\omega_s} d\omega' \frac{1}{\omega_M} e^{-\omega'/\omega_M} 4\hat{g}_B^-(\omega',\mu) \\
& + \int_0^{\omega_s} d\omega_1 \int_{\omega_s-\omega_1}^{\infty} \frac{d\omega_2}{\omega_2} e^{-\omega_s/\omega_M}\Bigg[\rho_{\parallel,\rm LP}^{(T+\tilde{T})}\left(\frac{\omega_s-\omega_1}{\omega_2},\omega_1,\omega_2,\mu\right) \\
& + \frac{m}{n\cdot p}\rho_{\parallel,\rm NLP}^{(T+\tilde{T})}\left(\frac{\omega_s-\omega_1}{\omega_2},\omega_1,\omega_2,\mu\right)\Bigg] \\
& + \int_0^{\omega_s} d\omega' \int_0^{\omega'} d\omega_1 \int_{\omega'-\omega_1}^{\infty} \frac{d\omega_2}{\omega_2}\frac{1}{\omega_M} e^{-\omega'/\omega_M}\Bigg[\rho_{\parallel,\rm LP}^{(T+\tilde{T})}\left(\frac{\omega'-\omega_1}{\omega_2},\omega_1,\omega_2,\mu\right) \\
& + \frac{m}{n\cdot p}\rho_{\parallel,\rm NLP}^{(T+\tilde{T})}\left(\frac{\omega'-\omega_1}{\omega_2},\omega_1,\omega_2,\mu\right)\Bigg]\Bigg\}. \qquad (5.19)
\end{aligned}
$$

Adding up the two-particle leading-twist NLL and higher-twist results of the $B \to A$ form factors, we explicitly express the final LCSR at large hadronic recoil as

$$
F^i_{B\to A}(q^2) = F^{i,\rm NLL}_{B\to A}(q^2) + F^{i,\rm HT}_{B\to A}(q^2), \qquad (5.20)
$$

where one can extract $F^{i,\rm NLL}_{B\to A}$ $(i = A, V_0, V_1, \mathcal{V}_{12}, T_1, T_2, \mathcal{T}_{23})$ from (4.1) and $F^{i,\rm HT}_{B\to A}$ from (5.13)–(5.19). The calligraphic form factor $\mathcal{V}_{12}$ represents a linear combination of $V_1$ and $V_2$ given in (5.16) as well as $\mathcal{T}_{23}$ given in (5.19).

## VI. NUMERICAL ANALYSIS

In this section, we intend to explore the phenomenological aspects of the $B \to A$ form factors. To begin with, we collect some necessary inputs including $B$-meson DAs up to the twist-six accuracy, intrinsic sum rule parameters, the decay constants of $B$-meson and $p$-wave axial-vector mesons, etc. Then the LCDA models and the values of the Borel parameter are discussed in detail. Utilizing the $z$-series parametrization, we extrapolate the QCD calculations to the whole kinematic region and display the fitted results of the $B \to A$ form factors. We also present the predictions for the branching ratios of the exclusive $B \to A\ell\bar{\nu}_\ell$ and $B \to A\nu_\ell\bar{\nu}_\ell$ processes and relevant observables including transverse asymmetries, forward-backward asymmetries, and lepton-flavor universality observables employing the given mixing angles $\theta_K = -34°$, $\theta_{1_{P_1}} = 28°$, $\theta_{3_{P_1}} = 23°$.

### A. Theory inputs

We will continue with the discussion of ingredients entering the derived LCSR for the $B \to A$ form factors, the intrinsic sum rule parameters, the $B$-meson DAs up to the twist-six accuracy, and the shape parameters of the $B$-meson with LCDAs included. We explicitly present the expressions of the three-parameter model for the two-particle and three-particle $B$-meson DAs up to the twist-six accuracy in Appendix A. For the sake of convenience, the inverse moments for the leading-twist $B$-meson LCDA can be defined as [66]

$$
\begin{aligned}
\frac{1}{\lambda_B(\mu)} &= \int_0^\infty \frac{d\omega}{\omega}\phi_B^+(\omega,\mu), \\
\frac{\hat{\sigma}_n(\mu)}{\lambda_B(\mu)} &= \int_0^\infty \frac{d\omega}{\omega}\ln^n \frac{{\rm e}^{-\gamma_E}\lambda_B(\mu)}{\omega}\phi_B^+(\omega,\mu), \qquad (6.1)
\end{aligned}
$$

where $\gamma_E$ is the Euler-Mascheroni constant. According to the definitions for the logarithmic inverse moments of the leading-twist $B$-meson DA in (6.1), the expressions for three nonperturbative shape parameters ($\lambda_B$, $\hat{\sigma}_1$, and $\hat{\sigma}_2$) are finally displayed in terms of $\alpha$, $\beta$, and the dimensionful parameter $\omega_0$ as follows [66–68]:

$$
\begin{aligned}
\lambda_B(\mu_0) &= \frac{\alpha-1}{\beta-1}\omega_0, \\
\hat{\sigma}_1(\mu_0) &= \psi(\beta-1) - \psi(\alpha-1) + \ln\frac{\alpha-1}{\beta-1}, \\
\hat{\sigma}_2(\mu_0) &= \hat{\sigma}_1^2(\mu) + \psi'(\alpha-1) - \psi'(\beta-1) + \frac{\pi^2}{6}, \qquad (6.2)
\end{aligned}
$$

with $\psi(x)$ denoting the digamma function. The employed values of the shape parameters ($\lambda_B$, $\hat{\sigma}_1$, and $\hat{\sigma}_2$) and other inputs are shown in Table I.

Based on the requirements in [42], we obtain the central values and ranges of the Borel mass $M^2$ and the effective threshold $s_0$ in order to calculate each $B \to A$ form factor. Applying this Borel window, we find the continuum contributions during the LL calculations of the $B \to A$ form factors lie between 9% and 12%. The value range of factorization scale $\mu$ entering the derived LCSR for the $B \to A$ form factors at NLL is from 1 to 2 GeV around the central value $\mu = 1.5$ GeV. We will vary the two hard scales $\mu_{h_1}$ and $\mu_{h_2}$ independently in the range $[m_b/2, 2m_b]$ (see [66–68] for further discussions). Form factor $f^A_{B\to a_1}(0)$ dependence on the Borel parameters $M^2$ and $s_0$ is displayed in Fig. 4.

We then proceed to consider the input masses of the singlet and the octet for axial-vector mesons. It is similar to the vector-meson $\omega_1 - \omega_8$ mixing in that the nonstrange axial-vector mesons $f_1$ and $f_8$ are also mixed together because of the same C parities as well as the singlet $h_1$ and the pure octet $h_8$. The $K_{1A} - K_{1B}$ mixing, conversely, can happen due to the flavor difference between the strange and nonstrange light quark. The mixing matrices of these axial-vector mesons ($f_{1,8}$, $h_{1,8}$, and $K_{1A,1B}$) are given by [1,6]

$$\begin{pmatrix} |f_1(1285)\rangle \\ |f_1(1420)\rangle \end{pmatrix} = \begin{pmatrix} \cos\theta_{^3P_1} & \sin\theta_{^3P_1} \\ -\sin\theta_{^3P_1} & \cos\theta_{^3P_1} \end{pmatrix} \begin{pmatrix} |f_1\rangle \\ |f_8\rangle \end{pmatrix}, \quad (6.3)$$

$$\begin{pmatrix} |h_1(1170)\rangle \\ |h_1(1415)\rangle \end{pmatrix} = \begin{pmatrix} \cos\theta_{^1P_1} & \sin\theta_{^1P_1} \\ -\sin\theta_{^1P_1} & \cos\theta_{^1P_1} \end{pmatrix} \begin{pmatrix} |h_1\rangle \\ |h_8\rangle \end{pmatrix}, \quad (6.4)$$

$$\begin{pmatrix} |K_1(1270)\rangle \\ |K_1(1400)\rangle \end{pmatrix} = \begin{pmatrix} \sin\theta_{K_1} & \cos\theta_{K_1} \\ \cos\theta_{K_1} & -\sin\theta_{K_1} \end{pmatrix} \begin{pmatrix} |K_{1A}\rangle \\ |K_{1B}\rangle \end{pmatrix}, \quad (6.5)$$

where $\theta_{^3P_1}$, $\theta_{^1P_1}$, and $\theta_{K_1}$ are the mixing angles, and the left-hand sides of equations are the physical mass eigenstates. It is straightforward to write down the relations of squared masses by taking advantage of the mixing matrices in Eqs. (6.3)–(6.5) and the mass squared matrices as (see the Appendix of [15] for a detailed derivation)

$$\begin{aligned} m_{f_1}^2 &= m_{f_1(1285)}^2 \cos^2\theta_{^3P_1} + m_{f_1(1420)}^2 \sin^2\theta_{^3P_1}, \\ m_{f_8}^2 &= m_{f_1(1285)}^2 \sin^2\theta_{^3P_1} + m_{f_1(1420)}^2 \cos^2\theta_{^3P_1}, \end{aligned} \quad (6.6)$$

$$\begin{aligned} m_{h_1}^2 &= m_{h_1(1170)}^2 \cos^2\theta_{^1P_1} + m_{h_1(1415)}^2 \sin^2\theta_{^1P_1}, \\ m_{h_8}^2 &= m_{h_1(1170)}^2 \sin^2\theta_{^1P_1} + m_{h_1(1415)}^2 \cos^2\theta_{^1P_1}, \end{aligned} \quad (6.7)$$

$$\begin{aligned} m_{K_{1A}}^2 &= m_{K_1(1400)}^2 \cos^2\theta_{K_1} + m_{K_1(1270)}^2 \sin^2\theta_{K_1}, \\ m_{K_{1B}}^2 &= m_{K_1(1400)}^2 \sin^2\theta_{K_1} + m_{K_1(1270)}^2 \cos^2\theta_{K_1}. \end{aligned} \quad (6.8)$$

TABLE I. Theory inputs used for the numerical analysis including the masses of the relevant mesons and quarks, the $B$-meson and axial-vector-meson decay constants, the LCSR parameters ($M^2$ and $s_0$), and the hadronic parameters ($\lambda_B$, $\hat{\sigma}_{1,2}$, and $\lambda_{E,H}$), etc. The quark masses in the table are in the $\overline{\mathrm{MS}}$ scheme at a renormalization scale of 2 GeV.

| Parameter | Value | Ref. | Parameter | Value | Ref. |
|---|---|---|---|---|---|
| $m_{B_d}$ | $5279.72 \pm 0.08$ MeV | [3] | $m_{B_s}$ | $5366.93 \pm 0.10$ MeV | [3] |
| $m_{B_{d1}(1^-)}$ | $5324.75 \pm 0.20$ MeV | [3] | $m_{B_{s1}(1^-)}$ | $5415.40 \pm 1.40$ MeV | [3] |
| $m_{B_{d2}(1^+)}$ | $5726.10 \pm 1.20$ MeV | [3] | $m_{B_{s2}(1^+)}$ | $5828.73 \pm 0.20$ MeV | [3] |
| $m_{a_1(1260)}$ | $1230 \pm 40$ MeV | [3] | $m_{b_1(1235)}$ | $1229.5 \pm 3.2$ MeV | [3] |
| $m_{f_1(1285)}$ | $1281.8 \pm 0.5$ MeV | [3] | $m_{h_1(1170)}$ | $1166 \pm 6$ MeV | [3] |
| $m_{f_1(1420)}$ | $1428.4 \pm 1.5$ MeV | [3] | $m_{h_1(1415)}$ | $1409 \pm 9$ MeV | [3] |
| $m_{K_1(1270)}$ | $1253 \pm 7$ MeV | [3] | $m_{K_1(1400)}$ | $1403 \pm 7$ MeV | [3] |
| $\bar{m}_u$ (2 GeV) | $2.130 \pm 0.041$ MeV | [69] | $\bar{m}_s$ (2 GeV) | $93.85 \pm 0.75$ MeV | [70] |
| $\bar{m}_b$ (2 GeV) | $4.195 \pm 0.014$ GeV | [69] | | | |
| $f_{B_d}\vert_{N_f=2+1+1}$ | $190.0 \pm 1.3$ MeV | [71] | $f_{B_s}\vert_{N_f=2+1+1}$ | $230.3 \pm 1.3$ MeV | [71] |
| $f_{a_1,\parallel}$ | $238 \pm 10$ MeV | [1] | $f_{b_1,\perp}$(1 GeV) | $180 \pm 8$ MeV | [1] |
| $f_{f_1,\parallel}$ | $245 \pm 13$ MeV | [1] | $f_{h_1,\perp}$(1 GeV) | $180 \pm 12$ MeV | [1] |
| $f_{f_8,\parallel}$ | $239 \pm 13$ MeV | [1] | $f_{h_8,\perp}$(1 GeV) | $190 \pm 10$ MeV | [1] |
| $f_{K_{1A},\parallel}$ | $250 \pm 13$ MeV | [1] | $f_{K_{1B},\perp}$(1 GeV) | $190 \pm 10$ MeV | [1] |
| $s_0$ | $4.5 \pm 0.3\ \mathrm{GeV}^2$ | | $M^2$ | $6.0 \pm 0.3\ \mathrm{GeV}^2$ | |
| $\lambda_{B_d}$ | $0.35 \pm 0.15$ GeV | [67,68,72–75] | $\{\hat{\sigma}_1, \hat{\sigma}_2\}$ | $\{0.7, 6.0\}$<br>$\{0.0, \pi^2/6\}$<br>$\{-0.7, -6.0\}$ | [67,68] |
| $\lambda_{B_s}$ | $0.40 \pm 0.15$ GeV | [67,68] | | | |
| $(\lambda_E^2/\lambda_H^2)$ | $0.5 \pm 0.1$ | | $(2\lambda_E^2 + \lambda_H^2)$ | $0.25 \pm 0.15\ \mathrm{GeV}^2$ | |
| $\mu = \mu_{hc}$ | $1.5 \pm 0.5$ GeV | | $\mu_0$ | 1 GeV | |
| $\mu_{h_1}$ | $[m_b/2, 2m_b]$ | | $\mu_{h_2}$ | $[m_b/2, 2m_b]$ | |
| $\tau_{B_d}$ | $1.517 \pm 0.004$ ps | [3] | $m_\mu$ | 105.658 MeV | [3] |
| $\tau_{B_s}$ | $1.527 \pm 0.011$ ps | [3] | $m_\tau$ | $1776.93 \pm 0.09$ MeV | [3] |

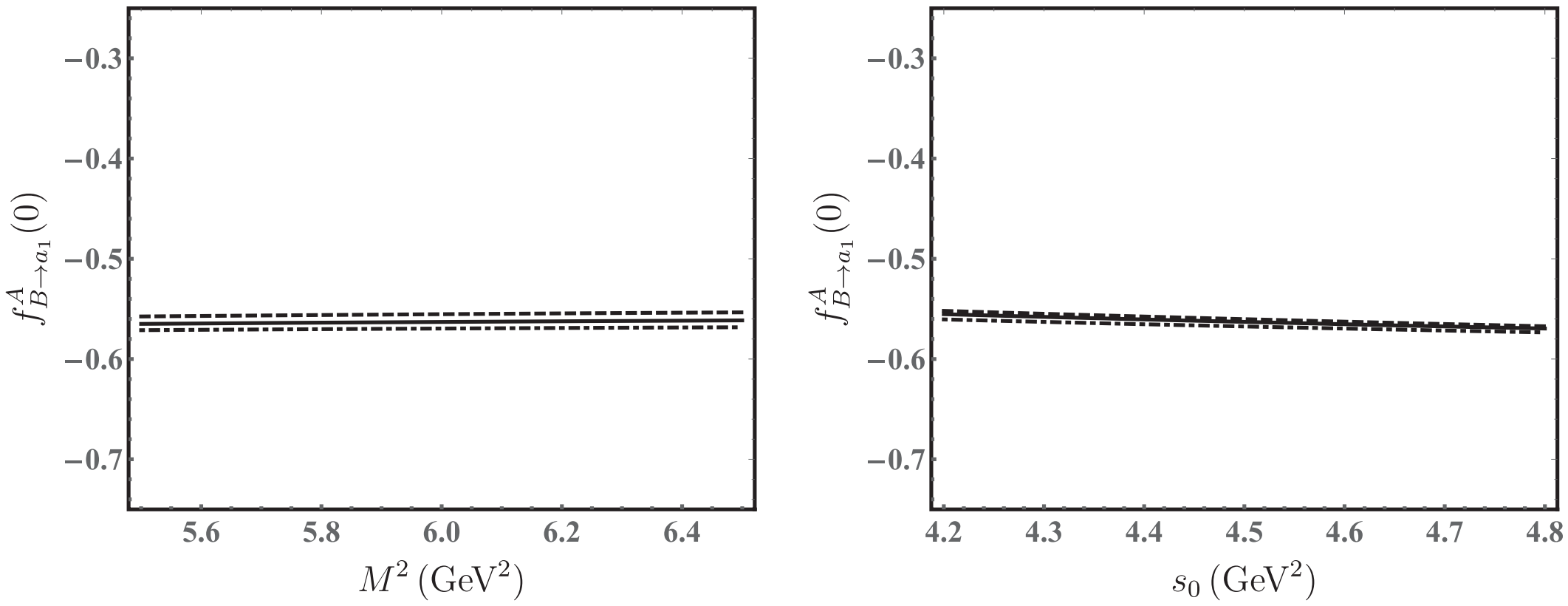

FIG. 4. The Borel mass dependence (left panel) and effective threshold dependence (right panel) of the $B \to a_1$ form factor $A$ at $q^2 = 0$. The solid, dotted, and dashed lines correspond to $s_0 = s_{0\mathrm{min}}, s_{0\mathrm{cen}}, s_{0\mathrm{max}}$ (left panel) and $M^2 = M^2_{\mathrm{min}}, M^2_{\mathrm{cen}}, M^2_{\mathrm{max}}$ (right panel), respectively.

Substituting the values of the mixing angles $\theta_K = -34°$, $\theta_{1_{P_1}} = 28°$, $\theta_{3_{P_1}} = 23°$ and the masses of the mass-eigenstate mesons into the above equations, respectively, we can obtain the singlet and octet masses including $m_{f_{1,8}}$, $m_{h_{1,8}}$, and $m_{K_{1A,1B}}$ to further calculate.

### B. Predictions for $B \to A$ form factors

We present the fitted results of the semileptonic $B \to A$ form factors at the maximum recoil point $q^2 = 0$ in Table II. Utilizing the parametrized mixing relations in terms of the $B \to A$ form factors in [76], the values of the form factors in the $B \to f_1(1285, 1420)$, $h_1(1170, 1415)$, $K_1(1270, 1400)$ processes are displayed in Table III. We also collect some numerical results from other methods shown in Table IV and compare these results with our own [see [77] for the results of the $B \to K_1(1270, 1400)$ form factors using the three-point QCD sum rules approach]. It is worth noting that the mixing angles used in [2] are $\theta_K = -45°(45°)$, $\theta_{1_{P_1}} = 45°(10°)$, $\theta_{3_{P_1}} = 50°(38°)$, which is different from the mixing angles $\theta_K = -34°$, $\theta_{1_{P_1}} = 28°$, $\theta_{3_{P_1}} = 23°$ applied in this work.

After investigating the results of the final sum rules for the $B \to A$ form factors, we now readily discuss in detail the NLL contributions and the higher-twist corrections to these form factors computed with the $B$-meson LCDAs. The corrections to the $B \to A$ form factors at NLL can yield an approximately 20%–25% reduction to the LL QCD calculations, in agreement with the previous observations for the $B \to \pi, K$ [44], and $B \to V$ [46] form factors. Furthermore, the total higher-twist contributions generate approximately $\mathcal{O}(10\%)$ corrections to the NLL QCD predictions. In particular, the three-particle corrections to the $B \to A$ form factors $\mathcal{V}_0$ and $\mathcal{V}_{12}$ give rise to an approximately 4% reduction to the two-particle contributions. On the contrary, the three-particle predictions to other $B \to A$ form factors can cause an approximately 10% enhancement to two-particle corrections.

TABLE II. $B \to a_1, b_1, f_1, f_8, h_1, h_8, K_{1A}, K_{1B}$ and $B_s \to f_1, f_8, h_1, h_8, K_{1A}, K_{1B}$ form factors with higher-twist corrections at the maximum recoil point $q^2 = 0$.

| Processes | $\mathcal{V}_0(0)$ | $\mathcal{V}_{12}(0)$ | $\mathcal{T}_{23}(0)$ | $\mathcal{V}_1(0)$ | $\mathcal{A}(0)$ | $\mathcal{T}_1(0)$ | $\mathcal{T}_2(0)$ |
|---|---|---|---|---|---|---|---|
| $B \to a_1$ | 0.48(7) | 0.22(3) | 0.21(3) | −0.56(9) | −0.33(6) | −0.44(7) | −0.44(7) |
| $B \to b_1$ | 0.64(9) | 0.30(4) | 0.28(4) | −0.74(12) | −0.44(7) | −0.59(9) | −0.59(9) |
| $B \to f_1$ | 0.48(7) | 0.24(4) | 0.22(4) | −0.57(10) | −0.32(5) | −0.45(7) | −0.45(7) |
| $B \to f_8$ | 0.52(8) | 0.28(4) | 0.26(4) | −0.63(11) | −0.34(6) | −0.48(8) | −0.48(8) |
| $B \to h_1$ | 0.63(10) | 0.29(5) | 0.27(5) | −0.73(13) | −0.44(8) | −0.58(10) | −0.58(10) |
| $B \to h_8$ | 0.63(10) | 0.33(5) | 0.30(5) | −0.76(13) | −0.42(7) | −0.59(10) | −0.59(10) |
| $B \to K_{1A}$ | 0.48(8) | 0.25(4) | 0.23(4) | −0.58(10) | −0.32(6) | −0.44(7) | −0.44(7) |
| $B \to K_{1B}$ | 0.62(9) | 0.31(4) | 0.29(4) | −0.74(11) | −0.42(7) | −0.57(9) | −0.57(9) |
| $B_s \to f_1$ | 0.54(8) | 0.26(4) | 0.25(4) | −0.63(11) | −0.36(6) | −0.50(8) | −0.50(8) |
| $B_s \to f_8$ | 0.58(9) | 0.30(5) | 0.28(5) | −0.70(11) | −0.38(6) | −0.53(8) | −0.53(8) |
| $B_s \to h_1$ | 0.71(11) | 0.32(5) | 0.30(5) | −0.82(14) | −0.49(8) | −0.65(11) | −0.65(11) |
| $B_s \to h_8$ | 0.71(12) | 0.36(6) | 0.34(6) | −0.85(15) | −0.47(8) | −0.65(11) | −0.65(11) |
| $B_s \to K_{1A}$ | 0.54(9) | 0.27(5) | 0.25(5) | −0.64(12) | −0.36(7) | −0.50(9) | −0.50(9) |
| $B_s \to K_{1B}$ | 0.69(11) | 0.34(5) | 0.31(5) | −0.81(14) | −0.47(8) | −0.64(10) | −0.64(10) |

TABLE III. $B(B_s) \to f_1(1285, 1420)$, $B(B_s) \to h_1(1170, 1415)$, and $B(B_s) \to K_1(1270, 1400)$ form factors for axial-vector mesons at the maximum recoil point $q^2 = 0$ with higher-twist corrections and the given mixing angles $\theta_K = -34°$, $\theta_{1_{P_1}} = 28°$, $\theta_{3_{P_1}} = 23°$.

| | $B \to f_1(1285)$ | $B^- \to f_1(1420)$ | $B_s \to f_1(1285)$ | $B_s \to f_1(1420)$ |
|---|---|---|---|---|
| $V_0(0)$ | 0.68(8) | 0.30(8) | 0.76(9) | 0.33(8) |
| $V_2(0)$ | −0.71(9) | −0.32(10) | −0.80(10) | −0.36(10) |
| $V_1(0)$ | −0.78(10) | −0.36(11) | −0.86(11) | −0.40(11) |
| $A(0)$ | −0.43(6) | −0.18(6) | −0.48(6) | −0.21(6) |
| $T_1(0)$ | −0.60(7) | −0.26(8) | −0.67(8) | −0.30(8) |
| $T_2(0)$ | −0.59(7) | −0.26(8) | −0.66(8) | −0.29(8) |
| $T_3(0)$ | −0.90(11) | −0.41(12) | −1.00(12) | −0.46(13) |
| | $B \to h_1(1170)$ | $B \to h_1(1415)$ | $B_s \to h_1(1170)$ | $B_s \to h_1(1415)$ |
| $V_0(0)$ | 0.93(10) | 0.28(10) | 1.04(12) | 0.32(11) |
| $V_2(0)$ | −0.93(11) | −0.29(12) | −1.04(13) | −0.33(14) |
| $V_1(0)$ | −1.02(13) | −0.33(13) | −1.14(14) | −0.37(15) |
| $A(0)$ | −0.57(7) | −0.16(8) | −0.64(8) | −0.18(9) |
| $T_1(0)$ | −0.79(10) | −0.24(10) | −0.88(11) | −0.27(11) |
| $T_2(0)$ | −0.78(9) | −0.24(10) | −0.88(11) | −0.27(11) |
| $T_3(0)$ | −1.26(14) | −0.40(15) | −1.31(16) | −0.43(17) |
| | $B \to K_1(1270)$ | $B \to K_1(1400)$ | $B_s \to K_1(1270)$ | $B_s \to K_1(1400)$ |
| $V_0(0)$ | 0.24(9) | 0.71(8) | 0.27(11) | 0.79(9) |
| $V_2(0)$ | −0.27(10) | −0.84(10) | −0.30(12) | −0.94(11) |
| $V_1(0)$ | −0.29(11) | −0.88(10) | −0.31(13) | −0.98(12) |
| $A(0)$ | −0.17(6) | −0.50(6) | −0.19(8) | −0.56(7) |
| $T_1(0)$ | −0.23(8) | −0.69(8) | −0.25(10) | −0.77(9) |
| $T_2(0)$ | −0.23(8) | −0.69(8) | −0.25(10) | −0.77(9) |
| $T_3(0)$ | −0.34(13) | −1.04(12) | −0.37(15) | −1.16(14) |

TABLE IV. The comparison between the fitted results of $B \to A$ form factors obtained using different approaches at the maximum recoil point $q^2 = 0$ with the given $K_1(1270) - K_1(1400)$ mixing angle $\theta_K = -34°$.

| Processes | Methods | $V_0(0)$ | $V_1(0)$ | $V_2(0)$ | $A(0)$ | $T_1(0)$ | $T_2(0)$ | $T_3(0)$ |
|---|---|---|---|---|---|---|---|---|
| $\bar{B}_0 \to a_1^+(1260)$ | This work | 0.48(7) | −0.56(9) | −0.53(8) | −0.33(6) | −0.44(7) | −0.44(7) | −0.65(10) |
| | LCSR [20] | 0.30 | 0.68 | 0.31 | 0.42 | 0.44 | 0.44 | 0.41 |
| | LCSR [18] | 0.30 | 0.37 | 0.42 | 0.48 | | | |
| | pQCD [2] | 0.34 | 0.43 | 0.13 | 0.26 | 0.34 | 0.34 | 0.30 |
| | LCSR [81] | 0.11 | 0.73 | 0.41 | 0.41 | | | |
| | LCSR [16] | 0.29 | 0.67 | 0.31 | 0.41 | | | |
| $\bar{B}_0 \to b_1^+(1235)$ | This work | 0.64(9) | −0.74(12) | −0.70(11) | −0.44(7) | −0.59(9) | −0.59(9) | −0.87(13) |
| | LCSR [18] | −0.39 | −0.20 | −0.09 | −0.25 | | | |
| | pQCD [2] | 0.45 | 0.33 | 0.03 | 0.19 | 0.27 | 0.27 | 0.18 |
| | LCSR [81] | −0.05 | −0.29 | −0.17 | −0.17 | | | |
| $\bar{B}_s^0 \to K_1^+(1270)$ | This work | 0.27(11) | −0.31(13) | −0.30(12) | −0.19(8) | −0.25(10) | −0.25(10) | −0.37(15) |
| | AdS/QCD [82] | 0.29 | −0.45 | −0.39 | −0.60 | −0.37 | −0.36 | −0.22 |
| $\bar{B}_s^0 \to K_1^+(1400)$ | This work | 0.79(9) | −0.98(12) | −0.94(11) | −0.56(7) | −0.77(9) | −0.77(9) | −1.16(14) |
| | AdS/QCD [82] | −0.29 | 0.13 | 0.20 | 0.11 | 0.11 | 0.10 | 0.14 |

As the validity of the light-cone operator product expansion technique is only demonstrated at the large hadronic recoil, it is of necessity to extrapolate the LCSR predictions of the $B \to A$ form factors to the large $q^2$ region. Therefore, we apply the model-independent Bourrely-Caprini-Lellouch (BCL) parametrization and present the definition of the variable $z$ in the whole complex plane as [78]

TABLE V. Theory inputs of the resonance masses of $B$ mesons with different quantum numbers entering the BCL parametrization for the $B \to A$ form factors, where the calligraphic form factors implicate the linear combinations of the conventionally defined form factors.

| Form factors | $J^P$ | $b \to d$ (GeV) | $b \to s$ (GeV) |
| --- | --- | --- | --- |
| $\mathcal{A}(q^2)$, $\mathcal{T}_1(q^2)$ | $1^+$ | 5.726 | 5.829 |
| $\mathcal{V}_0(q^2)$ | $0^+$ | | |
| $\mathcal{V}_1(q^2)$, $\mathcal{V}_{12}(q^2)$, $\mathcal{T}_2(q^2)$, $\mathcal{T}_{23}(q^2)$ | $1^-$ | 5.325 | 5.415 |

$$z(q^2, t_0) = \frac{\sqrt{t_+ - q^2} - \sqrt{t_+ - t_0}}{\sqrt{t_+ - q^2} + \sqrt{t_+ - t_0}}, \tag{6.9}$$

with parameters $t_+$ and $t_0$ given as [44,79]

$$t_+ = (m_B + m_A)^2, \qquad t_0 = (m_B - m_A)^2. \tag{6.10}$$

Using this $z$ parametrization of production, the variable $z$ can conformally map the cut $q^2$ plane onto a unit disk $|z(q^2, t_0)| < 1$. The BCL $z$-series expansion can be derived as [78]

$$F^i_{B\to A}(q^2) = \frac{F^i_{B\to A}(0)}{1 - q^2/m^2_{i,\mathrm{pole}}} \times \left\{1 + \sum_{k=1}^{N} b^i_k \left[z(q^2, t_0)^k - z(0, t_0)^k\right]\right\}, \tag{6.11}$$

where the $m_{i,\mathrm{pole}}$ denote the resonance masses of $B$ mesons with different quantum numbers. We summarize the resonance masses with $J^P = 1^+$ and $J^P = 1^-$ from the Particle Data Group [3] in Table V. In the absence of available data of the resonance masses with $J^P = 0^+$, the fitting results of form factor $V_0$ are obtained without corresponding resonance masses. Considering there are no available LQCD results to constrain the uncertainties of the $B \to A$ form factors in the large recoil region (see [80] for further discussions on the combination of the LCSR and LQCD results), we eventually enforce the truncation of $N = 1$ because the error of fitting values with $N = 2$ is too large in the large $q^2$ region to guarantee the reliability of the $B \to A$ form factors in this range.

We will concretely discuss the $\chi^2$ fitting of the BCL parameters next. Six data points for each $B \to A$ form factor are calculated using LCSR with $B$-meson LCDAs corresponding to $q^2 = -6, -4, -2, 0, 2, 4\ \mathrm{GeV}^2$, respectively. Two relations, namely, $(m_B - m_A)V_1(0) - (m_B + m_A)V_2(0) = 2m_A V_0(0)$ and $T_1(0) = T_2(0)$, obtained from the definitions of the $B \to A$ form factors are also taken into account. After applying the combined fitting, we can obtain fourteen minimal $\chi^2$ with a range between 8 and 13 for 28 d.o.f., which suggests that it is sufficient to obtain excellent fitted results using this fitting scheme. Tables X–XXIII show the fitting results of the BCL series coefficients $b^i_k$ and relative correlation matrices predicted from the $B$-meson LCSR in Appendix B, and Figs. 10–23 display the momentum-transfer dependence of these form factors in Appendix C.

### C. Semileptonic $B \to A\ell\bar{\nu}_\ell$ decays

It is straightforward to write down the effective Hamiltonian for the $b \to u$ transition

$$\mathcal{H}_{\mathrm{eff}}(b \to u\ell\bar{\nu}_\ell) = \frac{G_F}{\sqrt{2}} V_{ub}\bar{u}\gamma_\mu(1-\gamma_5)b\bar{\ell}\gamma^\mu(1-\gamma_5)\nu_\ell + \mathrm{H.c.}, \tag{6.12}$$

where we can readily write down the differential decay rate of $B \to A\ell\bar{\nu}_\ell$ as [2]

$$\frac{d\Gamma_L(B \to A\ell\bar{\nu}_\ell)}{dq^2} = \left(\frac{q^2 - m^2_\ell}{q^2}\right)^2 \frac{\sqrt{\lambda(m^2_B, m^2_A, q^2)} G^2_F V^2_{ub}}{384 m^3_B \pi^3} \frac{1}{q^2} \left\{3m^2_\ell \lambda(m^2_B, m^2_A, q^2) V^2_0(q^2) + (m^2_\ell + 2q^2) \left| \frac{1}{2m_A} \left[(m^2_B - m^2_A - q^2)(m_B - m_A)V_1(q^2) - \frac{\lambda(m^2_B, m^2_A, q^2)}{m_B - m_A} V_2(q^2)\right] \right|^2 \right\}, \tag{6.13}$$

$$\frac{d\Gamma_\pm(B \to A\ell\bar{\nu}_\ell)}{dq^2} = \left(\frac{q^2 - m^2_\ell}{q^2}\right)^2 \frac{\sqrt{\lambda(m^2_B, m^2_A, q^2)} G^2_F V^2_{ub}}{384 m^3_B \pi^3} \left\{(m^2_\ell + 2q^2)\lambda(m^2_B, m^2_A, q^2) \left| \frac{A(q^2)}{m_B - m_A} \mp \frac{(m_B - m_A)V_1(q^2)}{\sqrt{\lambda(m^2_B, m^2_A, q^2)}} \right|^2 \right\}, \tag{6.14}$$

with $\lambda(m^2_B, m^2_A, q^2) = (m^2_B + m^2_A - q^2)^2 - 4m^2_B m^2_A$. The subscripts $L$ and $\pm$ in (6.13) and (6.14) represent longitudinal and transverse polarization branching ratios, respectively. In Eqs. (6.13) and (6.14), our integral range is from $m^2_\ell$ to $(m_B - m_A)^2$, so these decay widths can keep finite at $q^2 \to m^2_\ell$ due to the suppression of the factor $(q^2 - m^2_\ell)^2/q^4$, and these decay widths will go to zero at $q^2 = (m_B - m_A)^2$ because the kinematic coefficient $\lambda(m_B^2, m_A^2, q^2) = 0$ at the upper limit $(m_B - m_A)^2$ of the integral. We fail to extract the values of the CKM matrix element $V_{ub}$ because there are no available experimental results at present. The differential decay widths divided by $|V_{ub}|^2$ and the normalized differential decay rates of the exclusive $B \to A\ell\bar{\nu}_\ell$ $(\ell = \mu, \tau)$ with the form factors calculated from LCSR are presented in Figs. 5–7. We also display the numerical integration results of the $B \to A\ell\bar{\nu}_\ell$ $(\ell = \mu, \tau)$ branching ratios in Tables VI and VII.

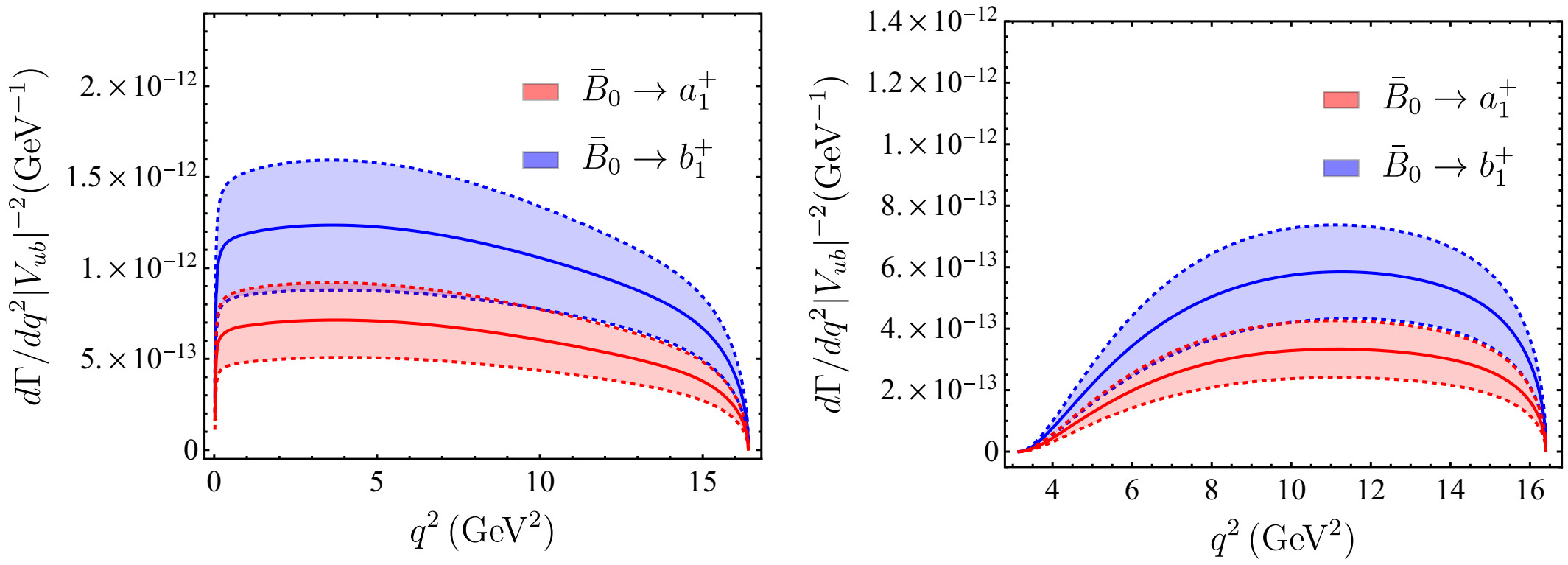

FIG. 5. The differential decay widths of $\bar{B}_0 \to a_1^+(1260)\ell\bar{\nu}_\ell$ and $\bar{B}_0 \to b_1^+(1235)\ell\bar{\nu}_\ell$. The leptons generated in the left (right) panel correspond to $\mu$ $(\tau)$ and $\bar{\nu}_\mu$ $(\bar{\nu}_\tau)$.

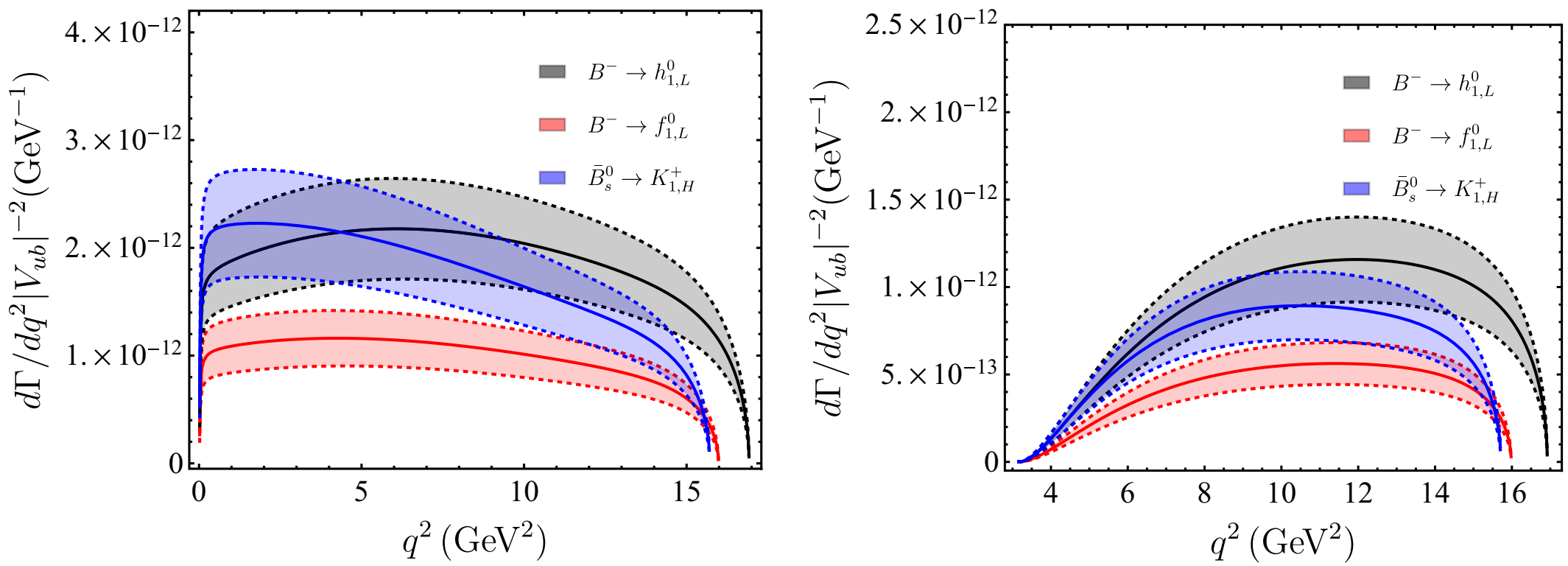

FIG. 6. The differential decay widths of $B^- \to h_1^0(1170)\ell\bar{\nu}_\ell$, $B^- \to f_1^0(1285)\ell\bar{\nu}_\ell$, and $\bar{B}_s^0 \to K_1^+(1400)\ell\bar{\nu}_\ell$. $h_{1,L}^0$, $f_{1,L}^0$, and $K_{1,H}^+$ represent $h_1^0(1170)$, $f_1^0(1285)$, and $K_1^+(1400)$, respectively.

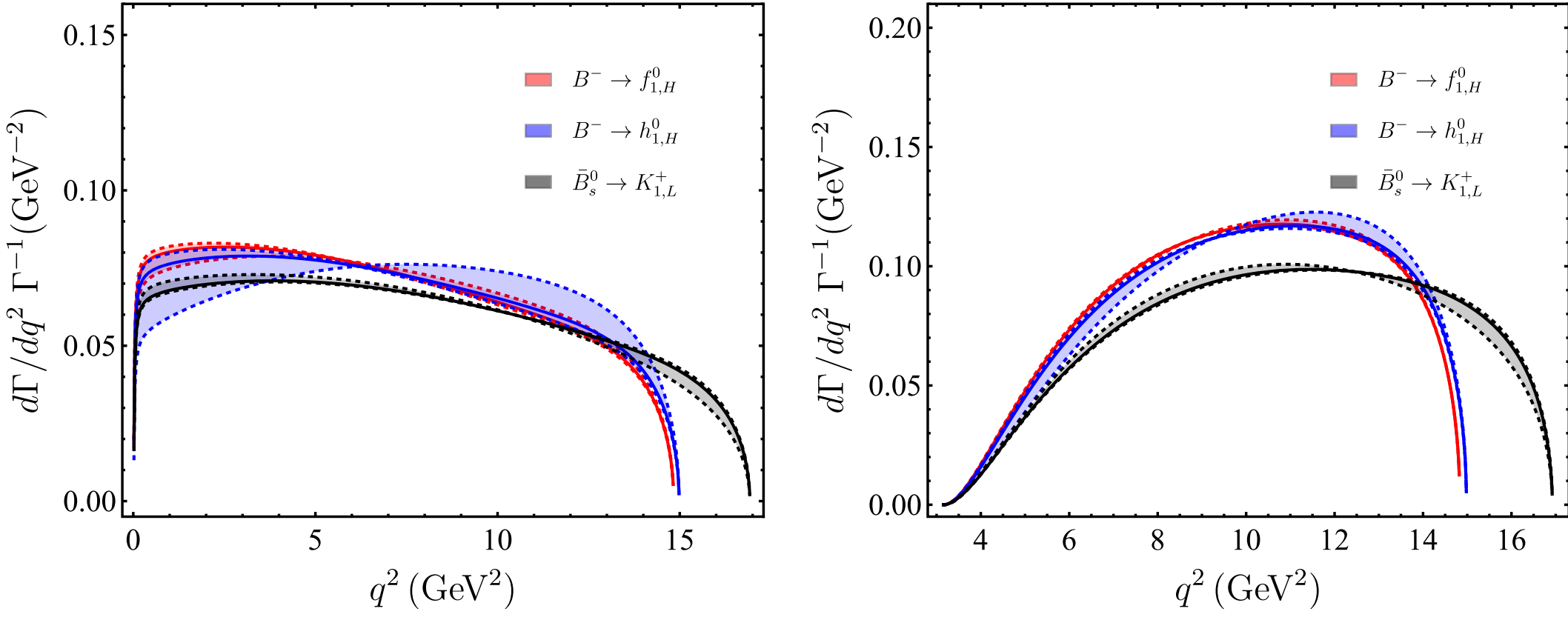

FIG. 7. The normalized differential decay widths of $B^- \to h_1^0(1415)\ell\bar{\nu}_\ell$, $B^- \to f_1^0(1420)\ell\bar{\nu}_\ell$, and $\bar{B}_s^0 \to K_1^+(1270)\ell\bar{\nu}_\ell$. $h_{1,H}^0$, $f_{1,H}^0$, and $K_{1,L}^+$ represent $h_1^0(1415)$, $f_1^0(1420)$, and $K_1^+(1270)$, respectively.

TABLE VI. The summary of branching ratios for $B \to A\mu\bar{\nu}_\mu$ calculated with the given mixing angles $\theta_K = -34°$, $\theta_{1_{P_1}} = 28°$, $\theta_{3_{P_1}} = 23°$, where $\mathcal{BR}_{\rm L}$ and $\mathcal{BR}_\pm$ are the contributions to the branching ratios from longitudinal and transverse polarization, respectively, and $\mathcal{BR}_{\rm T} = \mathcal{BR}_+ + \mathcal{BR}_-$.

| Processes | $10^4 \times \mathcal{BR}_{\rm L}$ | $10^5 \times \mathcal{BR}_+$ | $10^4 \times \mathcal{BR}_-$ | $10^4 \times \mathcal{BR}_{\rm total}$ | $\mathcal{BR}_{\rm L}/\mathcal{BR}_{\rm T}$ |
|---|---|---|---|---|---|
| $\bar{B}_0 \to a_1^+(1260)$ | 1.18(36) | 0.63(30) | 1.98(61) | 3.23(98) | 0.58(5) |
| $\bar{B}_0 \to b_1^+(1235)$ | 2.02(65) | 1.10(52) | 3.50(102) | 5.63(169) | 0.56(5) |
| $B^- \to f_1^0(1285)$ | 1.82(44) | 1.36(48) | 3.35(82) | 5.31(129) | 0.52(4) |
| $B^- \to f_1^0(1420)$ | 0.32(18) | 0.24(18) | 0.54(30) | 0.88(49) | 0.56(10) |
| $B^- \to h_1^0(1170)$ | 3.30(82) | 2.89(103) | 6.90(171) | 10.49(256) | 0.46(4) |
| $B^- \to h_1^0(1415)$ | 0.25(22) | 0.24(23) | 0.46(35) | 0.73(57) | 0.51(13) |
| $\bar{B}_s^0 \to K_1^+(1270)$ | 0.43(30) | 0.21(32) | 0.70(54) | 1.15(85) | 0.60(51) |
| $\bar{B}_s^0 \to K_1^+(1400)$ | 3.68(90) | 1.53(57) | 5.25(131) | 9.09(224) | 0.68(4) |

TABLE VII. The summary of the branching ratios for $B \to A\tau\bar{\nu}_\tau$ calculated with the given mixing angles $\theta_K = -34°$, $\theta_{1_{P_1}} = 28°$, $\theta_{3_{P_1}} = 23°$, where $\mathcal{BR}_{\rm L}$ and $\mathcal{BR}_\pm$ are the contributions to the branching ratios from longitudinal and transverse polarization, respectively, and $\mathcal{BR}_{\rm T} = \mathcal{BR}_+ + \mathcal{BR}_-$.

| Processes | $10^4 \times \mathcal{BR}_{\rm L}$ | $10^5 \times \mathcal{BR}_+$ | $10^4 \times \mathcal{BR}_-$ | $10^4 \times \mathcal{BR}_{\rm total}$ | $\mathcal{BR}_{\rm L}/\mathcal{BR}_{\rm T}$ |
|---|---|---|---|---|---|
| $\bar{B}_0 \to a_1^+(1260)$ | 0.16(4) | 0.40(18) | 0.88(27) | 1.08(32) | 0.17(2) |
| $\bar{B}_0 \to b_1^+(1235)$ | 0.26(9) | 0.70(32) | 1.56(45) | 1.89(55) | 0.16(2) |
| $B^- \to f_1^0(1285)$ | 0.24(5) | 0.83(28) | 1.45(35) | 1.77(42) | 0.15(1) |
| $B^- \to f_1^0(1420)$ | 0.03(2) | 0.14(10) | 0.22(12) | 0.26(14) | 0.15(3) |
| $B^- \to h_1^0(1170)$ | 0.49(11) | 1.81(62) | 3.17(77) | 3.84(92) | 0.15(1) |
| $B^- \to h_1^0(1415)$ | 0.03(2) | 0.14(13) | 0.19(14) | 0.23(17) | 0.14(3) |
| $\bar{B}_s^0 \to K_1^+(1270)$ | 0.06(4) | 0.14(20) | 0.32(24) | 0.40(30) | 0.18(29) |
| $\bar{B}_s^0 \to K_1^+(1400)$ | 0.44(10) | 0.97(35) | 2.24(55) | 2.77(68) | 0.19(1) |

According to the expressions for the differential decay rates of $B \to A\ell\bar{\nu}_\ell$, we can compute the lepton-flavor universality observables with the combined fitting results of the $B \to A$ form factors by using the definition

$$\mathcal{R}_A = \frac{\Gamma(B \to A\tau\bar{\nu}_\tau)}{\Gamma(B \to A\mu\bar{\nu}_\mu)} = \frac{\int_{m_\tau^2}^{q^2_{\rm max}} dq^2 d\Gamma(B \to A\tau\bar{\nu}_\tau)/dq^2}{\int_{m_\mu^2}^{q^2_{\rm max}} dq^2 d\Gamma(B \to A\mu\bar{\nu}_\mu)/dq^2}. \tag{6.15}$$

In order to better understand the phenomenological aspect of the $B \to A\ell\bar{\nu}_\ell$ transverse polarized branching ratios, the lepton transverse polarization asymmetries are also calculated with the combined BCL $z$-series expansion fitting of the $B \to A$ form factors obtained from LCSR, whose expressions are given as

$$\begin{aligned}\mathcal{A}_T^\ell &= \frac{\Gamma_-(B \to A\ell\bar{\nu}_\ell) - \Gamma_+(B \to A\ell\bar{\nu}_\ell)}{\Gamma_-(B \to A\ell\bar{\nu}_\ell) + \Gamma_+(B \to A\ell\bar{\nu}_\ell)} \\ &= \frac{\int_{m_\ell^2}^{q^2_{\rm max}} dq^2 d\Gamma_-(B \to A\ell\bar{\nu}_\ell)/dq^2 - \int_{m_\ell^2}^{q^2_{\rm max}} dq^2 d\Gamma_+(B \to A\ell\bar{\nu}_\ell)/dq^2}{\int_{m_\ell^2}^{q^2_{\rm max}} dq^2 d\Gamma_-(B \to A\ell\bar{\nu}_\ell)/dq^2 + \int_{m_\ell^2}^{q^2_{\rm max}} dq^2 d\Gamma_+(B \to A\ell\bar{\nu}_\ell)/dq^2}.\end{aligned} \tag{6.16}$$

According to [46], we can derive the angular distribution of the $B \to A\ell\bar{\nu}_\ell$ decays as

$$\frac{d^2\Gamma(B \to A\ell\bar{\nu}_\ell)}{dq^2 d\cos\theta} = \frac{G_F^2 |V_{ub}|^2 q^2}{256\pi^3 m_B^3} \lambda^{1/2}(m_B^2, m_A^2, q^2) \left\{ \sin^2\theta |H_0(q^2)|^2 + (1-\cos\theta)^2 \frac{|H_+(q^2)|^2}{2} + (1+\cos\theta)^2 \frac{|H_-(q^2)|^2}{2} \right\}, \tag{6.17}$$

TABLE VIII. The summary of the numerical results of the observables including lepton-flavor universality observables $\mathcal{R}_A$, lepton transverse asymmetries $\mathcal{A}_T^\ell$, and forward-backward asymmetries $\mathcal{A}_{FB}^\ell$.

| Processes | $\mathcal{R}_A$ | $\mathcal{A}_T^\mu$ | $\mathcal{A}_T^\tau$ | $\mathcal{A}_{FB}^\mu$ | $\mathcal{A}_{FB}^\tau$ |
| --- | --- | --- | --- | --- | --- |
| $\bar{B}_0 \to a_1^+(1260)$ | 0.334(11) | 0.937(21) | 0.912(27) | 0.452(13) | 0.528(15) |
| $\bar{B}_0 \to b_1^+(1235)$ | 0.336(11) | 0.937(20) | 0.912(26) | 0.457(11) | 0.533(13) |
| $B^- \to f_1^0(1285)$ | 0.333(8) | 0.920(18) | 0.890(23) | 0.452(10) | 0.528(12) |
| $B^- \to f_1^0(1420)$ | 0.302(22) | 0.907(52) | 0.872(63) | 0.441(25) | 0.521(30) |
| $B^- \to h_1^0(1170)$ | 0.364(9) | 0.919(20) | 0.892(24) | 0.462(11) | 0.533(12) |
| $B^- \to h_1^0(1415)$ | 0.322(41) | 0.868(139) | 0.832(144) | 0.432(75) | 0.507(86) |
| $\bar{B}_s^0 \to K_1^+(1270)$ | 0.344(32) | 0.914(107) | 0.890(115) | 0.434(60) | 0.506(70) |
| $\bar{B}_s^0 \to K_1^+(1400)$ | 0.304(7) | 0.942(14) | 0.915(19) | 0.442(9) | 0.522(10) |

with the helicity amplitudes $H_i(q^2)(i = \pm, 0)$ given as

$$H_\pm(q^2) = (m_B - m_A)\left[\frac{\lambda^{1/2}(m_B^2, m_A^2, q^2)}{(m_B - m_A)^2}A(q^2) \mp V_1(q^2)\right],$$
$$H_0(q^2) = \frac{m_B - m_A}{2m_A\sqrt{q^2}}\left[(m_B^2 - m_A^2 - q^2)V_1(q^2) - \frac{\lambda(m_B^2, m_A^2, q^2)}{(m_B - m_A)^2}V_2(q^2)\right]. \tag{6.18}$$

Employing the relations (6.17) and (6.18), it is straightforward to write down the expression of lepton forward-backward asymmetry in the semileptonic $B \to A\ell\bar{\nu}_\ell$ decay mode

$$\mathcal{A}_{FB}^\ell(q^2) = \left[\int_0^1 d\cos\theta \frac{d^2\Gamma}{dq^2 d\cos\theta} - \int_{-1}^0 d\cos\theta \frac{d^2\Gamma}{dq^2 d\cos\theta}\right]\left[\frac{d\Gamma(B \to A\ell\bar{\nu}_\ell)}{dq^2}\right]^{-1}$$
$$= \frac{3}{4}\frac{|H_-|^2 - |H_+|^2}{|H_0|^2 + |H_-|^2 + |H_+|^2}. \tag{6.19}$$

The numerical predictions of the observables including lepton-flavor universality observables, lepton transverse polarization asymmetries, and forward-backward asymmetries taking advantage of the BCL $z$-series expansion combined fitted $B \to A$ form factors are presented in Table VIII. We also display the $q^2$ dependence of leptonic forward-backward asymmetry for the $B \to A\ell\bar{\nu}_\ell$ processes in Fig. 8.

Finally, we can summarize our investigation of several physical observables as follows:

(i) Most of the branching ratios are of the order $10^{-4}$. The branching ratios of $B \to A\mu\bar{\nu}_\mu$ are larger than the corresponding branching ratios of the $B \to A\tau\bar{\nu}_\tau$ decays. It is obvious to figure out the relations from the results of the $B \to A\mu\bar{\nu}_\mu$ branching ratios: $\mathcal{BR}_- > \mathcal{BR}_\mathrm{L} \gg \mathcal{BR}_+$. After reviewing Table VII, we can find the relations from the $B \to A\tau\bar{\nu}_\tau$ branching ratios: $\mathcal{BR}_- > \mathcal{BR}_\mathrm{L} > \mathcal{BR}_+$. Because of the chosen mixing angles $\theta_K = -34°$, $\theta_{1_{P_1}} = 28°$, and $\theta_{3_{P_1}} = 23°$, the branching ratios of the $B^- \to f_1^0(1420)/h_1^0(1415)$ and $\bar{B}_s^0 \to K_1^+(1270)$ decays are notably smaller than the branching ratios of semileptonic $B^- \to f_1^0(1285)/h_1^0(1170)$ and $\bar{B}_s^0 \to K_1^+(1400)$, respectively (see [2,18] for more results of the $B \to A\ell\bar{\nu}_\ell$ branching ratios and related discussions).

(ii) The ratios $\mathcal{BR}_\mathrm{L}/\mathcal{BR}_\mathrm{T}$ for the $B \to A\mu\bar{\nu}_\mu$ processes in Table VI are approximately 0.4–0.7. However, the numerical results $\mathcal{BR}_\mathrm{L}/\mathcal{BR}_\mathrm{T}$ for the $B \to A\tau\bar{\nu}_\tau$ decays are predicted to be close to 0.16.

(iii) The lepton-flavor universality observables $\mathcal{R}_{a_1(1260)}$, $\mathcal{R}_{b_1(1235)}$, $\mathcal{R}_{f_1(1285)}$, and $\mathcal{R}_{h_1(1415)}$ are close to 0.330, and the values of $\mathcal{R}_{f_1(1420)}$ and $\mathcal{R}_{K_1(1400)}$ are approximately 0.300 (see [83] for $\mathcal{R}_{a_1(1260)}^{\mathrm{SM}} = 0.341$). The lepton transverse polarization asymmetries of $B \to A\mu\bar{\nu}_\mu$ are larger than those of the corresponding $B \to A\tau\bar{\nu}_\tau$. On the contrary, the theory predicted value $\mathcal{A}_{FB}^\mu$ is smaller than $\mathcal{A}_{FB}^\tau$.

### D. Semileptonic $B \to A\nu_\ell\bar{\nu}_\ell$ decays

We will proceed to explore the phenomenological aspects of semileptonic $B \to A\nu_\ell\bar{\nu}_\ell$ decays employing the form factors predicted from LCSR with higher-twist corrections. The semileptonic differential decay rate of $B \to A\nu_\ell\bar{\nu}_\ell$ is derived as [46,84]

$$\frac{d\Gamma(B\to A\nu_\ell\bar{\nu}_\ell)}{dq^2}=\frac{G_F^2\alpha_{\rm em}^2}{256\pi^5}\frac{\lambda^{3/2}(m_B^2,m_A^2,q^2)}{m_B^3\sin^4\theta_W}|V_{tb}V_{ts}^*|^2\left[X_t\left(\frac{m_t^2}{m_W^2},\frac{m_H^2}{m_t^2},\sin\theta_W,\mu\right)\right]^2[H_A(q^2)+H_{V_1}(q^2)+H_{V_{12}}(q^2)], \tag{6.20}$$

where the expressions of the three invariant functions $H_i(q^2)$ are

$$H_A(q^2)=\frac{2q^2}{(m_B-m_A)^2}[A(q^2)]^2,\qquad H_{V_1}(q^2)=\frac{2q^2(m_B-m_A)^2}{\lambda(m_B^2,m_A^2,q^2)}[V_1(q^2)]^2,$$
$$H_{V_{12}}(q^2)=\frac{64m_A^2m_B^2}{\lambda(m_B^2,m_A^2,q^2)}[V_{12}(q^2)]^2. \tag{6.21}$$

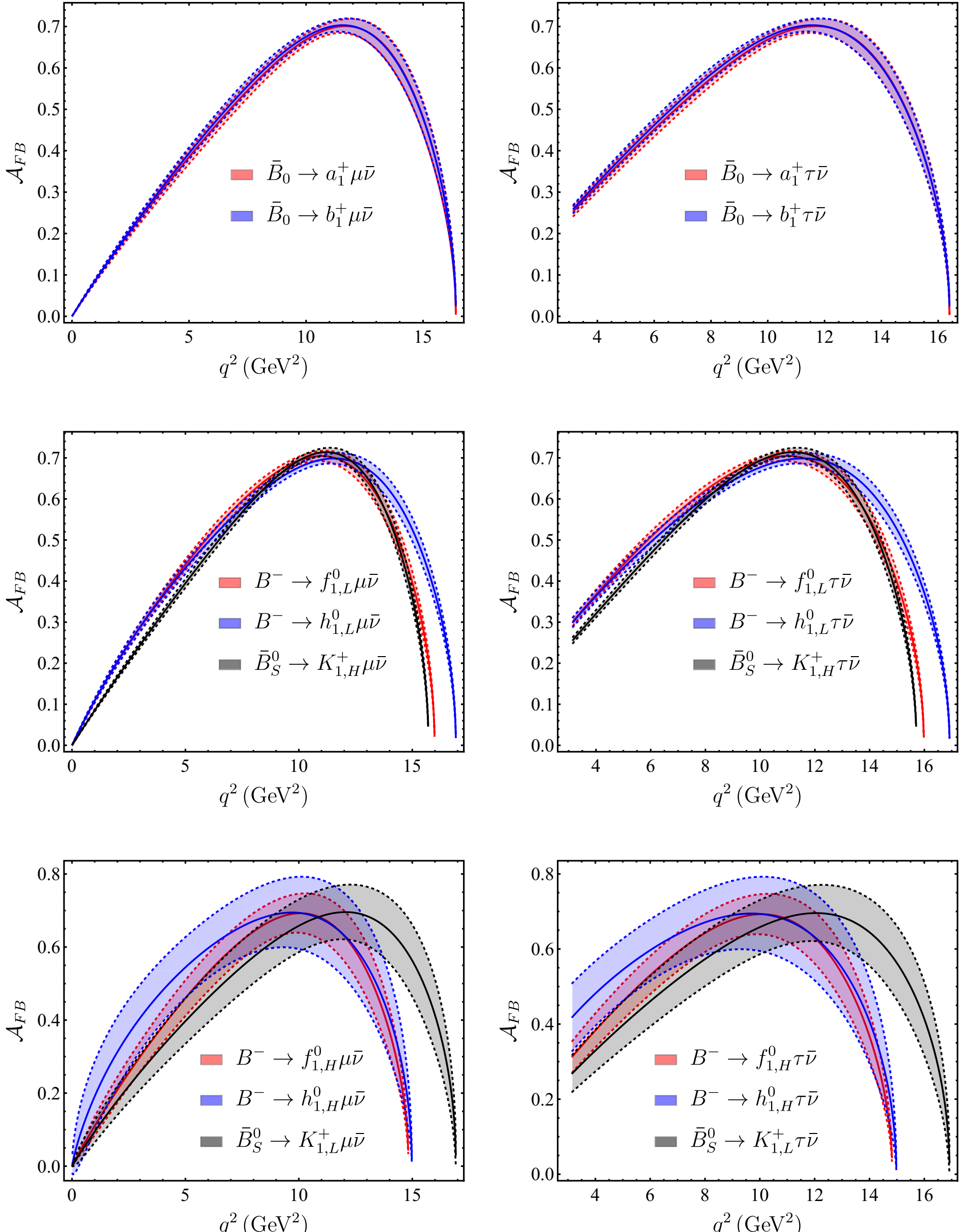

FIG. 8. The numerical predictions of the forward-backward asymmetry utilizing the BCL $z$-series expansion fitted $B \to A$ form factors from LCSR with higher-twist corrections.

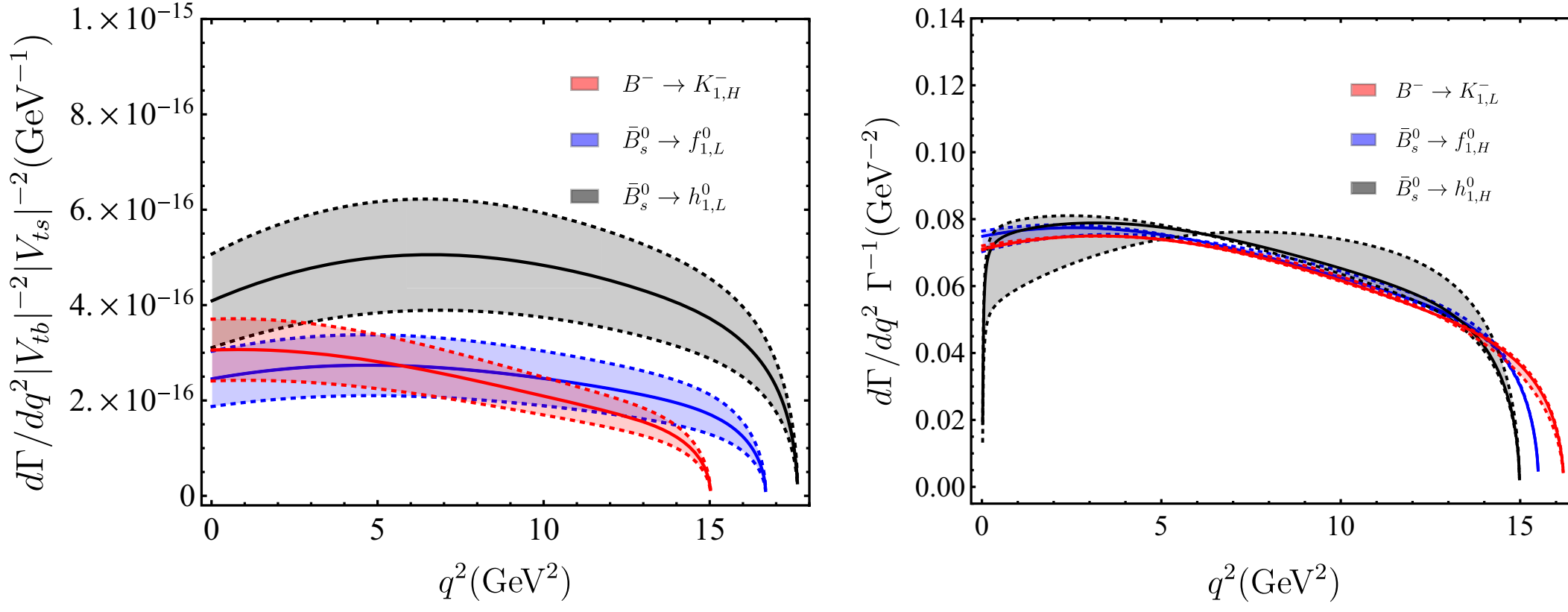

FIG. 9. The differential decay widths (left panel) of $\bar{B}^0_s \to f^0_1(1285)\nu\bar{\nu}_\ell$, $\bar{B}^0_s \to h^0_1(1170)\nu\bar{\nu}_\ell$, and $B^- \to K^+_1(1400)\nu\bar{\nu}_\ell$ and the normalized differential decay widths (right panel) of $\bar{B}^0_s \to f^0_1(1420)\nu\bar{\nu}_\ell$, $\bar{B}^0_s \to h^0_1(1415)\nu\bar{\nu}_\ell$, and $B^- \to K^+_1(1270)\nu\bar{\nu}_\ell$ are shown, where $h^0_{1,(L,H)}$, $f^0_{1,(L,H)}$, $K^+_{1,(L,H)}$ correspond to $h^0_1(1170, 1415)$, $f^0_1(1285, 1420)$, $K^+_1(1270, 1400)$, respectively.

We adopt the helicity form factor $V_{12}(q^2)$ from [85]

$$V_{12}(q^2) = \frac{(m_B - m_A)^2(m_B^2 - m_A^2 - q^2)V_1(q^2) - \lambda(m_B^2, m_A^2, q^2)V_2(q^2)}{16 m_B m_A^2 (m_B - m_A)}, \tag{6.22}$$

where we replace $(m_B + m_V)$ with $(m_B - m_A)$ as well as (6.21) because of the difference mentioned in [2] between the definitions of the $B \to A$ form factors and those of $B \to V$ form factors.

In order to obtain higher-precision theory predictions, we take advantage of the short-distance Wilson coefficient $X_t$ with the NLO QCD correction and two-loop electroweak correction [86–90].

We display the differential decay rates of rare exclusive $\bar{B}^0_s \to f^0_1(1285)\nu\bar{\nu}_\ell$, $\bar{B}^0_s \to h^0_1(1170)\nu\bar{\nu}_\ell$, and $B^- \to K^+_1(1400)\nu\bar{\nu}_\ell$ divided by $|V_{tb}|^2|V_{ts}|^2$ and the normalized differential decay widths of $\bar{B}^0_s \to f^0_1(1420)\nu\bar{\nu}_\ell$, $\bar{B}^0_s \to h^0_1(1415)\nu\bar{\nu}_\ell$, and $B^- \to K^+_1(1270)\nu\bar{\nu}_\ell$ utilizing the BCL parametrized $B \to A$ form factors with the mixing angles $\theta_K = -34°$, $\theta_{1_{P_1}} = 28°$, $\theta_{3_{P_1}} = 23°$ in Fig. 9. The numerical predictions of the branching ratios for semileptonic $B \to A\nu\bar{\nu}_\ell$ are summarized in Table IX, from which it is prominently clear that the $B \to A\nu\bar{\nu}_\ell$ branching ratios for $f_1(1285, 1420)$, $h_1(1170, 1415)$, and $K_1(1270, 1400)$ are also sensitive to the mixing angles. We expect that more LQCD and experimental results in the future will shed more light on the phenomenological investigations of semileptonic $B \to A\nu\bar{\nu}_\ell$.

TABLE IX. The summary of branching ratios for $B \to A\nu\bar{\nu}$ calculated with the given mixing angles $\theta_K = -34°$, $\theta_{1_{P_1}} = 28°$, $\theta_{3_{P_1}} = 23°$.

| Processes | $10^5 \times \mathcal{BR}$ |
| --- | --- |
| $B^- \to K^-_1(1270)$ | 0.18(11) |
| $B^- \to K^-_1(1400)$ | 1.38(29) |
| $\bar{B}^0_s \to f^0_1(1285)$ | 1.60(36) |
| $\bar{B}^0_s \to f^0_1(1420)$ | 0.27(14) |
| $\bar{B}^0_s \to h^0_1(1170)$ | 3.18(73) |
| $\bar{B}^0_s \to h^0_1(1415)$ | 0.22(18) |

## VII. CONCLUSION

In this study, we have calculated the NLO QCD contributions of the $B \to A$ form factors with higher-twist corrections from LCSR applying $B$-meson LCDAs up to twist-six accuracy. First, we constructed the LCSR for the $B \to A$ form factors at tree level by adopting the vacuum-to-$B$-meson correlation functions, which defined with axial-vector-meson interpolating currents. After observing the leading-twist contributions of the $B \to A$ form factors at tree level, it turned out that only the $B$-meson LCDA $\phi^-_B(\omega, \mu)$ contributes to the results. At one-loop level, the $B \to A$ correlation function is factorized into $B$-meson LCDAs and short-distance coefficients, among which the hard coefficients for A0-type currents and $\mu$-dependent jet functions from $\Pi^{(T+\tilde{T})}_{\mu,\parallel}$, $\Pi^{(V-A)}_{\delta\mu,\perp}$ and $\Pi^{(T+\tilde{T})}_{\delta\mu,\perp}$ are equivalent to the corresponding ones of $B \to V$ in SCET [46]. In addition, $\Pi^{(V-A)}_{\mu,\parallel}$ and $\Pi^{(T+\tilde{T})}_{\mu,\parallel}$ coincide with the results of $B \to \pi, K$ [44] under the limit $m_q \to 0$ up to NLL accuracy. It has also been verified that the $B \to A$ form factors exhibit scale independence. To obtain the NLL RG improved $B \to A$ form factors, we performed a resummation of large logarithms by applying the standard renormalization group method. Furthermore, we computed the higher-twist corrections to the $B \to A$ form factors including two-particle and three-particle contributions with higher-twist $B$-meson DAs up to the twist-six accuracy at tree level. We noticed the relations $\rho_{\rm LP}^{B\to A} = -\rho_{\rm LP}^{B\to V}$ and $\rho_{\rm NLP}^{B\to A} = \rho_{\rm NLP}^{B\to V}$ between our results and relative ones of $B \to V$ in [46] by deriving the invariant functions entering the three-particle higher-twist corrections.

We also explored the NLL and higher-twist corrections for the $B \to A$ form factors by substituting theory inputs into the final sum rules for these form factors. Corrections to the $B \to A$ form factors in NLL can yield an approximately 20%–25% reduction in the LL QCD calculations (see [44,46] for a consistent conclusion). Furthermore, the total higher-twist contributions generate approximately $\mathcal{O}(10\%)$ corrections to NLL QCD predictions. In particular, the genuine three-particle corrections to the $B \to A$ form factors $\mathcal{V}_0$ and $\mathcal{V}_{12}$ give rise to an approximately 4% reduction in the two-particle contributions. On the contrary, the genuine three-particle predictions to other $B \to A$ form factors can cause an approximately 10% enhancement to two-particle corrections.

For the sake of the extrapolation of the LCSR predictions of the $B \to A$ form factors to the whole $q^2$ region, we derived the BCL $z$-series expanded formulas of these form factors. Considering the correlations between the input LCSR data points, we then performed the $\chi^2$ fitting of the BCL parametrization and provided the results of seven sets of BCL parameters, as well as their correlation matrix. We expect future LQCD data points that can be combined with this method to reduce the uncertainties of the form factors in the large $q^2$ region.

Taking advantage of the theory predictions of the $B \to A$ form factors obtained from BCL parametrization, we investigated the phenomenological aspects of the semileptonic $B \to A\ell\bar{\nu}_\ell$ decays by calculating the differential decay widths, the branching ratios, and other observables including transverse asymmetries, forward-backward asymmetries, and lepton-flavor universality observables with the given mixing angles $\theta_K = -34°$, $\theta_{1_{P_1}} = 28°$, $\theta_{3_{P_1}} = 23°$. The differential decay widths and the branching ratios of rare exclusive $B \to A\nu_\ell\bar{\nu}_\ell$ decays have also been predicated with these mixing angles. We expect these theory predictions to be verified in future experiments.

## ACKNOWLEDGMENTS

We would like to thank Bo-Xuan Shi, Xue-Chen Zhao, and Yong-Kang Huang for helpful discussions. This work acknowledges support from the National Natural Science Foundation of China with Grants No. 12075125 and No. 12475097.

## DATA AVAILABILITY

The data are not publicly available. The data are available from the authors upon reasonable request.

## APPENDIX A: $B$-MESON DISTRIBUTION AMPLITUDES

The explicit expressions of the three-parameter model for twist-two and twist-three $B$-meson LCDAs used in numerical analysis are given as [66]

$$
\begin{aligned}
\phi_B^+(\omega,\mu) &= U_\phi(\mu,\mu_0)\frac{1}{\omega^{p+1}}\frac{\Gamma(\beta)}{\Gamma(\alpha)}\mathcal{G}(\omega;0,2,1),\\
\phi_B^{-\rm WW}(\omega,\mu) &= U_\phi(\mu,\mu_0)\frac{1}{\omega^{p+1}}\frac{\Gamma(\beta)}{\Gamma(\alpha)}\mathcal{G}(\omega;0,1,1),\\
\phi_B^{-\rm t3}(\omega,\mu) &= -\frac{1}{6}U_\phi^{\rm t3}(\mu,\mu_0)\mathcal{N}(\lambda_E^2-\lambda_H^2)\frac{\omega_0^2}{\omega^{p+3}}\frac{\Gamma(\beta)}{\Gamma(\alpha)}\Bigg\{\mathcal{G}(\omega;0,3,3)\\
&\quad + (\beta-\alpha)\left[\frac{\omega}{\omega_0}\mathcal{G}(\omega;0,2,2)-\beta\frac{\omega}{\omega_0}\mathcal{G}(\omega;1,2,2)-\mathcal{G}(\omega;1,3,3)\right]\Bigg\},
\end{aligned}
\tag{A1}
$$

where $p = \frac{\Gamma_{\rm cusp}^{(0)}}{2\beta_0}\ln[\alpha_s(\mu)/\alpha_s(\mu_0)]$. The twist-three two-particle LCDA can be written as $\phi_B^-(\omega,\mu) = \phi_B^{-\rm WW}(\omega,\mu) + \phi_B^{-\rm t3}(\omega,\mu)$ and

$$
\mathcal{G}(\omega;l,m,n) \equiv G_{23}^{21}\left(\frac{\omega}{\omega_0}\Bigg|\begin{matrix}1,\beta+l\\ p+m,\alpha,p+n\end{matrix}\right)
\tag{A2}
$$

denotes the Meijer $G$-function. The evolution factor $U_\phi(\mu,\mu_0)$ and $U_\phi^{\rm t3}(\mu,\mu_0)$ at one-loop level is shown as

$$U_{\phi}(\mu,\mu_0) = \exp\left\{-\frac{\Gamma_{\text{cusp}}^{(0)}}{4\beta_0^2}\left(\frac{4\pi}{\alpha_s(\mu_0)}\left[\ln r - 1 + \frac{1}{r}\right] - \frac{\beta_1}{2\beta_0}\ln^2 r + \left(\frac{\Gamma_{\text{cusp}}^{(1)}}{\Gamma_{\text{cusp}}^{(0)}} - \frac{\beta_1}{\beta_0}\right)[r - 1 - \ln r]\right)\right\}(\mathrm{e}^{2\gamma_E}\mu_0)^{\frac{\Gamma_{\text{cusp}}^{(0)}}{2\beta_0}\ln r} r^{\frac{\gamma_{\text{t2}}^{(0)}}{2\beta_0}},$$

$$U_{\phi}^{\text{t3}}(\mu,\mu_0) = U_{\phi}(\mu,\mu_0)\Big|_{\gamma_{t2}^{(0)} \to \gamma_{t2}^{(0)} + \gamma_{t3}^{(0)}}, \tag{A3}$$

where $r = \alpha_s(\mu)/\alpha_s(\mu_0)$. The cusp anomalous dimensions $\Gamma_{\text{cusp}}^{(i)}$ can be extended to various orders and

$$\gamma_{\text{t2}}^{(0)} = -2C_F, \qquad \gamma_{\text{t3}}^{(0)} = 2N_c. \tag{A4}$$

The higher-twist $B$-meson DAs up to the twist-six accuracy used in the phenomenological analysis are presented as

$$\hat{g}_B^-(\omega) = \frac{1}{4}\left\{(3\omega - 2\bar{\Lambda})\mathbb{F}(\omega;1) + 3\mathbb{F}(\omega;2) + \frac{1}{3}\mathcal{N}(\lambda_E^2 - \lambda_H^2)\omega\left[\omega(\bar{\Lambda} - \omega)\mathbb{F}(\omega;-1) - \left(2\bar{\Lambda} - \frac{3}{2}\omega\right)\mathbb{F}(\omega;0)\right]\right\},$$

$$\phi_4(\omega_1,\omega_2) = \frac{1}{2}\mathcal{N}(\lambda_E^2 + \lambda_H^2)\omega_2^2\mathbb{F}(\omega_1 + \omega_2;-1),$$

$$\psi_5(\omega_1,\omega_2) = -\mathcal{N}\lambda_E^2\omega_2\mathbb{F}(\omega_1 + \omega_2;0), \qquad \tilde{\psi}_5(\omega_1,\omega_2) = -\mathcal{N}\lambda_H^2\omega_2\mathbb{F}(\omega_1 + \omega_2;0),$$

$$\phi_6(\omega_1,\omega_2) = \mathcal{N}(\lambda_E^2 - \lambda_H^2)\mathbb{F}(\omega_1 + \omega_2;1), \tag{A5}$$

where

$$\mathcal{N} = \frac{1}{3}\frac{\beta(\beta+1)}{\alpha(\alpha+1)}\frac{1}{\omega_0^2}, \qquad \bar{\Lambda} = \frac{3}{2}\frac{\alpha}{\beta}\omega_0,$$

$$\mathbb{F}(\omega;n) \equiv \omega_0^{n-1}U(\beta - \alpha, 2 - n - \alpha, \omega/\omega_0)\frac{\Gamma(\beta)}{\Gamma(\alpha)}e^{-\omega/\omega_0}, \tag{A6}$$

with $U(a,b,z)$ being the hypergeometric $U$ function.

## APPENDIX B: FITTED RESULTS OF THE SEMILEPTONIC $B \to A$ FORM FACTORS

TABLE X. Fitted results of the BCL parameters for the $B \to a_1$ form factors.

| $B \to a_1$ form factors | | Correlation matrix | | | | | | | | | | | |
|---|---|---|---|---|---|---|---|---|---|---|---|---|---|
| $p$ | Values | $b_1^{\mathcal{V}_0}$ | $b_0^{\mathcal{V}_{12}}$ | $b_1^{\mathcal{V}_{12}}$ | $b_0^{\mathcal{T}_{23}}$ | $b_1^{\mathcal{T}_{23}}$ | $b_0^{\mathcal{V}_1}$ | $b_1^{\mathcal{V}_1}$ | $b_0^{\mathcal{A}}$ | $b_1^{\mathcal{A}}$ | $b_0^{\mathcal{T}_1}$ | $b_1^{\mathcal{T}_1}$ | $b_1^{\mathcal{T}_2}$ |
| $b_1^{\mathcal{V}_0}$ | −13.637(760) | 1.000 | 0.505 | 0.382 | 0.513 | 0.317 | −0.524 | 0.328 | −0.524 | 0.328 | −0.529 | 0.327 | 0.327 |
| $b_0^{\mathcal{V}_{12}}$ | 0.224(35) | | 1.000 | 0.527 | 0.983 | 0.513 | −0.989 | 0.531 | −0.989 | 0.531 | −0.993 | 0.519 | 0.520 |
| $b_1^{\mathcal{V}_{12}}$ | −8.604(906) | | | 1.000 | 0.524 | 0.345 | −0.540 | 0.356 | −0.540 | 0.356 | −0.551 | 0.360 | 0.361 |
| $b_0^{\mathcal{T}_{23}}$ | 0.209(34) | | | | 1.000 | 0.491 | −0.987 | 0.532 | −0.987 | 0.532 | −0.984 | 0.509 | 0.510 |
| $b_1^{\mathcal{T}_{23}}$ | −8.875(915) | | | | | 1.000 | −0.520 | 0.335 | −0.520 | 0.334 | −0.522 | 0.326 | 0.326 |
| $b_0^{\mathcal{V}_1}$ | −0.429(75) | | | | | | 1.000 | −0.523 | 0.990 | −0.538 | 0.990 | −0.521 | −0.522 |
| $b_1^{\mathcal{V}_1}$ | −3.498(1065) | | | | | | | 1.000 | −0.538 | 0.345 | −0.540 | 0.338 | 0.338 |
| $b_0^{\mathcal{A}}$ | −0.430(75) | | | | | | | | 1.000 | −0.521 | 0.990 | −0.521 | −0.522 |
| $b_1^{\mathcal{A}}$ | −4.272(1056) | | | | | | | | | 1.000 | −0.540 | 0.337 | 0.338 |
| $b_0^{\mathcal{T}_1}$ | −0.445(75) | | | | | | | | | | 1.000 | −0.519 | −0.521 |
| $b_1^{\mathcal{T}_1}$ | −3.912(1022) | | | | | | | | | | | 1.000 | 0.351 |
| $b_1^{\mathcal{T}_2}$ | −3.153(1030) | | | | | | | | | | | | 1.000 |

TABLE XI. Fitted results of the BCL parameters for the $B \to f_1$ form factors.

| $B \to f_1$ form factors | | Correlation matrix | | | | | | | | | | | |
|---|---|---|---|---|---|---|---|---|---|---|---|---|---|
| $p$ | Values | $b_1^{\mathcal{V}_0}$ | $b_0^{\mathcal{V}_{12}}$ | $b_1^{\mathcal{V}_{12}}$ | $b_0^{\mathcal{T}_{23}}$ | $b_1^{\mathcal{T}_{23}}$ | $b_0^{\mathcal{V}_1}$ | $b_1^{\mathcal{V}_1}$ | $b_0^{\mathcal{A}}$ | $b_1^{\mathcal{A}}$ | $b_0^{\mathcal{T}_1}$ | $b_1^{\mathcal{T}_1}$ | $b_1^{\mathcal{T}_2}$ |
| $b_1^{\mathcal{V}_0}$ | −13.921(779) | 1.000 | 0.499 | 0.385 | 0.508 | 0.320 | −0.520 | 0.333 | −0.520 | 0.333 | −0.525 | 0.332 | 0.332 |
| $b_0^{\mathcal{V}_{12}}$ | 0.239(38) | | 1.000 | 0.525 | 0.983 | 0.509 | −0.989 | 0.531 | −0.989 | 0.531 | −0.993 | 0.520 | 0.520 |
| $b_1^{\mathcal{V}_{12}}$ | −8.790(933) | | | 1.000 | 0.522 | 0.349 | −0.539 | 0.362 | −0.539 | 0.362 | −0.550 | 0.367 | 0.368 |
| $b_0^{\mathcal{T}_{23}}$ | 0.223(37) | | | | 1.000 | 0.487 | −0.987 | 0.532 | −0.987 | 0.532 | −0.984 | 0.510 | 0.511 |
| $b_1^{\mathcal{T}_{23}}$ | −9.064(938) | | | | | 1.000 | −0.518 | 0.340 | −0.518 | 0.340 | −0.520 | 0.332 | 0.332 |
| $b_0^{\mathcal{V}_1}$ | −0.430(75) | | | | | | 1.000 | −0.525 | 0.990 | −0.540 | 0.990 | −0.524 | −0.524 |
| $b_1^{\mathcal{V}_1}$ | −3.577(1099) | | | | | | | 1.000 | −0.540 | 0.353 | −0.542 | 0.345 | 0.346 |
| $b_0^{\mathcal{A}}$ | −0.431(75) | | | | | | | | 1.000 | −0.523 | 0.990 | −0.523 | −0.524 |
| $b_1^{\mathcal{A}}$ | −4.368(1090) | | | | | | | | | 1.000 | −0.542 | 0.345 | 0.346 |
| $b_0^{\mathcal{T}_1}$ | −0.445(76) | | | | | | | | | | 1.000 | −0.521 | −0.523 |
| $b_1^{\mathcal{T}_1}$ | −4.001(1055) | | | | | | | | | | | 1.000 | 0.359 |
| $b_1^{\mathcal{T}_2}$ | −3.226(1063) | | | | | | | | | | | | 1.000 |

TABLE XII. Fitted results of the BCL parameters for the $B \to f_8$ form factors.

| $B \to f_8$ form factors | | Correlation matrix | | | | | | | | | | | |
|---|---|---|---|---|---|---|---|---|---|---|---|---|---|
| $p$ | Values | $b_1^{\mathcal{V}_0}$ | $b_0^{\mathcal{V}_{12}}$ | $b_1^{\mathcal{V}_{12}}$ | $b_0^{\mathcal{T}_{23}}$ | $b_1^{\mathcal{T}_{23}}$ | $b_0^{\mathcal{V}_1}$ | $b_1^{\mathcal{V}_1}$ | $b_0^{\mathcal{A}}$ | $b_1^{\mathcal{A}}$ | $b_0^{\mathcal{T}_1}$ | $b_1^{\mathcal{T}_1}$ | $b_1^{\mathcal{T}_2}$ |
| $b_1^{\mathcal{V}_0}$ | −14.295(810) | 1.000 | 0.496 | 0.379 | 0.507 | 0.313 | −0.517 | 0.327 | −0.517 | 0.327 | −0.521 | 0.326 | 0.327 |
| $b_0^{\mathcal{V}_{12}}$ | 0.275(44) | | 1.000 | 0.525 | 0.983 | 0.502 | −0.989 | 0.527 | −0.989 | 0.526 | −0.993 | 0.520 | 0.521 |
| $b_1^{\mathcal{V}_{12}}$ | −9.041(967) | | | 1.000 | 0.524 | 0.341 | −0.540 | 0.355 | −0.539 | 0.354 | −0.549 | 0.359 | 0.360 |
| $b_0^{\mathcal{T}_{23}}$ | 0.257(42) | | | | 1.000 | 0.482 | −0.987 | 0.530 | −0.987 | 0.530 | −0.984 | 0.512 | 0.513 |
| $b_1^{\mathcal{T}_{23}}$ | −9.307(966) | | | | | 1.000 | −0.512 | 0.333 | −0.512 | 0.333 | −0.512 | 0.325 | 0.326 |
| $b_0^{\mathcal{V}_1}$ | −0.460(81) | | | | | | 1.000 | −0.521 | 0.990 | −0.536 | 0.990 | −0.524 | −0.525 |
| $b_1^{\mathcal{V}_1}$ | −3.679(1137) | | | | | | | 1.000 | −0.536 | 0.346 | −0.537 | 0.339 | 0.340 |
| $b_0^{\mathcal{A}}$ | −0.461(81) | | | | | | | | 1.000 | −0.519 | 0.990 | −0.524 | −0.525 |
| $b_1^{\mathcal{A}}$ | −4.491(1127) | | | | | | | | | 1.000 | −0.536 | 0.339 | 0.340 |
| $b_0^{\mathcal{T}_1}$ | −0.476(82) | | | | | | | | | | 1.000 | −0.521 | −0.522 |
| $b_1^{\mathcal{T}_1}$ | −4.123(1096) | | | | | | | | | | | 1.000 | 0.354 |
| $b_1^{\mathcal{T}_2}$ | −3.327(1105) | | | | | | | | | | | | 1.000 |

TABLE XIII. Fitted results of the BCL parameters for the $B \to K_{1A}$ form factors.

| $B \to K_{1A}$ form factors | | Correlation matrix | | | | | | | | | | | |
|---|---|---|---|---|---|---|---|---|---|---|---|---|---|
| $p$ | Values | $b_1^{\mathcal{V}_0}$ | $b_0^{\mathcal{V}_{12}}$ | $b_1^{\mathcal{V}_{12}}$ | $b_0^{\mathcal{T}_{23}}$ | $b_1^{\mathcal{T}_{23}}$ | $b_0^{\mathcal{V}_1}$ | $b_1^{\mathcal{V}_1}$ | $b_0^{\mathcal{A}}$ | $b_1^{\mathcal{A}}$ | $b_0^{\mathcal{T}_1}$ | $b_1^{\mathcal{T}_1}$ | $b_1^{\mathcal{T}_2}$ |
| $b_1^{\mathcal{V}_0}$ | −14.083(816) | 1.000 | 0.475 | 0.368 | 0.490 | 0.302 | −0.499 | 0.312 | −0.499 | 0.311 | −0.502 | 0.310 | 0.311 |
| $b_0^{\mathcal{V}_{12}}$ | 0.248(41) | | 1.000 | 0.504 | 0.983 | 0.479 | −0.989 | 0.497 | −0.989 | 0.496 | −0.993 | 0.493 | 0.494 |
| $b_1^{\mathcal{V}_{12}}$ | −8.896(974) | | | 1.000 | 0.508 | 0.330 | −0.521 | 0.340 | −0.521 | 0.340 | −0.529 | 0.345 | 0.345 |
| $b_0^{\mathcal{T}_{23}}$ | 0.232(39) | | | | 1.000 | 0.462 | −0.987 | 0.504 | −0.987 | 0.503 | −0.985 | 0.489 | 0.490 |
| $b_1^{\mathcal{T}_{23}}$ | −9.154(970) | | | | | 1.000 | −0.491 | 0.320 | −0.490 | 0.320 | −0.490 | 0.312 | 0.313 |
| $b_0^{\mathcal{V}_1}$ | −0.430(78) | | | | | | 1.000 | −0.492 | 0.990 | −0.508 | 0.990 | −0.499 | −0.500 |
| $b_1^{\mathcal{V}_1}$ | −3.591(1137) | | | | | | | 1.000 | −0.508 | 0.330 | −0.509 | 0.323 | 0.323 |
| $b_0^{\mathcal{A}}$ | −0.430(78) | | | | | | | | 1.000 | −0.490 | 0.990 | −0.499 | −0.500 |
| $b_1^{\mathcal{A}}$ | −4.389(1127) | | | | | | | | | 1.000 | −0.508 | 0.322 | 0.323 |
| $b_0^{\mathcal{T}_1}$ | −0.444(79) | | | | | | | | | | 1.000 | −0.494 | −0.496 |
| $b_1^{\mathcal{T}_1}$ | −4.030(1096) | | | | | | | | | | | 1.000 | 0.338 |
| $b_1^{\mathcal{T}_2}$ | −3.249(1105) | | | | | | | | | | | | 1.000 |

TABLE XIV. Fitted results of the BCL parameters for the $B_s \to f_1$ form factors.

| $B_s \to f_1$ form factors | | Correlation matrix | | | | | | | | | | | |
|---|---|---|---|---|---|---|---|---|---|---|---|---|---|
| $p$ | Values | $b_1^{\mathcal{V}_0}$ | $b_0^{\mathcal{V}_{12}}$ | $b_1^{\mathcal{V}_{12}}$ | $b_0^{\mathcal{T}_{23}}$ | $b_1^{\mathcal{T}_{23}}$ | $b_0^{\mathcal{V}_1}$ | $b_1^{\mathcal{V}_1}$ | $b_0^{\mathcal{A}}$ | $b_1^{\mathcal{A}}$ | $b_0^{\mathcal{T}_1}$ | $b_1^{\mathcal{T}_1}$ | $b_1^{\mathcal{T}_2}$ |
| $b_1^{\mathcal{V}_0}$ | −14.004(768) | 1.000 | 0.403 | 0.291 | 0.416 | 0.224 | −0.426 | 0.240 | −0.426 | 0.239 | −0.430 | 0.237 | 0.237 |
| $b_0^{\mathcal{V}_{12}}$ | 0.264(42) | | 1.000 | 0.399 | 0.983 | 0.400 | −0.989 | 0.431 | −0.989 | 0.430 | −0.993 | 0.416 | 0.417 |
| $b_1^{\mathcal{V}_{12}}$ | −8.879(881) | | | 1.000 | 0.403 | 0.232 | −0.416 | 0.247 | −0.416 | 0.246 | −0.425 | 0.249 | 0.249 |
| $b_0^{\mathcal{T}_{23}}$ | 0.245(41) | | | | 1.000 | 0.377 | −0.987 | 0.437 | −0.987 | 0.436 | −0.984 | 0.410 | 0.411 |
| $b_1^{\mathcal{T}_{23}}$ | −9.157(914) | | | | | 1.000 | −0.409 | 0.240 | −0.409 | 0.240 | −0.409 | 0.229 | 0.230 |
| $b_0^{\mathcal{V}_1}$ | −0.479(85) | | | | | | 1.000 | −0.422 | 0.990 | −0.440 | 0.990 | −0.420 | −0.421 |
| $b_1^{\mathcal{V}_1}$ | −3.740(1070) | | | | | | | 1.000 | −0.441 | 0.256 | −0.441 | 0.245 | 0.246 |
| $b_0^{\mathcal{A}}$ | −0.480(85) | | | | | | | | 1.000 | −0.420 | 0.990 | −0.420 | −0.421 |
| $b_1^{\mathcal{A}}$ | −4.527(1061) | | | | | | | | | 1.000 | −0.440 | 0.245 | 0.245 |
| $b_0^{\mathcal{T}_1}$ | −0.497(86) | | | | | | | | | | 1.000 | −0.414 | −0.415 |
| $b_1^{\mathcal{T}_1}$ | −4.142(1018) | | | | | | | | | | | 1.000 | 0.264 |
| $b_1^{\mathcal{T}_2}$ | −3.372(1026) | | | | | | | | | | | | 1.000 |

TABLE XV. Fitted results of the BCL parameters for the $B_s \to f_8$ form factors.

| $B_s \to f_8$ form factors | | Correlation matrix | | | | | | | | | | | |
|---|---|---|---|---|---|---|---|---|---|---|---|---|---|
| $p$ | Values | $b_1^{\mathcal{V}_0}$ | $b_0^{\mathcal{V}_{12}}$ | $b_1^{\mathcal{V}_{12}}$ | $b_0^{\mathcal{T}_{23}}$ | $b_1^{\mathcal{T}_{23}}$ | $b_0^{\mathcal{V}_1}$ | $b_1^{\mathcal{V}_1}$ | $b_0^{\mathcal{A}}$ | $b_1^{\mathcal{A}}$ | $b_0^{\mathcal{T}_1}$ | $b_1^{\mathcal{T}_1}$ | $b_1^{\mathcal{T}_2}$ |
| $b_1^{\mathcal{V}_0}$ | −14.377(757) | 1.000 | 0.411 | 0.304 | 0.426 | 0.237 | −0.436 | 0.249 | −0.436 | 0.249 | −0.439 | 0.247 | 0.248 |
| $b_0^{\mathcal{V}_{12}}$ | 0.304(47) | | 1.000 | 0.407 | 0.982 | 0.400 | −0.988 | 0.426 | −0.988 | 0.425 | −0.993 | 0.417 | 0.418 |
| $b_1^{\mathcal{V}_{12}}$ | −9.119(867) | | | 1.000 | 0.411 | 0.244 | −0.425 | 0.254 | −0.424 | 0.254 | −0.433 | 0.258 | 0.258 |
| $b_0^{\mathcal{T}_{23}}$ | 0.284(45) | | | | 1.000 | 0.381 | −0.987 | 0.434 | −0.987 | 0.433 | −0.983 | 0.412 | 0.413 |
| $b_1^{\mathcal{T}_{23}}$ | −9.393(891) | | | | | 1.000 | −0.412 | 0.249 | −0.412 | 0.248 | −0.411 | 0.238 | 0.238 |
| $b_0^{\mathcal{V}_1}$ | −0.514(86) | | | | | | 1.000 | −0.419 | 0.990 | −0.437 | 0.990 | −0.422 | −0.423 |
| $b_1^{\mathcal{V}_1}$ | −3.826(1042) | | | | | | | 1.000 | −0.438 | 0.260 | −0.436 | 0.249 | 0.250 |
| $b_0^{\mathcal{A}}$ | −0.514(87) | | | | | | | | 1.000 | −0.417 | 0.990 | −0.422 | −0.423 |
| $b_1^{\mathcal{A}}$ | −4.634(1032) | | | | | | | | | 1.000 | −0.435 | 0.249 | 0.249 |
| $b_0^{\mathcal{T}_1}$ | −0.532(87) | | | | | | | | | | 1.000 | −0.415 | −0.416 |
| $b_1^{\mathcal{T}_1}$ | −4.248(994) | | | | | | | | | | | 1.000 | 0.269 |
| $b_1^{\mathcal{T}_2}$ | −3.458(1003) | | | | | | | | | | | | 1.000 |

TABLE XVI. Fitted results of the BCL parameters for the $B_s \to K_{1A}$ form factors.

| $B_s \to K_{1A}$ form factors | | Correlation matrix | | | | | | | | | | | |
|---|---|---|---|---|---|---|---|---|---|---|---|---|---|
| $p$ | Values | $b_1^{\mathcal{V}_0}$ | $b_0^{\mathcal{V}_{12}}$ | $b_1^{\mathcal{V}_{12}}$ | $b_0^{\mathcal{T}_{23}}$ | $b_1^{\mathcal{T}_{23}}$ | $b_0^{\mathcal{V}_1}$ | $b_1^{\mathcal{V}_1}$ | $b_0^{\mathcal{A}}$ | $b_1^{\mathcal{A}}$ | $b_0^{\mathcal{T}_1}$ | $b_1^{\mathcal{T}_1}$ | $b_1^{\mathcal{T}_2}$ |
| $b_1^{\mathcal{V}_0}$ | −14.173(815) | 1.000 | 0.388 | 0.283 | 0.406 | 0.215 | −0.413 | 0.228 | −0.413 | 0.228 | −0.415 | 0.226 | 0.227 |
| $b_0^{\mathcal{V}_{12}}$ | 0.273(46) | | 1.000 | 0.388 | 0.983 | 0.377 | −0.988 | 0.405 | −0.988 | 0.405 | −0.993 | 0.400 | 0.401 |
| $b_1^{\mathcal{V}_{12}}$ | −8.992(935) | | | 1.000 | 0.397 | 0.224 | −0.407 | 0.235 | −0.407 | 0.235 | −0.413 | 0.238 | 0.238 |
| $b_0^{\mathcal{T}_{23}}$ | 0.255(44) | | | | 1.000 | 0.359 | −0.987 | 0.416 | −0.987 | 0.416 | −0.985 | 0.399 | 0.400 |
| $b_1^{\mathcal{T}_{23}}$ | −9.254(957) | | | | | 1.000 | −0.390 | 0.230 | −0.389 | 0.230 | −0.388 | 0.220 | 0.220 |
| $b_0^{\mathcal{V}_1}$ | −0.479(89) | | | | | | 1.000 | −0.399 | 0.990 | −0.418 | 0.990 | −0.407 | −0.407 |
| $b_1^{\mathcal{V}_1}$ | −3.762(1123) | | | | | | | 1.000 | −0.418 | 0.243 | −0.416 | 0.233 | 0.233 |
| $b_0^{\mathcal{A}}$ | −0.480(89) | | | | | | | | 1.000 | −0.397 | 0.990 | −0.406 | −0.407 |
| $b_1^{\mathcal{A}}$ | −4.563(1113) | | | | | | | | | 1.000 | −0.416 | 0.233 | 0.233 |
| $b_0^{\mathcal{T}_1}$ | −0.496(90) | | | | | | | | | | 1.000 | −0.398 | −0.399 |
| $b_1^{\mathcal{T}_1}$ | −4.186(1075) | | | | | | | | | | | 1.000 | 0.253 |
| $b_1^{\mathcal{T}_2}$ | −3.404(1084) | | | | | | | | | | | | 1.000 |

TABLE XVII. Fitted results of the BCL parameters for the $B \to b_1$ form factors.

| $B \to b_1$ form factors | | Correlation matrix | | | | | | | | | | | |
|---|---|---|---|---|---|---|---|---|---|---|---|---|---|
| $p$ | Values | $b_1^{\mathcal{V}_0}$ | $b_0^{\mathcal{V}_{12}}$ | $b_1^{\mathcal{V}_{12}}$ | $b_0^{\mathcal{T}_{23}}$ | $b_1^{\mathcal{T}_{23}}$ | $b_0^{\mathcal{V}_1}$ | $b_1^{\mathcal{V}_1}$ | $b_0^{\mathcal{A}}$ | $b_1^{\mathcal{A}}$ | $b_0^{\mathcal{T}_1}$ | $b_1^{\mathcal{T}_1}$ | $b_1^{\mathcal{T}_2}$ |
| $b_1^{\mathcal{V}_0}$ | −13.614(713) | 1.000 | 0.490 | −0.105 | 0.504 | 0.333 | −0.515 | 0.342 | −0.515 | 0.342 | −0.518 | 0.342 | 0.342 |
| $b_0^{\mathcal{V}_{12}}$ | 0.297(43) | | 1.000 | −0.509 | 0.981 | 0.518 | −0.988 | 0.536 | −0.988 | 0.536 | −0.993 | 0.533 | 0.534 |
| $b_1^{\mathcal{V}_{12}}$ | −7.845(761) | | | 1.000 | −0.485 | −0.170 | 0.476 | −0.180 | 0.477 | −0.180 | 0.471 | −0.171 | −0.171 |
| $b_0^{\mathcal{T}_{23}}$ | 0.278(42) | | | | 1.000 | 0.504 | −0.986 | 0.544 | −0.986 | 0.543 | −0.983 | 0.527 | 0.528 |
| $b_1^{\mathcal{T}_{23}}$ | −8.859(859) | | | | | 1.000 | −0.532 | 0.362 | −0.532 | 0.362 | −0.531 | 0.353 | 0.354 |
| $b_0^{\mathcal{V}_1}$ | −0.571(92) | | | | | | 1.000 | −0.534 | 0.990 | −0.549 | 0.990 | −0.539 | −0.540 |
| $b_1^{\mathcal{V}_1}$ | −3.480(1009) | | | | | | | 1.000 | −0.549 | 0.371 | −0.549 | 0.363 | 0.364 |
| $b_0^{\mathcal{A}}$ | −0.572(92) | | | | | | | | 1.000 | −0.533 | 0.990 | −0.539 | −0.540 |
| $b_1^{\mathcal{A}}$ | −4.255(1001) | | | | | | | | | 1.000 | −0.548 | 0.363 | 0.364 |
| $b_0^{\mathcal{T}_1}$ | −0.590(93) | | | | | | | | | | 1.000 | −0.536 | −0.537 |
| $b_1^{\mathcal{T}_1}$ | −3.909(974) | | | | | | | | | | | 1.000 | 0.377 |
| $b_1^{\mathcal{T}_2}$ | −3.149(981) | | | | | | | | | | | | 1.000 |

TABLE XVIII. Fitted results of the BCL parameters for the $B \to h_1$ form factors.

| $B \to h_1$ form factors | | Correlation matrix | | | | | | | | | | | |
|---|---|---|---|---|---|---|---|---|---|---|---|---|---|
| $p$ | Values | $b_1^{\mathcal{V}_0}$ | $b_0^{\mathcal{V}_{12}}$ | $b_1^{\mathcal{V}_{12}}$ | $b_0^{\mathcal{T}_{23}}$ | $b_1^{\mathcal{T}_{23}}$ | $b_0^{\mathcal{V}_1}$ | $b_1^{\mathcal{V}_1}$ | $b_0^{\mathcal{A}}$ | $b_1^{\mathcal{A}}$ | $b_0^{\mathcal{T}_1}$ | $b_1^{\mathcal{T}_1}$ | $b_1^{\mathcal{T}_2}$ |
| $b_1^{\mathcal{V}_0}$ | −13.581(780) | 1.000 | 0.487 | 0.377 | 0.501 | 0.312 | −0.511 | 0.324 | −0.511 | 0.324 | −0.513 | 0.323 | 0.323 |
| $b_0^{\mathcal{V}_{12}}$ | 0.292(47) | | 1.000 | 0.525 | 0.983 | 0.496 | −0.989 | 0.518 | −0.989 | 0.517 | −0.993 | 0.514 | 0.515 |
| $b_1^{\mathcal{V}_{12}}$ | −8.581(937) | | | 1.000 | 0.527 | 0.343 | −0.542 | 0.356 | −0.542 | 0.355 | −0.550 | 0.360 | 0.361 |
| $b_0^{\mathcal{T}_{23}}$ | 0.273(46) | | | | 1.000 | 0.479 | −0.987 | 0.524 | −0.987 | 0.523 | −0.984 | 0.509 | 0.510 |
| $b_1^{\mathcal{T}_{23}}$ | −8.829(930) | | | | | 1.000 | −0.508 | 0.334 | −0.508 | 0.334 | −0.508 | 0.326 | 0.326 |
| $b_0^{\mathcal{V}_1}$ | −0.565(101) | | | | | | 1.000 | −0.514 | 0.990 | −0.529 | 0.990 | −0.520 | −0.521 |
| $b_1^{\mathcal{V}_1}$ | −3.469(1092) | | | | | | | 1.000 | −0.530 | 0.346 | −0.529 | 0.338 | 0.339 |
| $b_0^{\mathcal{A}}$ | −0.566(101) | | | | | | | | 1.000 | −0.513 | 0.990 | −0.520 | −0.521 |
| $b_1^{\mathcal{A}}$ | −4.241(1083) | | | | | | | | | 1.000 | −0.529 | 0.338 | 0.339 |
| $b_0^{\mathcal{T}_1}$ | −0.584(102) | | | | | | | | | | 1.000 | −0.516 | −0.517 |
| $b_1^{\mathcal{T}_1}$ | −3.896(1055) | | | | | | | | | | | 1.000 | 0.352 |
| $b_1^{\mathcal{T}_2}$ | −3.139(1064) | | | | | | | | | | | | 1.000 |

TABLE XIX. Fitted results of the BCL parameters for the $B \to h_8$ form factors.

| $B \to h_8$ form factors | | Correlation matrix | | | | | | | | | | | |
|---|---|---|---|---|---|---|---|---|---|---|---|---|---|
| $p$ | Values | $b_1^{\mathcal{V}_0}$ | $b_0^{\mathcal{V}_{12}}$ | $b_1^{\mathcal{V}_{12}}$ | $b_0^{\mathcal{T}_{23}}$ | $b_1^{\mathcal{T}_{23}}$ | $b_0^{\mathcal{V}_1}$ | $b_1^{\mathcal{V}_1}$ | $b_0^{\mathcal{A}}$ | $b_1^{\mathcal{A}}$ | $b_0^{\mathcal{T}_1}$ | $b_1^{\mathcal{T}_1}$ | $b_1^{\mathcal{T}_2}$ |
| $b_1^{\mathcal{V}_0}$ | −14.078(779) | 1.000 | 0.478 | 0.380 | 0.492 | 0.315 | −0.503 | 0.328 | −0.503 | 0.328 | −0.507 | 0.327 | 0.328 |
| $b_0^{\mathcal{V}_{12}}$ | 0.327(51) | | 1.000 | 0.488 | 0.983 | 0.474 | −0.989 | 0.500 | −0.989 | 0.499 | −0.993 | 0.493 | 0.494 |
| $b_1^{\mathcal{V}_{12}}$ | −8.865(925) | | | 1.000 | 0.491 | 0.340 | −0.508 | 0.353 | −0.508 | 0.353 | −0.517 | 0.358 | 0.359 |
| $b_0^{\mathcal{T}_{23}}$ | 0.305(49) | | | | 1.000 | 0.457 | −0.987 | 0.506 | −0.987 | 0.506 | −0.983 | 0.488 | 0.489 |
| $b_1^{\mathcal{T}_{23}}$ | −9.138(927) | | | | | 1.000 | −0.488 | 0.334 | −0.488 | 0.334 | −0.488 | 0.326 | 0.326 |
| $b_0^{\mathcal{V}_1}$ | −0.566(97) | | | | | | 1.000 | −0.497 | 0.990 | −0.513 | 0.990 | −0.501 | −0.502 |
| $b_1^{\mathcal{V}_1}$ | −3.580(1091) | | | | | | | 1.000 | −0.513 | 0.347 | −0.513 | 0.340 | 0.340 |
| $b_0^{\mathcal{A}}$ | −0.567(97) | | | | | | | | 1.000 | −0.496 | 0.990 | −0.501 | −0.502 |
| $b_1^{\mathcal{A}}$ | −4.381(1082) | | | | | | | | | 1.000 | −0.513 | 0.340 | 0.340 |
| $b_0^{\mathcal{T}_1}$ | −0.586(98) | | | | | | | | | | 1.000 | −0.497 | −0.499 |
| $b_1^{\mathcal{T}_1}$ | −4.018(1047) | | | | | | | | | | | 1.000 | 0.354 |
| $b_1^{\mathcal{T}_2}$ | −3.233(1056) | | | | | | | | | | | | 1.000 |

TABLE XX. Fitted results of the BCL parameters for the $B \to K_{1B}$ form factors.

| $B \to K_{1B}$ form factors | | Correlation matrix | | | | | | | | | | | |
|---|---|---|---|---|---|---|---|---|---|---|---|---|---|
| $p$ | Values | $b_1^{\mathcal{V}_0}$ | $b_0^{\mathcal{V}_{12}}$ | $b_1^{\mathcal{V}_{12}}$ | $b_0^{\mathcal{T}_{23}}$ | $b_1^{\mathcal{T}_{23}}$ | $b_0^{\mathcal{V}_1}$ | $b_1^{\mathcal{V}_1}$ | $b_0^{\mathcal{A}}$ | $b_1^{\mathcal{A}}$ | $b_0^{\mathcal{T}_1}$ | $b_1^{\mathcal{T}_1}$ | $b_1^{\mathcal{T}_2}$ |
| $b_1^{\mathcal{V}_0}$ | −13.878(772) | 1.000 | 0.508 | 0.411 | 0.518 | 0.348 | −0.531 | 0.358 | −0.531 | 0.357 | −0.536 | 0.357 | 0.357 |
| $b_0^{\mathcal{V}_{12}}$ | 0.306(44) | | 1.000 | 0.525 | 0.981 | 0.508 | −0.988 | 0.527 | −0.988 | 0.526 | −0.993 | 0.518 | 0.519 |
| $b_1^{\mathcal{V}_{12}}$ | −8.739(864) | | | 1.000 | 0.522 | 0.375 | −0.541 | 0.385 | −0.541 | 0.385 | −0.552 | 0.391 | 0.392 |
| $b_0^{\mathcal{T}_{23}}$ | 0.286(43) | | | | 1.000 | 0.492 | −0.986 | 0.532 | −0.986 | 0.531 | −0.982 | 0.510 | 0.511 |
| $b_1^{\mathcal{T}_{23}}$ | −9.014(867) | | | | | 1.000 | −0.522 | 0.365 | −0.521 | −0.521 | −0.523 | 0.356 | 0.356 |
| $b_0^{\mathcal{V}_1}$ | −0.555(88) | | | | | | 1.000 | −0.525 | 0.990 | −0.538 | 0.990 | −0.524 | −0.525 |
| $b_1^{\mathcal{V}_1}$ | −3.534(1016) | | | | | | | 1.000 | −0.539 | 0.375 | −0.541 | 0.367 | 0.368 |
| $b_0^{\mathcal{A}}$ | −0.556(88) | | | | | | | | 1.000 | −0.523 | 0.990 | −0.524 | −0.525 |
| $b_1^{\mathcal{A}}$ | −4.320(1007) | | | | | | | | | 1.000 | −0.540 | 0.366 | 0.367 |
| $b_0^{\mathcal{T}_1}$ | −0.574(89) | | | | | | | | | | 1.000 | −0.523 | −0.524 |
| $b_1^{\mathcal{T}_1}$ | −3.957(972) | | | | | | | | | | | 1.000 | 0.380 |
| $b_1^{\mathcal{T}_2}$ | −3.186(980) | | | | | | | | | | | | 1.000 |

TABLE XXI. Fitted results of the BCL parameters for the $B_s \to h_1$ form factors.

| $B_s \to h_1$ form factors | | Correlation matrix | | | | | | | | | | | |
|---|---|---|---|---|---|---|---|---|---|---|---|---|---|
| $p$ | Values | $b_1^{\mathcal{V}_0}$ | $b_0^{\mathcal{V}_{12}}$ | $b_1^{\mathcal{V}_{12}}$ | $b_0^{\mathcal{T}_{23}}$ | $b_1^{\mathcal{T}_{23}}$ | $b_0^{\mathcal{V}_1}$ | $b_1^{\mathcal{V}_1}$ | $b_0^{\mathcal{A}}$ | $b_1^{\mathcal{A}}$ | $b_0^{\mathcal{T}_1}$ | $b_1^{\mathcal{T}_1}$ | $b_1^{\mathcal{T}_2}$ |
| $b_1^{\mathcal{V}_0}$ | −13.670(724) | 1.000 | 0.401 | 0.303 | 0.418 | 0.237 | −0.428 | 0.252 | −0.428 | 0.251 | −0.430 | 0.249 | 0.249 |
| $b_0^{\mathcal{V}_{12}}$ | 0.323(50) | | 1.000 | 0.405 | 0.982 | 0.392 | −0.988 | 0.419 | −0.988 | 0.418 | −0.993 | 0.413 | 0.413 |
| $b_1^{\mathcal{V}_{12}}$ | −8.662(833) | | | 1.000 | 0.411 | 0.246 | −0.425 | 0.260 | −0.425 | 0.259 | −0.432 | 0.262 | 0.263 |
| $b_0^{\mathcal{T}_{23}}$ | 0.301(48) | | | | 1.000 | 0.376 | −0.987 | 0.430 | −0.987 | 0.429 | −0.984 | 0.411 | 0.412 |
| $b_1^{\mathcal{T}_{23}}$ | −8.919(853) | | | | | 1.000 | −0.408 | 0.254 | −0.408 | 0.254 | −0.405 | 0.242 | 0.243 |
| $b_0^{\mathcal{V}_1}$ | −0.631(107) | | | | | | 1.000 | −0.416 | 0.990 | −0.434 | 0.990 | −0.421 | −0.422 |
| $b_1^{\mathcal{V}_1}$ | −3.617(1001) | | | | | | | 1.000 | −0.435 | 0.269 | −0.432 | 0.257 | 0.258 |
| $b_0^{\mathcal{A}}$ | −0.632(107) | | | | | | | | 1.000 | −0.414 | 0.990 | −0.421 | −0.422 |
| $b_1^{\mathcal{A}}$ | −4.386(992) | | | | | | | | | 1.000 | −0.431 | 0.257 | 0.257 |
| $b_0^{\mathcal{T}_1}$ | −0.653(108) | | | | | | | | | | 1.000 | −0.413 | −0.414 |
| $b_1^{\mathcal{T}_1}$ | −4.023(958) | | | | | | | | | | | 1.000 | 0.275 |
| $b_1^{\mathcal{T}_2}$ | −3.271(966) | | | | | | | | | | | | 1.000 |

TABLE XXII. Fitted results of the BCL parameters for the $B_s \to h_8$ form factors.

| $B_s \to h_8$ form factors | | Correlation matrix | | | | | | | | | | | |
|---|---|---|---|---|---|---|---|---|---|---|---|---|---|
| $p$ | Values | $b_1^{\mathcal{V}_0}$ | $b_0^{\mathcal{V}_{12}}$ | $b_1^{\mathcal{V}_{12}}$ | $b_0^{\mathcal{T}_{23}}$ | $b_1^{\mathcal{T}_{23}}$ | $b_0^{\mathcal{V}_1}$ | $b_1^{\mathcal{V}_1}$ | $b_0^{\mathcal{A}}$ | $b_1^{\mathcal{A}}$ | $b_0^{\mathcal{T}_1}$ | $b_1^{\mathcal{T}_1}$ | $b_1^{\mathcal{T}_2}$ |
| $b_1^{\mathcal{V}_0}$ | −14.198(789) | 1.000 | 0.410 | 0.288 | 0.425 | 0.221 | −0.433 | 0.236 | −0.432 | 0.235 | −0.436 | 0.233 | 0.234 |
| $b_0^{\mathcal{V}_{12}}$ | 0.360(59) | | 1.000 | 0.400 | 0.983 | 0.395 | −0.989 | 0.427 | −0.989 | 0.426 | −0.993 | 0.420 | 0.420 |
| $b_1^{\mathcal{V}_{12}}$ | −9.003(902) | | | 1.000 | 0.406 | 0.227 | −0.416 | 0.240 | −0.416 | 0.239 | −0.424 | 0.243 | 0.243 |
| $b_0^{\mathcal{T}_{23}}$ | 0.336(57) | | | | 1.000 | 0.375 | −0.987 | 0.436 | −0.987 | 0.435 | −0.984 | 0.416 | 0.417 |
| $b_1^{\mathcal{T}_{23}}$ | −9.273(926) | | | | | 1.000 | −0.405 | 0.235 | −0.404 | 0.235 | −0.404 | 0.225 | 0.225 |
| $b_0^{\mathcal{V}_1}$ | −0.631(113) | | | | | | 1.000 | −0.418 | 0.990 | −0.436 | 0.990 | −0.423 | −0.424 |
| $b_1^{\mathcal{V}_1}$ | −3.789(1088) | | | | | | | 1.000 | −0.437 | 0.249 | −0.436 | 0.239 | 0.239 |
| $b_0^{\mathcal{A}}$ | −0.632(113) | | | | | | | | 1.000 | −0.416 | 0.990 | −0.423 | −0.424 |
| $b_1^{\mathcal{A}}$ | −4.586(1079) | | | | | | | | | 1.000 | −0.435 | 0.238 | 0.239 |
| $b_0^{\mathcal{T}_1}$ | −0.654(115) | | | | | | | | | | 1.000 | −0.416 | −0.417 |
| $b_1^{\mathcal{T}_1}$ | −4.206(1039) | | | | | | | | | | | 1.000 | 0.258 |
| $b_1^{\mathcal{T}_2}$ | −3.427(1048) | | | | | | | | | | | | 1.000 |

TABLE XXIII. Fitted results of the BCL parameters for the $B_s \to K_{1B}$ form factors.

| $B_s \to K_{1B}$ form factors | | Correlation matrix | | | | | | | | | | | |
|---|---|---|---|---|---|---|---|---|---|---|---|---|---|
| $p$ | Values | $b_1^{\mathcal{V}_0}$ | $b_0^{\mathcal{V}_{12}}$ | $b_1^{\mathcal{V}_{12}}$ | $b_0^{\mathcal{T}_{23}}$ | $b_1^{\mathcal{T}_{23}}$ | $b_0^{\mathcal{V}_1}$ | $b_1^{\mathcal{V}_1}$ | $b_0^{\mathcal{A}}$ | $b_1^{\mathcal{A}}$ | $b_0^{\mathcal{T}_1}$ | $b_1^{\mathcal{T}_1}$ | $b_1^{\mathcal{T}_2}$ |
| $b_1^{\mathcal{V}_0}$ | −13.975(756) | 1.000 | 0.415 | 0.301 | 0.429 | 0.235 | −0.438 | 0.250 | −0.438 | 0.250 | −0.442 | 0.247 | 0.248 |
| $b_0^{\mathcal{V}_{12}}$ | 0.336(53) | | 1.000 | 0.408 | 0.982 | 0.406 | −0.988 | 0.437 | −0.988 | 0.437 | −0.993 | 0.426 | 0.427 |
| $b_1^{\mathcal{V}_{12}}$ | −8.857(867) | | | 1.000 | 0.413 | 0.243 | −0.425 | 0.256 | −0.425 | 0.256 | −0.433 | 0.259 | 0.259 |
| $b_0^{\mathcal{T}_{23}}$ | 0.314(51) | | | | 1.000 | 0.386 | −0.987 | 0.445 | −0.987 | 0.445 | −0.984 | 0.421 | 0.422 |
| $b_1^{\mathcal{T}_{23}}$ | −9.131(897) | | | | | 1.000 | −0.416 | 0.251 | −0.416 | 0.251 | −0.224 | 0.240 | 0.240 |
| $b_0^{\mathcal{V}_1}$ | −0.616(108) | | | | | | 1.000 | −0.429 | 0.990 | −0.247 | 0.990 | −0.430 | −0.431 |
| $b_1^{\mathcal{V}_1}$ | −3.727(1052) | | | | | | | 1.000 | −0.448 | 0.265 | −0.447 | 0.254 | 0.254 |
| $b_0^{\mathcal{A}}$ | −0.617(108) | | | | | | | | 1.000 | −0.428 | 0.990 | −0.430 | −0.431 |
| $b_1^{\mathcal{A}}$ | −4.516(1043) | | | | | | | | | 1.000 | −0.446 | 0.254 | 0.254 |
| $b_0^{\mathcal{T}_1}$ | −0.639(109) | | | | | | | | | | 1.000 | −0.423 | −0.425 |
| $b_1^{\mathcal{T}_1}$ | −4.135(1001) | | | | | | | | | | | 1.000 | 0.272 |
| $b_1^{\mathcal{T}_2}$ | −3.365(1009) | | | | | | | | | | | | 1.000 |

## APPENDIX C: $q^2$ DEPENDENCE OF THE SEMILEPTONIC $B \to A$ FORM FACTORS

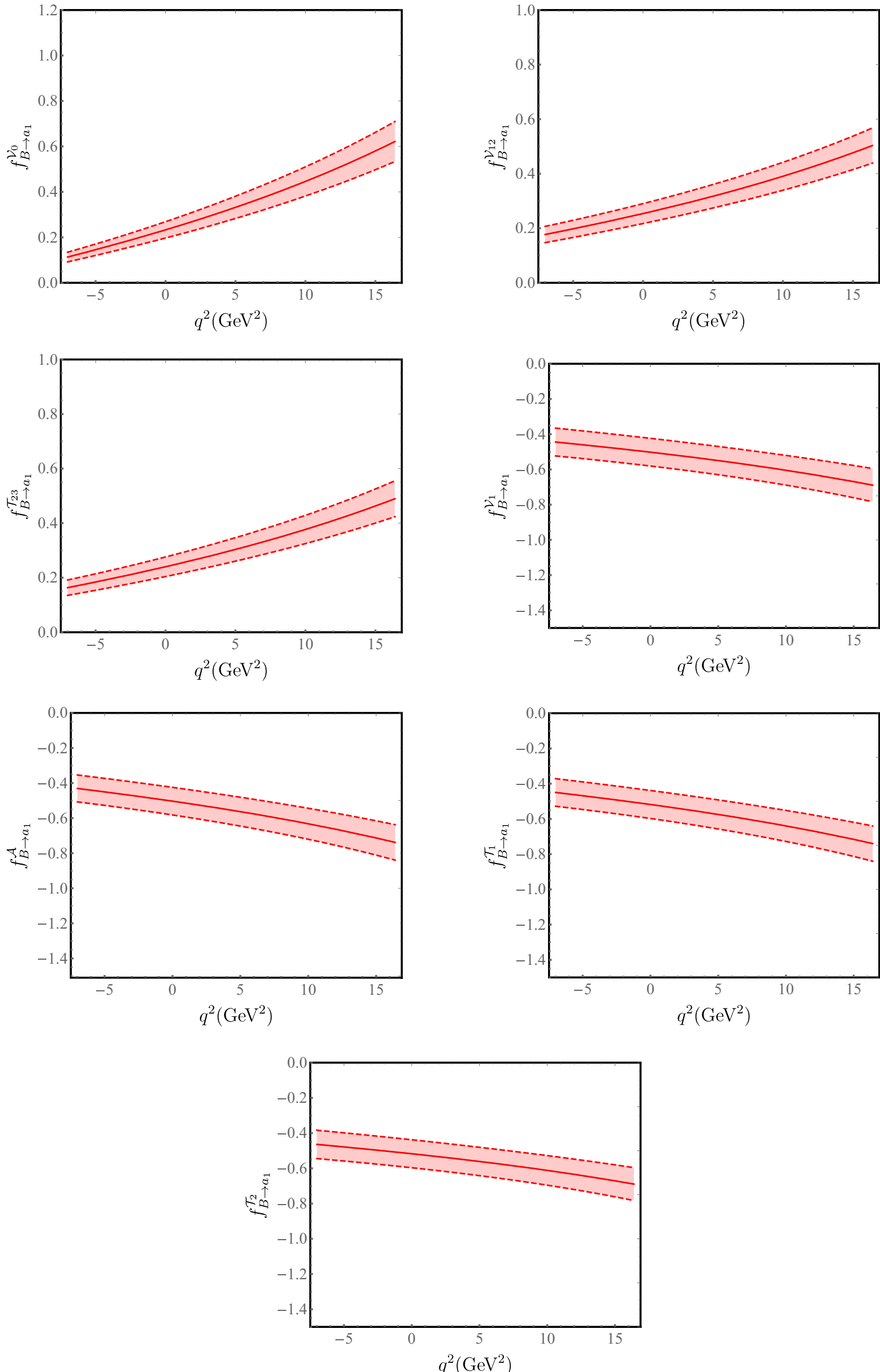

FIG. 10. The pink bands display the $q^2$ dependence of the $B \to a_1$ form factors employing the BCL parametrization, where the errors are computed using the Monte Carlo method. The calligraphic form factors represent the linear combinations of the defined form factors shown in Sec. IV.

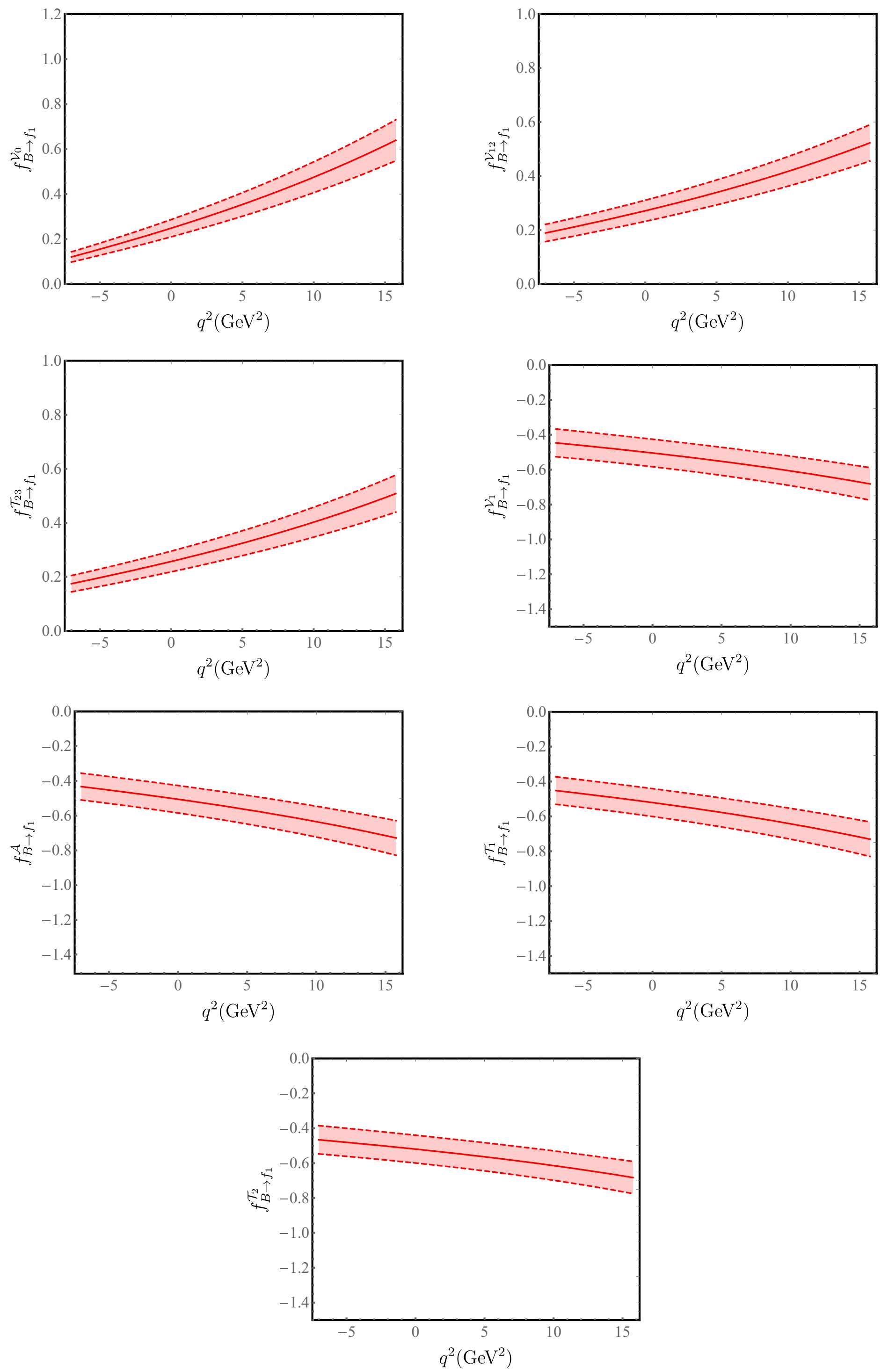

FIG. 11. The pink bands display the $q^2$ dependence of the $B \to f_1$ form factors employing the BCL parametrization, where the errors are computed using the Monte Carlo method. The calligraphic form factors represent the linear combinations of the defined form factors shown in Sec. IV.

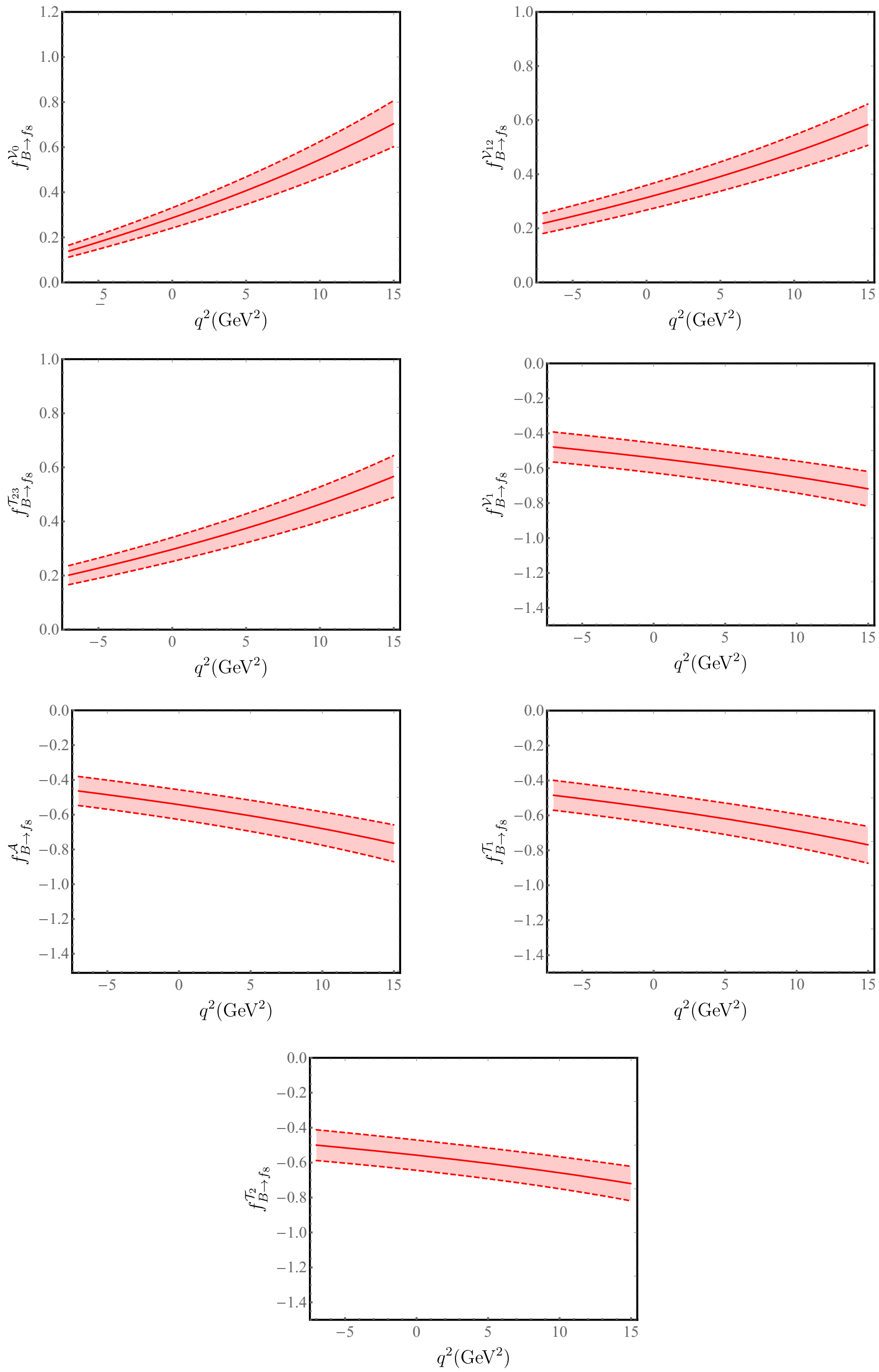

FIG. 12. The pink bands display the $q^2$ dependence of the $B \to f_8$ form factors employing the BCL parametrization, where the errors are computed using the Monte Carlo method. The calligraphic form factors represent the linear combinations of the defined form factors shown in Sec. IV.

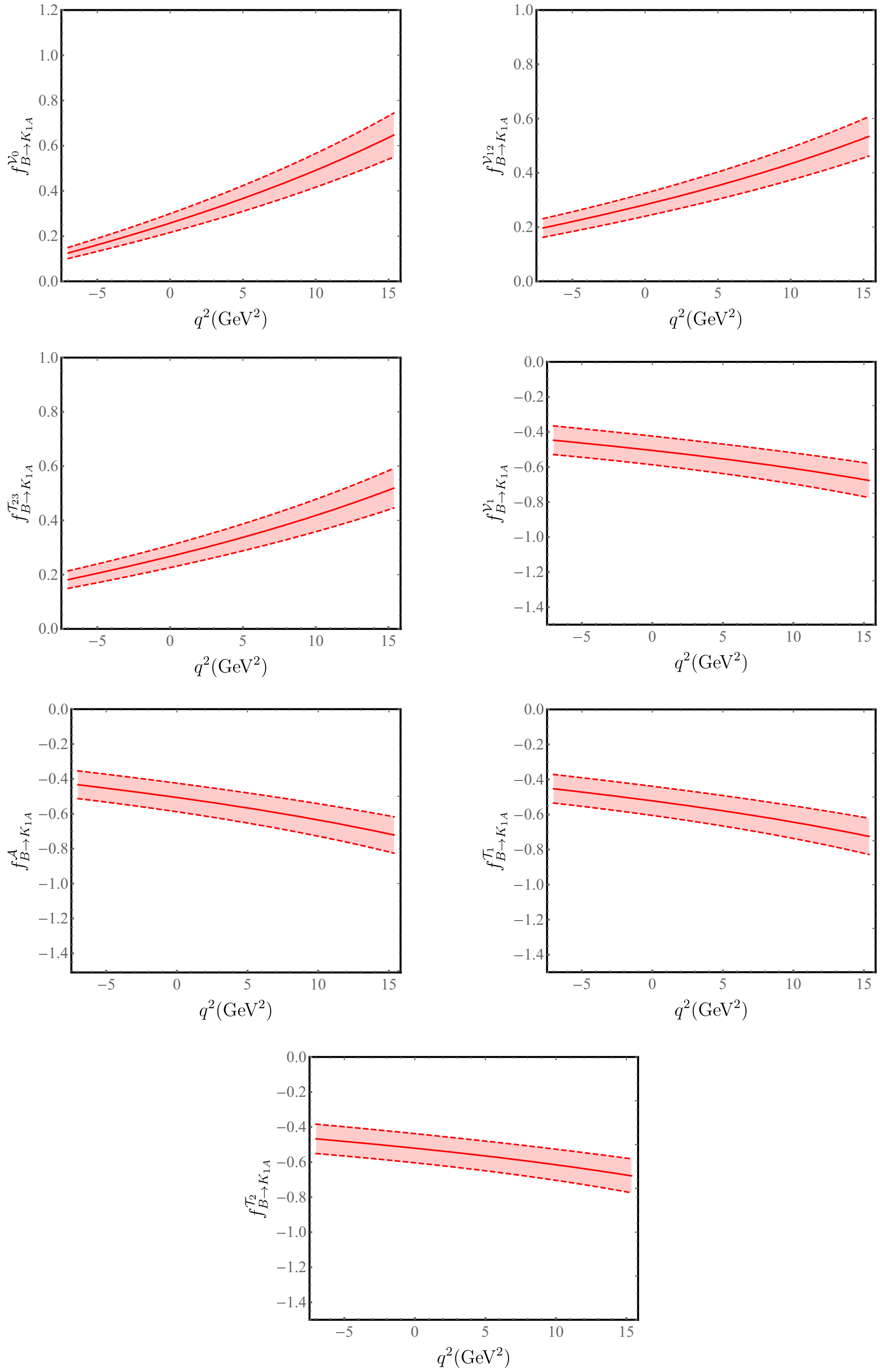

FIG. 13. The pink bands display the $q^2$ dependence of the $B \to K_{1A}$ form factors employing the BCL parametrization, where the errors are computed using the Monte Carlo method. The calligraphic form factors represent the linear combinations of the defined form factors shown in Sec. IV.

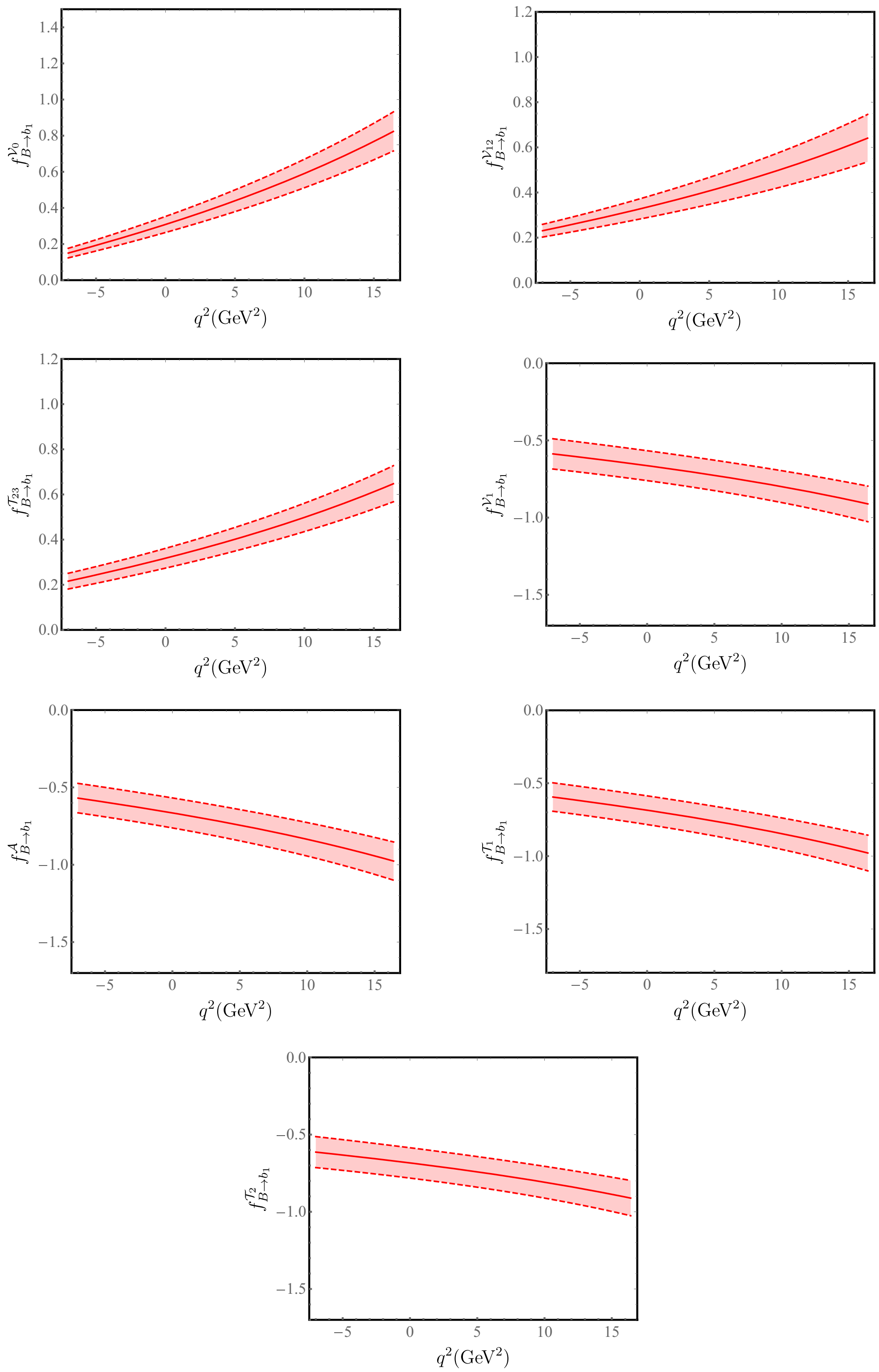

FIG. 14. The pink bands display the $q^2$ dependence of the $B \to b_1$ form factors employing the BCL parametrization, where the errors are computed using the Monte Carlo method. The calligraphic form factors represent the linear combinations of the defined form factors shown in Sec. IV.

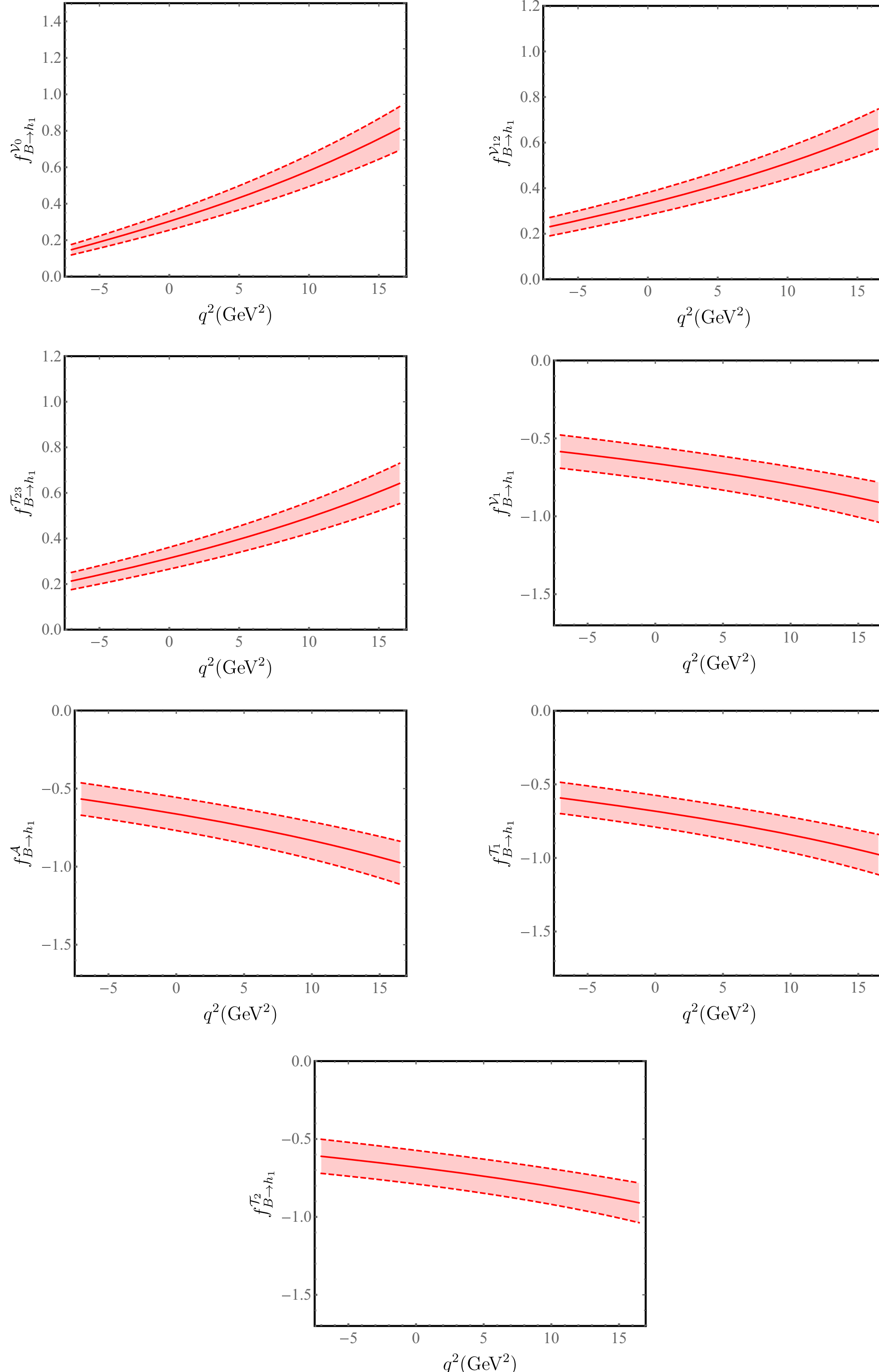

FIG. 15. The pink bands display the $q^2$ dependence of the $B \to h_1$ form factors employing the BCL parametrization, where the errors are computed using the Monte Carlo method. The calligraphic form factors represent the linear combinations of the defined form factors shown in Sec. IV.

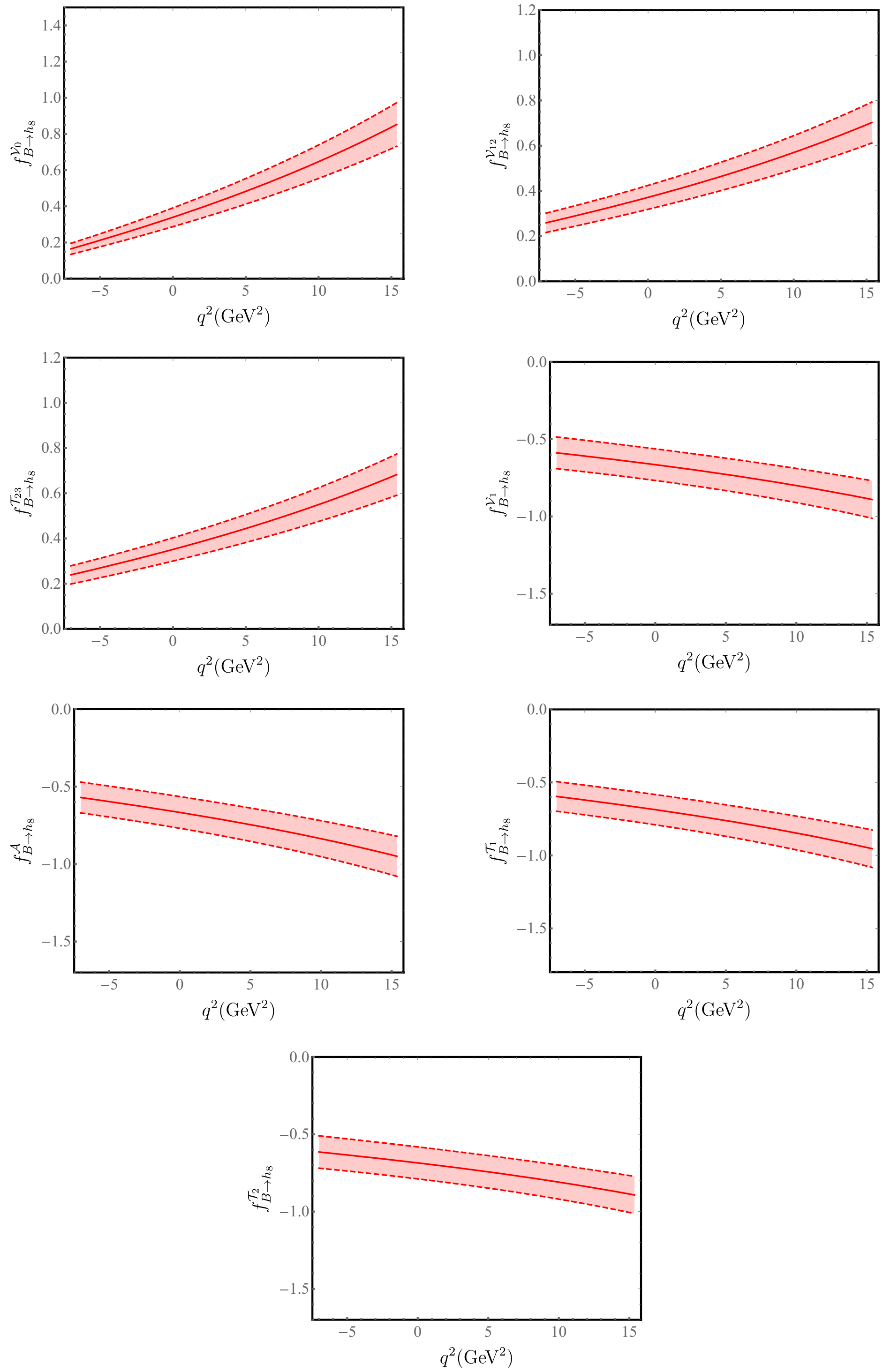

FIG. 16. The pink bands display the $q^2$ dependence of the $B \to h_8$ form factors employing the BCL parametrization, where the errors are computed using the Monte Carlo method. The calligraphic form factors represent the linear combinations of the defined form factors shown in Sec. IV.

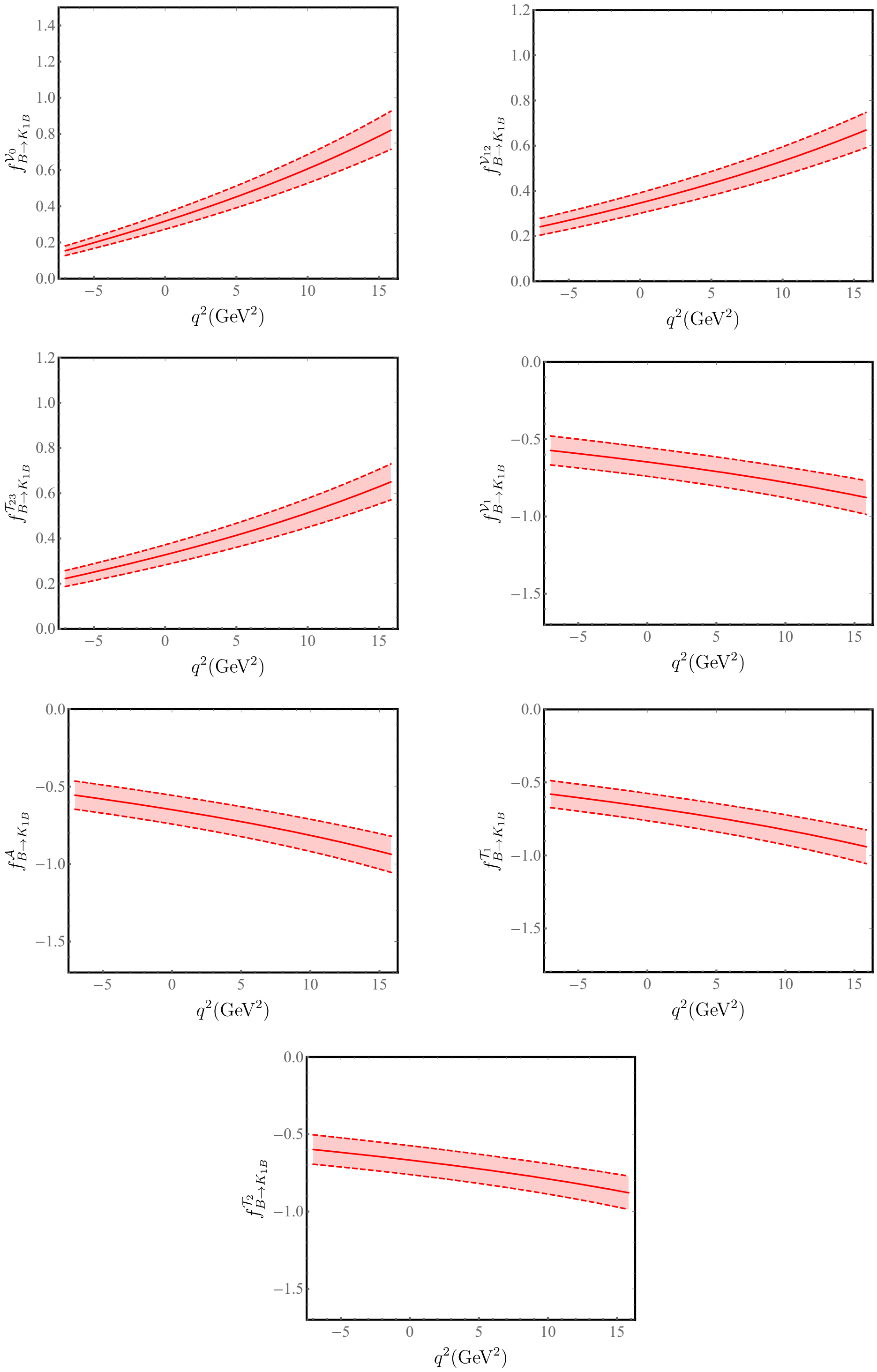

FIG. 17. The pink bands display the $q^2$ dependence of the $B \to K_{1B}$ form factors employing the BCL parametrization, where the errors are computed using the Monte Carlo method. The calligraphic form factors represent the linear combinations of the defined form factors shown in Sec. IV.

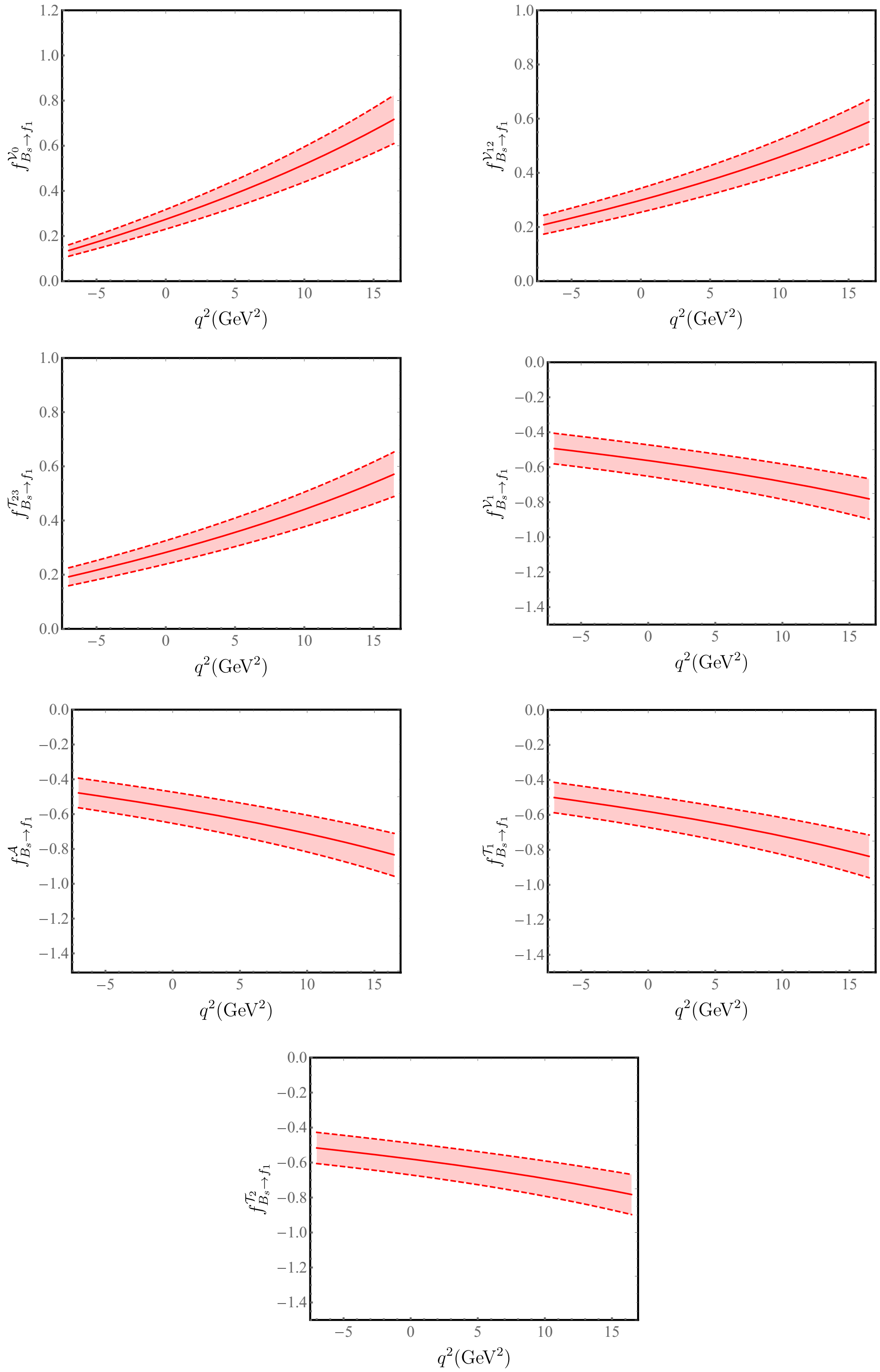

FIG. 18. The pink bands display the $q^2$ dependence of the $B_s \to f_1$ form factors employing the BCL parametrization, where the errors are computed using the Monte Carlo method. The calligraphic form factors represent the linear combinations of the defined form factors shown in Sec. IV.

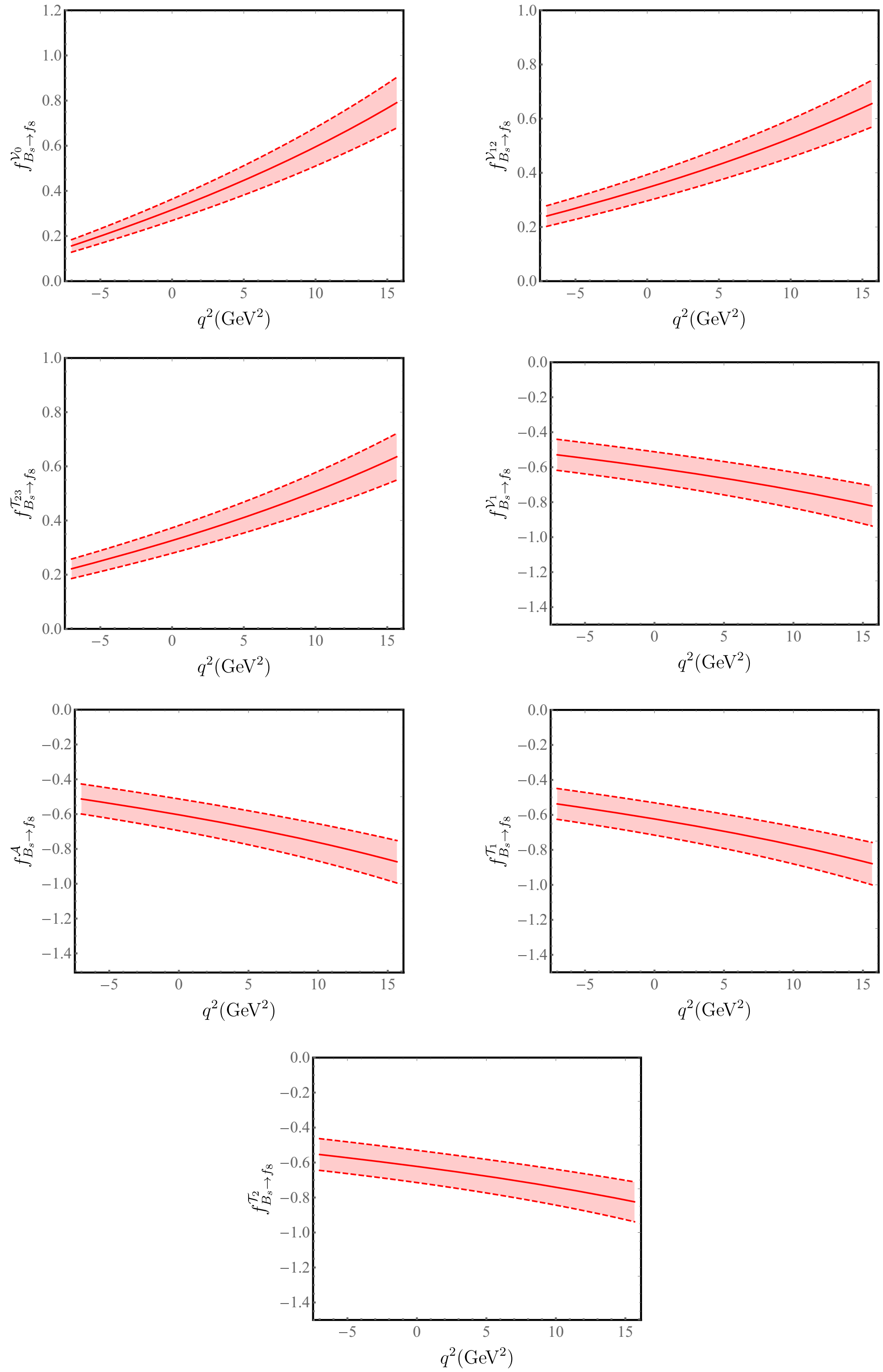

FIG. 19. The pink bands display the $q^2$ dependence of the $B_s \to f_8$ form factors employing the BCL parametrization, where the errors are computed using the Monte Carlo method. The calligraphic form factors represent the linear combinations of the defined form factors shown in Sec. IV.

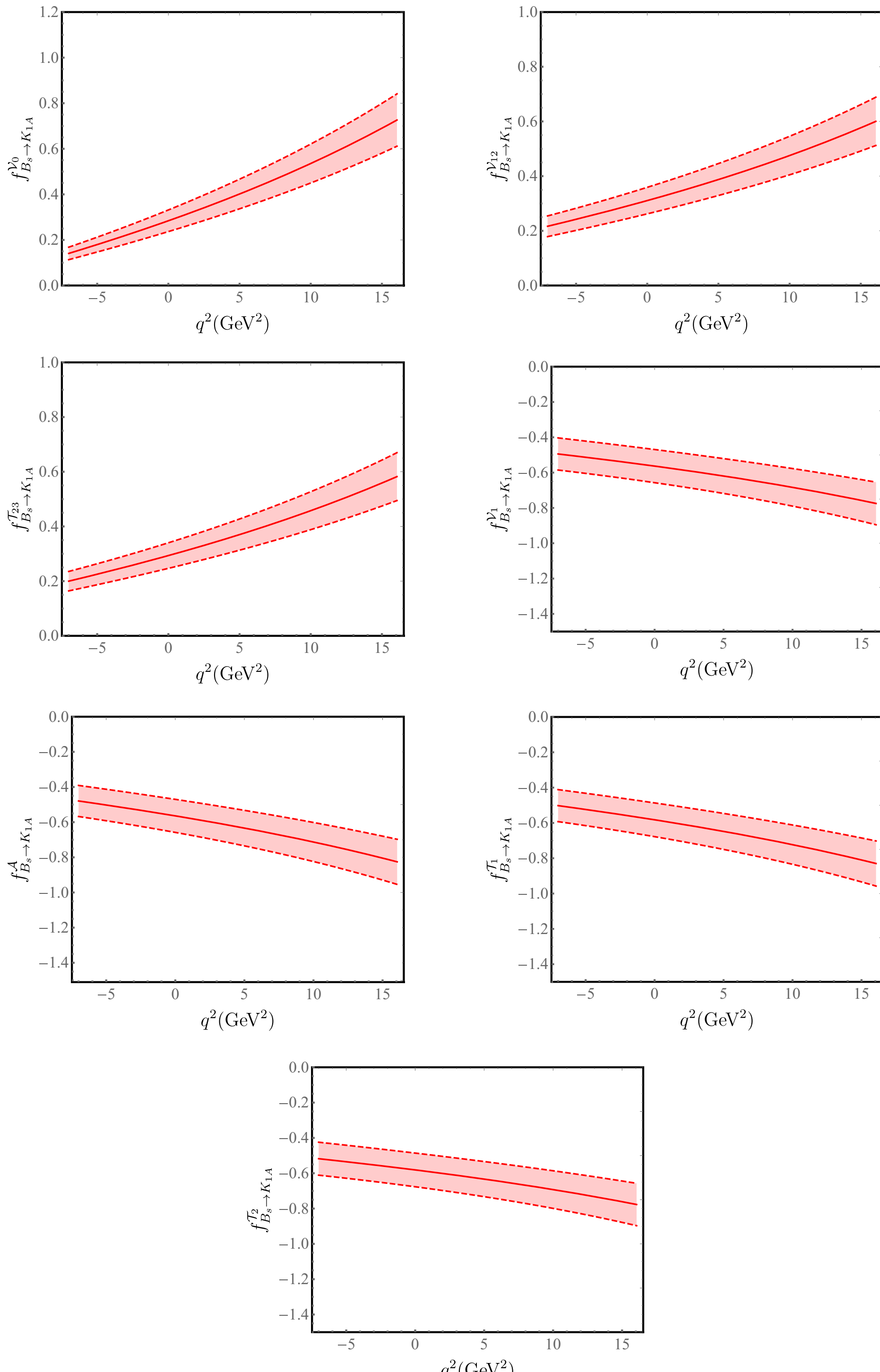

FIG. 20. The pink bands display the $q^2$ dependence of the $B_s \to K_{1A}$ form factors employing the BCL parametrization, where the errors are computed using the Monte Carlo method. The calligraphic form factors represent the linear combinations of the defined form factors shown in Sec. IV.

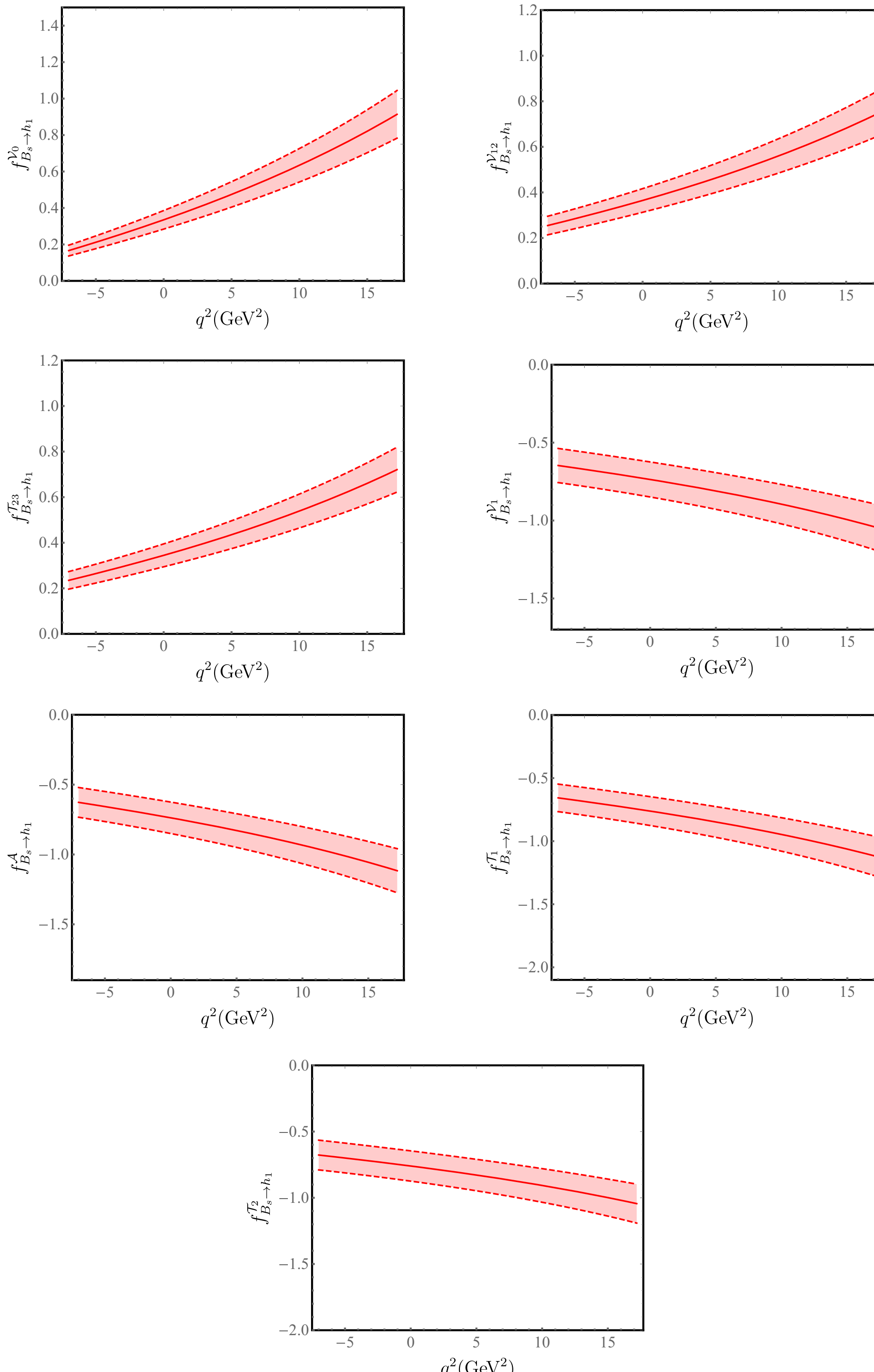

FIG. 21. The pink bands display the $q^2$ dependence of the $B_s \to h_1$ form factors employing the BCL parametrization, where the errors are computed using the Monte Carlo method. The calligraphic form factors represent the linear combinations of the defined form factors shown in Sec. IV.

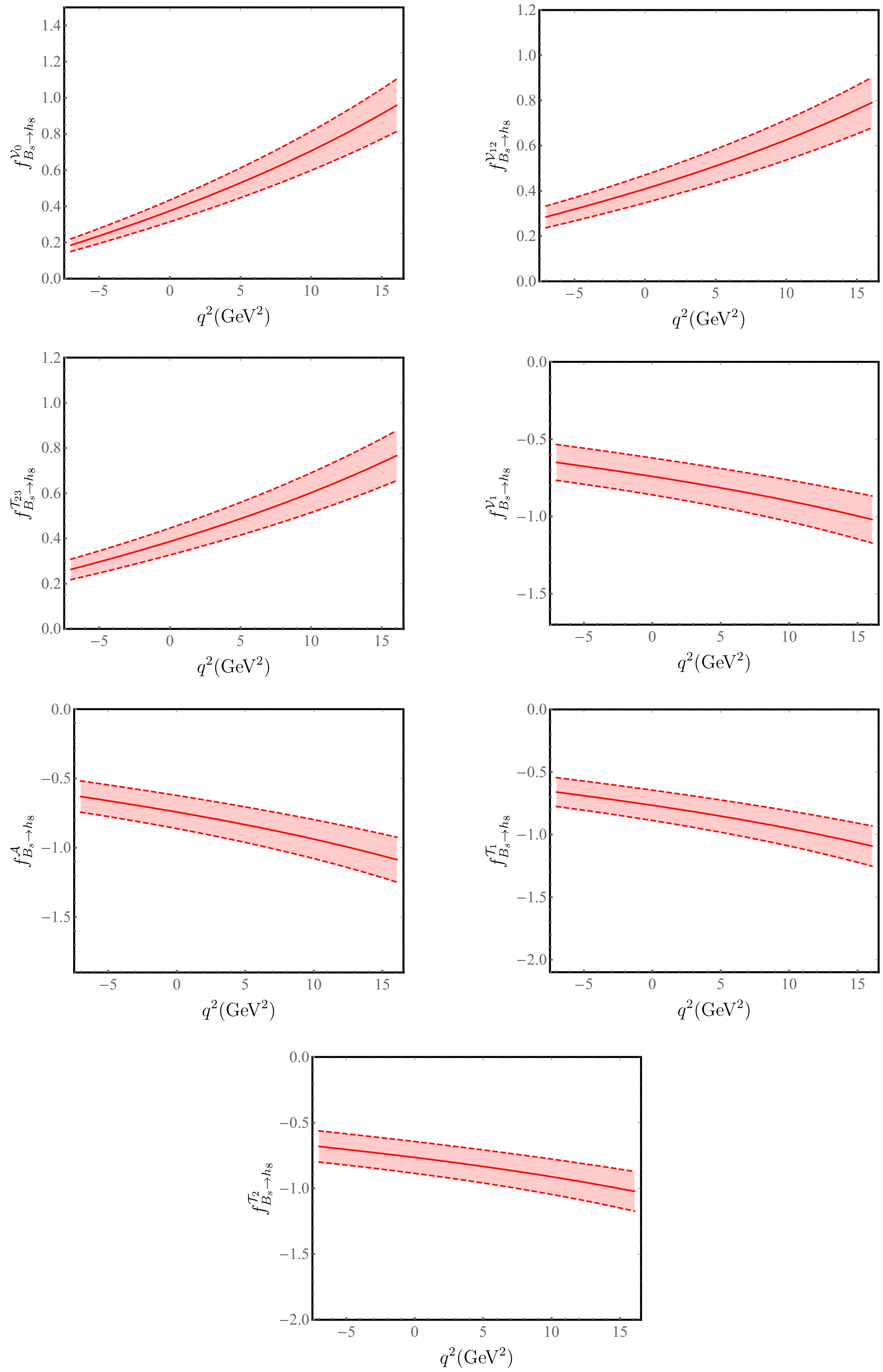

FIG. 22. The pink bands display the $q^2$ dependence of the $B_s \to h_8$ form factors employing the BCL parametrization, where the errors are computed using the Monte Carlo method. The calligraphic form factors represent the linear combinations of the defined form factors shown in Sec. IV.

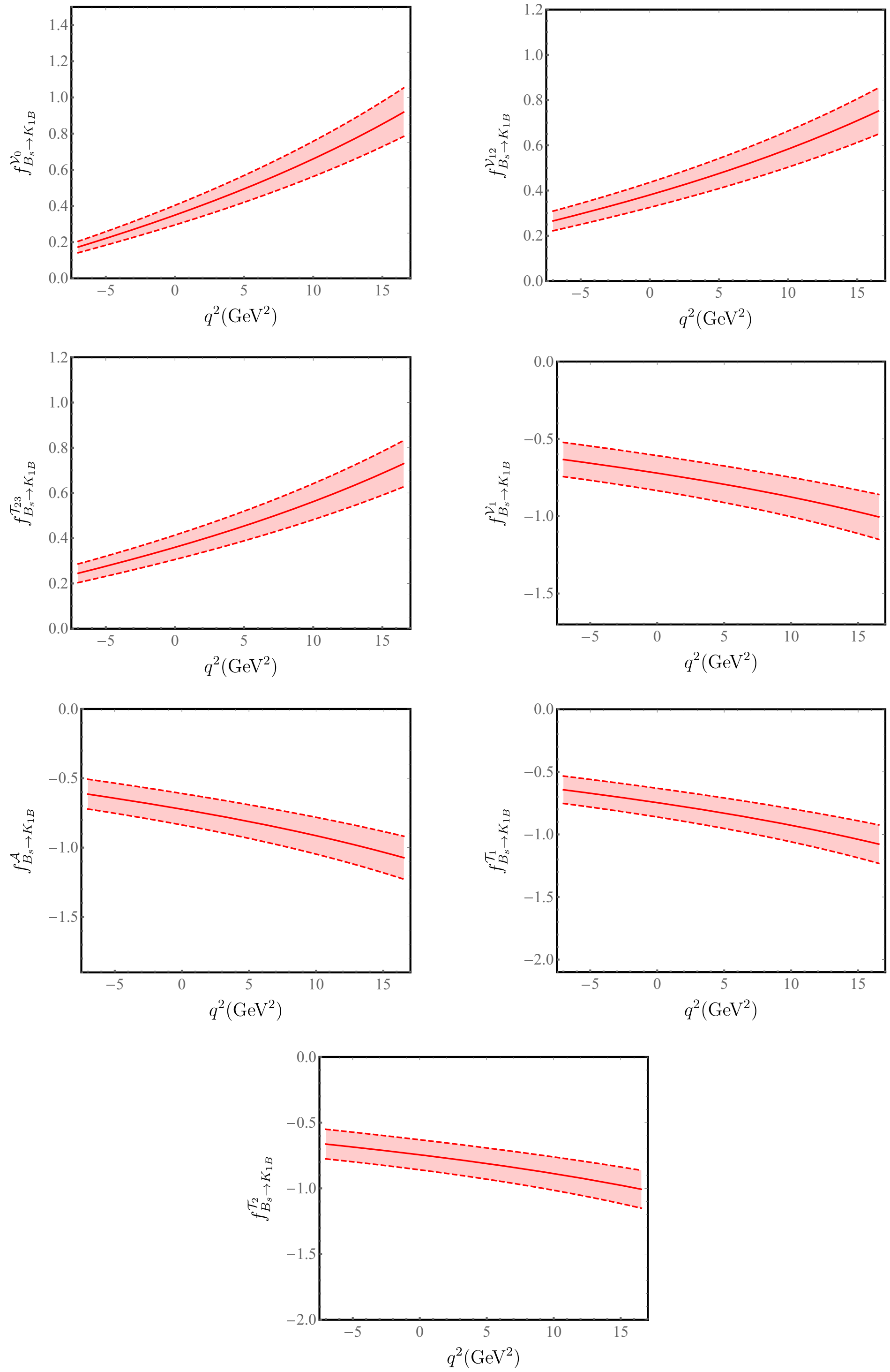

FIG. 23. The pink bands display the $q^2$ dependence of the $B_s \to K_{1B}$ form factors employing the BCL parametrization, where the errors are computed using the Monte Carlo method. The calligraphic form factors represent the linear combinations of the defined form factors shown in Sec. IV.